\documentclass[11pt,fleqn]{article}

\usepackage{a4}
\usepackage{amstext}
\usepackage{amsfonts}
\usepackage{amssymb}
\usepackage{color}
\usepackage{cite}
\usepackage{epsfig}
\usepackage{mathtools}
\usepackage{ulem}
\usepackage{caption}         
\usepackage{subcaption}   
\usepackage{framed}
\usepackage{multirow}
\usepackage{braket}
\usepackage{placeins}

\newcommand{\ltapprox}{\raisebox{-0.5ex}{$\,\stackrel{<}{\scriptstyle\sim}\,$}}

\begin{document}

% ********************
% ********************
% ********************
% ********************
% ********************

\begin{center}

{\huge \bf Hybrid static potential flux tubes}

{\huge \bf from SU(2) and SU(3) lattice gauge theory}

\vspace{0.5cm}

\textbf{Lasse M\"uller, Owe Philipsen, Christian Reisinger, Marc Wagner}

Goethe-Universit\"at Frankfurt am Main, Institut f\"ur Theoretische Physik, Max-von-Laue-Stra{\ss}e 1, D-60438 Frankfurt am Main, Germany

\vspace{0.5cm}

July 02, 2019

\end{center}

\begin{tabular*}{16cm}{l@{\extracolsep{\fill}}r} \hline \end{tabular*}

\vspace{-0.4cm}
\begin{center} \textbf{Abstract} \end{center}
\vspace{-0.4cm}

We compute chromoelectric and chromomagnetic flux densities for hybrid static potentials in SU(2) and SU(3) lattice gauge theory. In addition to the ordinary static potential with quantum numbers $\Lambda_\eta^\epsilon = \Sigma_g^+$, we present numerical results for seven hybrid static potentials corresponding to $\Lambda_\eta^{(\epsilon)} = \Sigma_u^+, \Sigma_g^-, \Sigma_u^-, \Pi_g, \Pi_u, \Delta_g, \Delta_u$, where the flux densities of five of them are studied for the first time in this work. We observe hybrid static potential flux tubes, which are significantly different from that of the ordinary static potential. They are reminiscent of vibrating strings, with localized peaks in the flux densities that can be interpreted as valence gluons. 

\begin{tabular*}{16cm}{l@{\extracolsep{\fill}}r} \hline \end{tabular*}

\thispagestyle{empty}

% ********************
% ********************
% ********************
% ********************
% ********************

\newpage

\setcounter{page}{1}

\section{Introduction}

The majority of mesons, i.e.\ hadrons with integer total angular momentum, are quark-antiquark pairs. It is, however, expected that some mesons, so-called exotic mesons, have a more complicated composition in terms of quarks and gluons. An important example are hybrid mesons, where gluons contribute to the quantum numbers $J^{P C}$ ($J$: total angular momentum; $P$: parity; $C$: charge conjugation) in a non-trivial way. In the quark model, where mesons are quark-antiquark pairs, quantum numbers are restricted to $P=(-1)^{L+1}$ and $C=(-1)^{L+S}$ with spin $S=0,1$ and orbital angular momentum $L=0,1,2,...$. Thus, mesons with $J^{PC} = 0^{+-} , 0^{--} , 1^{-+} , 2^{+-} , \ldots$, which are not allowed in the quark model, are obvious candidates for exotic mesons like hybrids. Moreover, a higher density of states than obtained by the quark model might also indicate hybrid mesons.

Experimentally observed examples, which could be hybrid mesons, are the $J^{PC} = 1^{-+}$ states $\pi_1(1400)$ and $\pi_1(1600)$. They could, however, also be tetraquarks, i.e.\ two quarks and two antiquarks without excited glue. For heavy-heavy mesons the situation seems to be even less clear. There are several exotic candidates, which could be hybrid mesons, but for none of them such an interpretation seems to be likely (see.\ e.g.\ the experimental review of exotic hadrons \cite{Olsen:2017bmm} and the discussion in section~VII.A of ref.\ \cite{Berwein:2015vca}). Thus, the search for gluonic excitations is an important part of the research program of current and future experiments, e.g.\ the GlueX experiment at the JLab accelerator or the PANDA experiment at the FAIR accelerator.

Also on the theoretical side there are many open questions concerning hybrid mesons (see e.g.\ the theoretical reviews \cite{Braaten:2014ita,Meyer:2015eta,Swanson:2015wgq,Lebed:2016hpi}). They are difficult to study, because in QCD total angular momentum $J$ and parity $P$ are not separately conserved for gluons on the one hand and for the quark-antiquark pair on the other hand. Only the overall $J^P$ are quantum numbers. For heavy hybrid mesons, e.g.\ composed of a $b$ and a $\bar{b}$ quark and gluons, a simplification and good approximation is to study the static limit. In that limit the quark positions are frozen, which allows to separate the treatment of gluons and quarks.

In this work we use SU(2) and SU(3) lattice gauge theory to study heavy hybrid mesons in the static limit. Since quite some time hybrid static potentials haven been computed by various groups, mainly with the intention to compute masses of heavy hybrid mesons using the Born-Oppenheimer approximation (see refs.\ \cite{Griffiths:1983ah,Campbell:1984fe,Campbell:1987nv,
Michael:1990az,Perantonis:1990dy,Juge:1997nc,Peardon:1997jr,
Juge:1997ir,Morningstar:1998xh,Michael:1998tr,
Juge:1999ie,Juge:1999aw,Michael:1999ge,Bali:2000vr,
Morningstar:2001nu,Juge:2002br,Michael:2003ai,Juge:2003qd,
Michael:2003xg,Bali:2003jq,Juge:2003ge,
Wolf:2014tta,Reisinger:2017btr,Reisinger:2018lne,Capitani:2018rox} and the recent review article \cite{Brandt:2016xsp}). We focus on a different problem, the computation of the gluonic flux densities for hybrid potential states, i.e.\ the structure of the flux tube, for several hybrid channels. While such flux tubes have been studied for the ordinary static potential using lattice gauge theory for quite some time (see refs.\ \cite{Fukugita:1983du,Flower:1985gs,
Wosiek:1987kx,DiGiacomo:1989yp,DiGiacomo:1990hc,
Cea:1992sd,Bali:1994de,Skala:1996ar,Bali:1997cp,
Cardoso:2013lla,Cea:2014uja,Cea:2015wjd,
Cea:2017ocq,Cea:2017bsa,Baker:2018mhw,
Baker:2018hyy}), this is a rather new direction for hybrid static potentials, where first results appeared only recently \cite{Bicudo:2018yhk,Mueller:2018fkg,Bicudo:2018jbb,Mueller:2018idu}. In this paper we substantially extend existing work by performing computations for seven hybrid static potential sectors characterized by quantum numbers $\Lambda_\eta^{(\epsilon)} = \Sigma_u^+,\Sigma_g^-,\Sigma_u^-,\Pi_g,\Pi_u,\Delta_g,\Delta_u$. Five of these sectors are studied for the first time, where preliminary results have been presented at a recent conference \cite{Mueller:2018idu}.

The paper is structured as follows. In section~\ref{SEC699} we discuss theoretical basics, including quantum numbers for hybrid static potentials, the construction of corresponding trial states and the computation of chromoelectric and chromomagnetic flux densities. Section~\ref{SEC549} contains a brief summary of our lattice setup. In section~\ref{SEC698} we present our numerical results. We start with a discussion of systematic errors and symmetries, before showing and interpreting our main results, the chromoelectric and chromomagnetic flux densities for the seven hybrid static potential sectors $\Lambda_\eta^{(\epsilon)} = \Sigma_u^+,\Sigma_g^-,\Sigma_u^-,\Pi_g,\Pi_u,\Delta_g,\Delta_u$. In section~\ref{SEC697} we conclude with a short summary and an outlook.

% ********************
% ********************
% ********************
% ********************
% ********************

\newpage

\section{\label{SEC699}Hybrid static potentials and flux tubes}

% ********************
% ********************
% ********************

\subsection{\label{SEC477}Hybrid static potential quantum numbers and trial states}

A hybrid static potential is the potential of a static quark $Q$ and a static antiquark $\bar{Q}$, where the gluons form non-trivial structures and, thus, contribute to the quantum numbers. Such potentials can be computed from temporal correlation functions of hybrid static potential trial states. After replacing the static quark operators by corresponding propagators, these correlation functions are similar to Wilson loops. Instead of straight spatial Wilson lines there are, however, parallel transporters with more complicated spatial structures. For a detailed discussion of such correlation functions see e.g.\ our recent work \cite{Capitani:2018rox}, where we have carried out a precision computation of hybrid static potentials using SU(3) lattice gauge theory.

In the following we consider a static quark and a static antiquark located at positions \\ $\mathbf{r}_Q=(0,0,+r/2)$ and $\mathbf{r}_{\bar{Q}}=(0,0,-r/2)$, respectively, i.e.\ they are separated along the $z$ axis. We omit the $x$ and the $y$ coordinates, i.e.\ $Q(+r/2) \equiv Q(0,0,+r/2)$ and $\bar{Q}(-r/2) \equiv \bar{Q}(0,0,-r/2)$.

Hybrid static potentials can be characterized by the following quantum numbers:
\begin{itemize}
\item $\Lambda = 0,1,2,\ldots$, the absolute value of the total angular momentum with respect to the $Q \bar{Q}$ separation axis, i.e.\ with respect to the $z$ axis.

\item $\eta = +,-$, the eigenvalue corresponding to the operator $\mathcal{P} \circ \mathcal{C}$, i.e.\ the combination of parity and charge conjugation.

\item $\epsilon = +,-$, the eigenvalue corresponding to the operator $\mathcal{P}_x$, which denotes the spatial reflection along the $x$ axis (an axis perpendicular to the $Q \bar{Q}$ separation axis).
\end{itemize}
It is common convention to write $\Lambda = \Sigma,\Pi,\Delta$ instead of $\Lambda = 0,1,2$ and $\eta = g,u$ (``gerade'', ``ungerade'') instead of $\eta = +,-$. Note that for absolute total angular momentum $\Lambda \geq 1$ the spectrum is degenerate with respect to $\epsilon = +$ and $\epsilon = -$, i.e.\ there are pairs of identical hybrid static potentials. Thus, the labeling of hybrid static potentials is typically $\Lambda^{\epsilon}_{\eta}$ for $\Lambda = 0 = \Sigma$ and $\Lambda_{\eta}$ for $\Lambda \geq 1$.

In \cite{Capitani:2018rox} we discussed hybrid static potential creation operators and trial states both in the continuum and in lattice gauge theory in detail and performed a comprehensive optimization of these operators in SU(3) lattice gauge theory. In this paper we use the information obtained during this optimization to define suitable hybrid static potential creation operators both for SU(2) and SU(3) lattice gauge theory. These operators are important building blocks of the 2-point and 3-point functions, which need to be computed for the investigation of hybrid static potential flux tubes (see section~\ref{SEC478}).

Our trial states, which have definite quantum numbers $\Lambda_\eta^\epsilon$, are
\begin{eqnarray}
| \Psi_{\Lambda_\eta^\epsilon}(r) \rangle \ \ = \ \ \bar{Q}(-r/2) a_{S;\Lambda_\eta^\epsilon}(-r/2,+r/2) Q(+r/2) \ket{\Omega}
\end{eqnarray}
with creation operators
\begin{eqnarray}
\nonumber & & \hspace{-0.7cm} a_{S;\Lambda_\eta^\epsilon}(-r/2,+r/2) \ \ = \\
\nonumber & & = \ \ \frac{1}{4} \sum_{k=0}^3 \textrm{exp}\bigg(\frac{i \pi \Lambda k}{2}\bigg) R\bigg(\frac{\pi k}{2}\bigg)
\Big(U(-r/2,r_1) \Big(S(r_1,r_2) + \epsilon S_{\mathcal{P}_x}(r_1,r_2)\Big) U(r_2,+r/2) + \\
\label{EQN011} & & \hspace{0.675cm} U(-r/2,-r_2) \Big(\eta S_{\mathcal{P} \circ \mathcal{C}}(-r_2,-r_1) + \eta \epsilon S_{(\mathcal{P} \circ \mathcal{C}) \mathcal{P}_x}(-r_2,-r_1)\Big) U(-r_1,+r/2)\Big) .
\end{eqnarray}
$U(-r/2,r_1) S(r_1,r_2) U(r_2,+r/2)$ is a product of link variables connecting the quark and the antiquark in a gauge invariant way, where both $U(-r/2,r_1)$ and $U(r_2,+r/2)$ are straight lines on the $z$ axis, while $S(r_1,r_2)$ has a more complicated shape. $S_{\mathcal{P} \circ \mathcal{C}}(-r_2,-r_1)$ is the spatial reflection of $S(r_1,r_2)$ combined with charge conjugation, $S_{\mathcal{P}_x}(r_1,r_2)$ is the spatial reflection of $S(r_1,r_2)$ along the $x$ axis and $S_{(\mathcal{P} \circ \mathcal{C}) \mathcal{P}_x}(-r_2,-r_1)$ is the combination of both operations.

$U(-r/2,r_1) S(r_1,r_2) U(r_2,+r/2)$ has been optimized in ref.\ \cite{Capitani:2018rox}, such that the overlap of $| \Psi_{\Lambda_\eta^\epsilon}(r) \rangle$ to the ground state in the $\Lambda_\eta^\epsilon$ sector is rather large. In contrast to ref.\ \cite{Capitani:2018rox} we use in this work only a single operator $S$ for each $\Lambda_\eta^\epsilon$ sector, not a linear combination obtained by a variational analysis. This reduces computation time to a feasible level, while the suppression of excited states in the 2-point and 3-point functions is still sufficiently strong. $U(-r/2,r_1) S(r_1,r_2) U(r_2,+r/2)$ is different for each of the eight $\Lambda_\eta^\epsilon$ sectors as well as for the two $Q \bar{Q}$ separations $r = 6 \, a , 10 \, a$ considered. For the $\Sigma_g^+$ sector, i.e.\ the ordinary static potential, it is just a straight line, while for $\Lambda_\eta^\epsilon = \Sigma_u^+,\Sigma_g^-,\Sigma_u^-,\Pi_g^-,\Pi_u^+,\Delta_g^\pm,\Delta_u^\pm$ details are collected in Table~\ref{TAB001}. For each $\Lambda_\eta^\epsilon$ sector we take that operator from the set of three to four operators we optimized in ref.\ \cite{Capitani:2018rox}, which minimizes the effective potential at $t = a$. Thus Table~\ref{TAB001} contains that part of the information shown in Table~1 to Table~7 of ref.\ \cite{Capitani:2018rox}, which is relevant in the context of this work.

\begin{table}[p]
\begin{center}
\includegraphics[width=14.0cm]{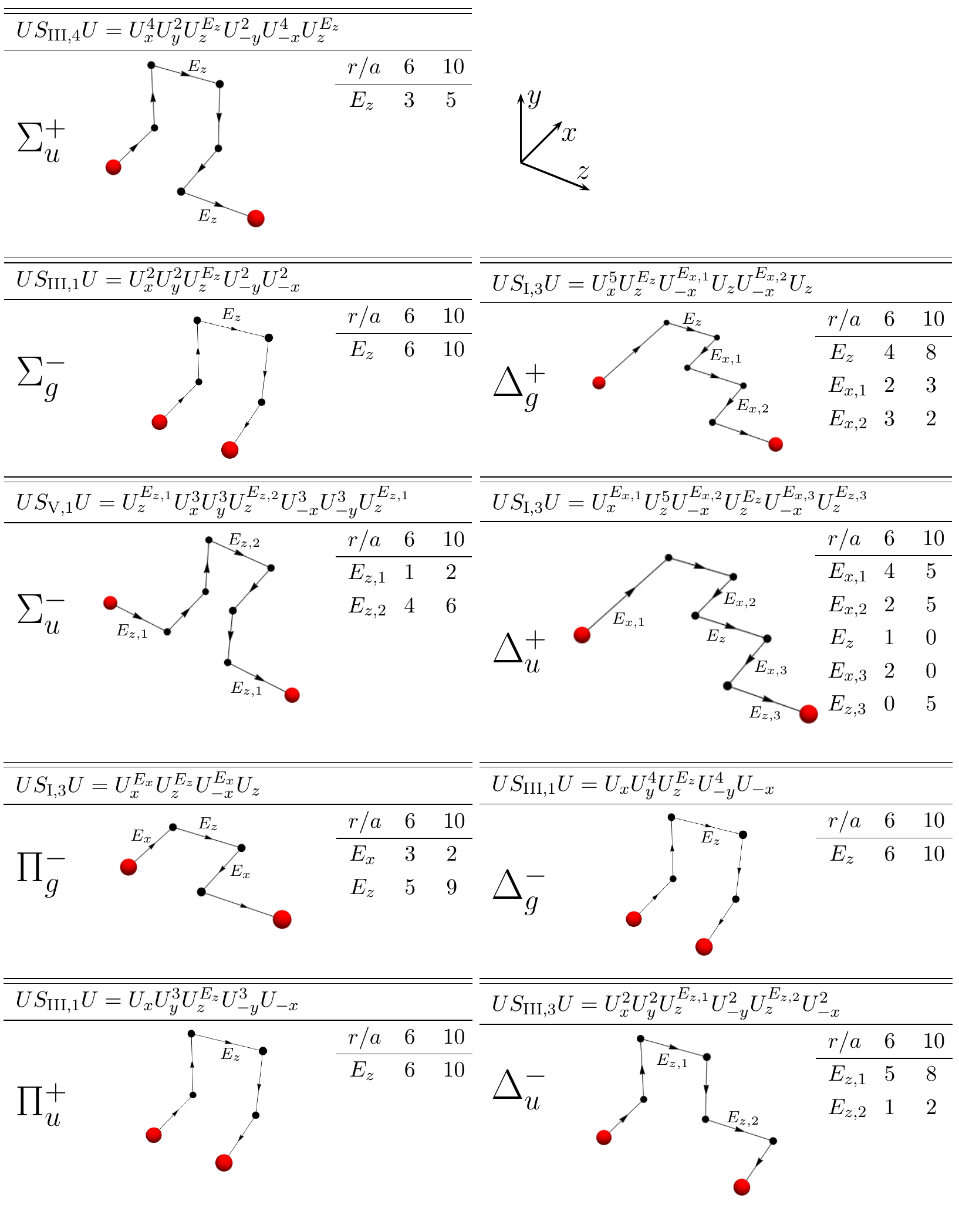}
\end{center}
\caption{\label{TAB001}Optimized creation operators for $\Lambda_\eta^\epsilon =\Sigma_u^+,\Sigma_g^-,\Sigma_u^-,\Pi_g^-,\Pi_u^+,\Delta_g^\pm,\Delta_u^\pm$. The notation in the caption of each of the tables follows ref.\ \protect\cite{Capitani:2018rox}. Note that, even though the $\Pi_\eta$ and the $\Delta_\eta$ hybrid potentials are degenerate with respect to $\epsilon$, the construction of creation operators via eq.\ (\ref{EQN011}) is not independent of $\epsilon$. One can obtain an optimized $\Pi_\eta^-$ operator from an optimized $\Pi_\eta^+$ operator by applying a $\pi/2$ rotation with respect to the $z$ axis. For $\Delta_\eta^\epsilon$ operators there is no analogous simple prescription. Therefore, we provide four different optimized $\Delta_\eta^\epsilon$ operators.}
\end{table}

% ********************
% ********************
% ********************

\subsection{\label{SEC478}Expectation values of squared field strength components}

The energy density of the gluon field is
\begin{eqnarray}
\label{EQN488} \mathcal{E}(\mathbf{x}) \ \ = \ \ \frac{1}{8 \pi} \bigg(\sum_{j=x,y,z} \sum_a E_j^a(\mathbf{x}) E_j^a(\mathbf{x}) + \sum_{j=x,y,z} \sum_a B_j^a(\mathbf{x}) B_j^a(\mathbf{x})\bigg) ,
\end{eqnarray}
where $E_j^a(\mathbf{x})$ and $B_j^a(\mathbf{x})$ denote the components of the chromoelectric and chromomagnetic field strengths with spatial indices $j$ and color indices $a$ ($a = 1,\ldots,3$ for gauge group SU(2) and $a = 1,\ldots,8$ for gauge group SU(3)). The main goal of this work is to compute the expectation values of the six gauge invariant terms $F_j^2(\mathbf{x}) = \sum_a F_j^a(\mathbf{x}) F_j^a(\mathbf{x})$ (no sum over $j$; $F_j^a = E_j^a$ or $F_j^a = B_j^a$) contributing to eq.\ (\ref{EQN488}) for states with a static quark-antiquark pair and quantum numbers $\Lambda_\eta^\epsilon$. These chromoelectric and chromomagnetic flux densities provide information about the shapes of hybrid static potential flux tubes and the gluonic energy distributions inside heavy-heavy hybrid mesons.

To compute the flux densities, we need the following quantities:
\begin{itemize}
\item 2-point correlation functions:
\begin{eqnarray}
\nonumber & & \hspace{-0.7cm} W_{\Lambda_\eta^\epsilon}(r,t_2,t_0) \ \ = \ \ \langle \Psi_{\Lambda_\eta^\epsilon}(r,t_2) | \Psi_{\Lambda_\eta^\epsilon}(r,t_0) \rangle \ \ = \\
\label{EQN711} & & = \ \ \sum_m \Big|\langle \Psi_{\Lambda_\eta^\epsilon}(r) | m_{\Lambda_\eta^\epsilon}(r) \rangle\Big|^2 e^{-(E_{m,\Lambda_\eta^\epsilon}(r) - E_\Omega) (t_2-t_0)} ,
\end{eqnarray}
where $t_2 > t_0$ and $| m_{\Lambda_\eta^\epsilon}(r) \rangle$ denotes the $m$-th energy eigenstate with a static quark $Q$ and a static antiquark $\bar{Q}$ at positions $(0,0,+r/2)$ and $(0,0,-r/2)$ and quantum numbers $\Lambda_\eta^\epsilon$. $E_{m,\Lambda_\eta^\epsilon}(r)$ is the corresponding energy eigenvalue, where \\ $E_{0,\Lambda_\eta^\epsilon}(r) < E_{1,\Lambda_\eta^\epsilon}(r) < E_{2,\Lambda_\eta^\epsilon}(r) < \ldots$ The static potential with quantum numbers $\Lambda_\eta^\epsilon$ is defined as $V_{\Lambda_\eta^\epsilon}(r) = E_{0,\Lambda_\eta^\epsilon}(r) - E_\Omega$.

\item 3-point correlation functions:
\begin{eqnarray}
\nonumber & & \hspace{-0.7cm} C_{\Lambda_\eta^\epsilon,F_j^2}(r,t_2,t_0;\mathbf{x},t_1) \ \ = \ \ \langle \Psi_{\Lambda_\eta^\epsilon}(r,t_2) | F_j^2(\mathbf{x},t_1) | \Psi_{\Lambda_\eta^\epsilon}(r,t_0) \rangle \ \ = \\
\nonumber & & = \ \ \sum_{m,n}
\langle \Psi_{\Lambda_\eta^\epsilon}(r) | m_{\Lambda_\eta^\epsilon}(r) \rangle
\langle m_{\Lambda_\eta^\epsilon}(r) | F_j^2(\mathbf{x}) | n_{\Lambda_\eta^\epsilon}(r) \rangle
\langle n_{\Lambda_\eta^\epsilon}(r) | \Psi_{\Lambda_\eta^\epsilon}(r) \rangle
e^{-(E_{m,\Lambda_\eta^\epsilon}(r) - E_\Omega) (t_2-t_1)} \\
 & & \hspace{0.675cm} e^{-(E_{n,\Lambda_\eta^\epsilon}(r) - E_\Omega) (t_1-t_0)} ,
\end{eqnarray}
where $t_2 > t_1 > t_0$.

\item Vacuum expectation values:
\begin{eqnarray}
\label{EQN713} \langle \Omega | F_j^2 | \Omega \rangle .
\end{eqnarray}
\end{itemize}
These quantities can be combined to expressions for the expectation values of $E_j^2(\mathbf{x})$ and $B_j^2(\mathbf{x})$ for static potential states with quantum numbers $\Lambda_\eta^\epsilon$ with the vacuum expectation value subtracted:
\begin{eqnarray}
\nonumber & & \hspace{-0.7cm} \Delta F_{j,\Lambda_\eta^\epsilon}^2(r;\mathbf{x}) \ \ = \ \ \langle 0_{\Lambda_\eta^\epsilon}(r) | F_j^2(\mathbf{x}) | 0_{\Lambda_\eta^\epsilon}(r) \rangle - \langle \Omega | F_j^2 | \Omega \rangle \ \ = \\
\label{EQN532} & & = \ \ \lim_{t_2-t_1,t_1-t_0 \rightarrow \infty} \underbrace{\frac{C_{\Lambda_\eta^\epsilon,F_j^2}(r,t_2,t_0;\mathbf{x},t_1)}{W_{\Lambda_\eta^\epsilon}(r,t_2,t_0)} - \langle \Omega | F_j^2 | \Omega \rangle}_{= \Delta F_{\textrm{eff};j,\Lambda_\eta^\epsilon}^2(r,t_2,t_0;\mathbf{x},t_1)}
\end{eqnarray}
(see also ref.\ \cite{Fukugita:1983du}, where this quantity was first defined and used to study flux densities for the ordinary static potential with quantum numbers $\Lambda_\eta^\epsilon = \Sigma_g^+$).

\begin{figure}[htb]
\begin{center}
\includegraphics[width=15.8cm]{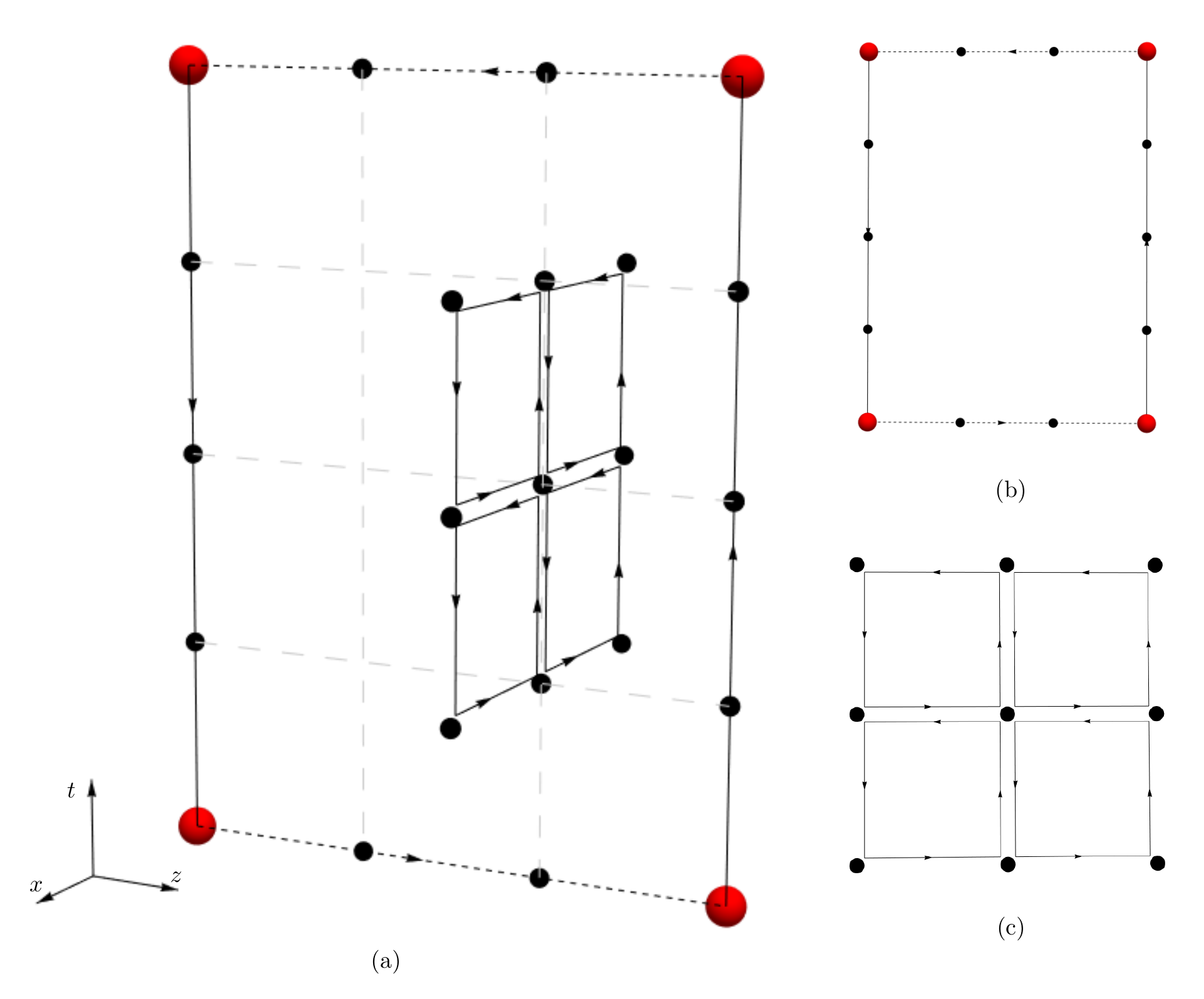}
\end{center}
\caption{\label{FIG020}Graphical illustration of lattice field theory quantities needed to determine $\Delta E_{j,\Lambda_\eta^\epsilon}^2(r;\mathbf{x})$ and $\Delta B_{j,\Lambda_\eta^\epsilon}^2(r;\mathbf{x})$. Red spheres, black dots, black solid lines and black dashed lines represent static (anti)quarks, lattice sites, gauge links and operators $a_{S;\Lambda_\eta^\epsilon}$, respectively. \textbf{(a)}~$\tilde{W}(r,t_2,t_0) \cdot P_{0 x}(\mathbf{x},t_1)$ for $r = 3 \, a$, $t_2 - t_0 = 4 \, a$, $t_1 = (t_2-t_0)/2$ (grey dashed lines parallel to the $t$ axis and the $z$ axis are drawn to guide the eye). \textbf{(b)}~The corresponding Wilson loop $\tilde{W}(r,t_2,t_0)$. \textbf{(c)}~The symmetrized plaquette $P_{\mu \nu}$.}
\end{figure}

The right hand side of eq.\ (\ref{EQN532}) can be evaluated using Euclidean lattice gauge theory path integrals,
\begin{eqnarray}
\label{EQN188} & & \hspace{-0.7cm} \Delta E_{\textrm{eff};j,\Lambda_\eta^\epsilon}^2(r,t_2,t_0;\mathbf{x},t_1) \ \ = \ \ +\bigg(\frac{\langle \tilde{W}(r,t_2,t_0) \cdot P_{0 j}(\mathbf{x},t_1) \rangle_U}{\langle \tilde{W}(r,t_2,t_0) \rangle_U} - \langle P_{0 j} \rangle_U\bigg) \\
\label{EQN189} & & \hspace{-0.7cm} \Delta B_{\textrm{eff};j,\Lambda_\eta^\epsilon}^2(r,t_2,t_0;\mathbf{x},t_1) \ \ = \ \ -\bigg(\frac{\langle \tilde{W}(r,t_2,t_0) \cdot |\epsilon_{j k l} / 2| P_{k l}(\mathbf{x},t_1) \rangle_U}{\langle \tilde{W}(r,t_2,t_0) \rangle_U} - \langle |\epsilon_{j k l} / 2| P_{k l} \rangle_U\bigg) ,
\end{eqnarray}
where $\langle \ldots \rangle_U$ denotes the path integral expectation value and
\begin{eqnarray}
\nonumber & & \hspace{-0.7cm} \tilde{W}(r,t_2,t_0) \ \ = \\
\label{EQN764} & & = \ \ \textrm{Tr}\Big(
  a_{S;\Lambda_\eta^\epsilon}(-r/2,+r/2;t_0)
  U(+r/2;t_0,t_2)
  \Big(a_{S;\Lambda_\eta^\epsilon}(-r/2,+r/2;t_2)\Big)^\dagger
  U(-r/2;t_2,t_0)
\Big)
\end{eqnarray}
(for $\Lambda_\eta^\epsilon = \Sigma_g^+$, i.e.\ the ordinary static potential, $\tilde{W}(r,t_2,t_0)$ is the standard Wilson loop). $P_{\mu \nu}$ is a symmetrized plaquette in the $\mu$-$\nu$ plane, also denoted as clover leaf. Eqs.\ (\ref{EQN188}) and (\ref{EQN189}) as well as the clover leaf are illustrated in Figure~\ref{FIG020}.

% ********************
% ********************
% ********************

\subsection{\label{SEC455}Angular dependence of flux densities}

As discussed in section~\ref{SEC477}, for absolute total angular momentum $\Lambda \geq 1$ the spectrum is degenerate with respect to $\epsilon$, i.e.\ $V_{\Lambda_\eta^+}(r) = V_{\Lambda_\eta^-}(r)$. In other words, the states $| 0_{\Lambda_\eta^+}(r) \rangle$ and $| 0_{\Lambda_\eta^-}(r) \rangle$ have the same energy. Their flux densities $\Delta F_{j,\Lambda_\eta^+}^2(r;\mathbf{x})$ and $\Delta F_{j,\Lambda_\eta^-}^2(r;\mathbf{x})$ are, however, not identical, but related by rotations, as we discuss in the following.

One can show that under rotations around the $z$ axis with angle $\alpha$ the states $| 0_{\Lambda_\eta^\epsilon}(r) \rangle$ transform according to
\begin{eqnarray}
\label{EQN239} R_z(\alpha) | 0_{\Lambda_\eta^\pm}(r) \rangle \ \ = \ \ \cos(\alpha \Lambda) | 0_{\Lambda_\eta^\pm}(r) \rangle + i \sin(\alpha \Lambda) | 0_{\Lambda_\eta^\mp}(r) \rangle
\end{eqnarray}
(see appendix~\ref{APP600}), while the field strength components transform as
\begin{eqnarray}
\label{EQN240} R_z^\dagger(\alpha) F_j^a(\mathbf{x}) R_z(\alpha) \ \ = \ \ R_{j k}(-\alpha) F_k^a(R(-\alpha) \mathbf{x}) \ \ = \ \ R_{j k}(-\alpha) F_k^a(\mathbf{x}_{-\alpha}) .
\end{eqnarray}
$R(\alpha)$ denotes the corresponding standard $3 \times 3$ rotation matrix, i.e.\
\begin{eqnarray}
R(\alpha) \ \ = \ \ \left(\begin{array}{ccc}
+\cos(\alpha) & -\sin(\alpha) & 0 \\
+\sin(\alpha) & +\cos(\alpha) & 0 \\
0 & 0 & 1
\end{array}\right) ,
\end{eqnarray}
and we have defined $\mathbf{x}_{-\alpha} = R(-\alpha) \mathbf{x}$. Now we consider the rotated flux densities \\ $\langle 0_{\Lambda_\eta^\pm}(r) | R_z^\dagger(\alpha) F_j^2(\mathbf{x}) R_z(\alpha) | 0_{\Lambda_\eta^\pm}(r) \rangle - \langle \Omega | F_j^2 | \Omega \rangle$ and rewrite them in two different ways, using first eq.\ (\ref{EQN239}),
\begin{eqnarray}
\nonumber & & \hspace{-0.7cm} \langle 0_{\Lambda_\eta^\pm}(r) | R_z^\dagger(\alpha) F_j^2(\mathbf{x}) R_z(\alpha) | 0_{\Lambda_\eta^\pm}(r) \rangle - \langle \Omega | F_j^2 | \Omega \rangle \ \ = \\
\nonumber & & = \ \ \cos^2(\alpha \Lambda) \Delta F_{j,\Lambda_\eta^\pm}^2(r;\mathbf{x}) + \sin^2(\alpha \Lambda) \Delta F_{j,\Lambda_\eta^\mp}^2(r;\mathbf{x}) \\
\label{EQN560} & & \hspace{0.675cm} + i \cos(\alpha \Lambda) \sin(\alpha \Lambda) \Big(\langle 0_{\Lambda_\eta^\pm}(r) | F_j^2(\mathbf{x}) | 0_{\Lambda_\eta^\mp}(r) \rangle - \langle 0_{\Lambda_\eta^\mp}(r) | F_j^2(\mathbf{x}) | 0_{\Lambda_\eta^\pm}(r) \rangle\Big) ,
\end{eqnarray}
then eq.\ (\ref{EQN240}),
\begin{eqnarray}
\nonumber & & \hspace{-0.7cm} \langle 0_{\Lambda_\eta^\pm}(r) | R_z^\dagger(\alpha) F_j^2(\mathbf{x}) R_z(\alpha) | 0_{\Lambda_\eta^\pm}(r) \rangle - \langle \Omega | F_j^2 | \Omega \rangle \ \ = \\
\nonumber & & = \ \ \left[\begin{array}{c}
c_\alpha^2 \Delta F_{x,\Lambda_\eta^\pm}^2(r;\mathbf{x}_{-\alpha}) + s_\alpha^2 \Delta F_{y,\Lambda_\eta^\pm}^2(r;\mathbf{x}_{-\alpha}) + 2 c_\alpha s_\alpha \langle 0_{\Lambda_\eta^\pm}(r) | F_x(\mathbf{x}_{-\alpha}) F_y(\mathbf{x}_{-\alpha}) | 0_{\Lambda_\eta^\pm}(r) \rangle \\
c_\alpha^2 \Delta F_{y,\Lambda_\eta^\pm}^2(r;\mathbf{x}_{-\alpha}) + s_\alpha^2 \Delta F_{x,\Lambda_\eta^\pm}^2(r;\mathbf{x}_{-\alpha}) - 2 c_\alpha s_\alpha \langle 0_{\Lambda_\eta^\pm}(r) | F_x(\mathbf{x}_{-\alpha}) F_y(\mathbf{x}_{-\alpha}) | 0_{\Lambda_\eta^\pm}(r) \rangle \\
\Delta F_{z,\Lambda_\eta^\pm}^2(r;\mathbf{x}_{-\alpha})
\end{array}\right]_j \\
\label{EQN561} & &
\end{eqnarray}
with the shorthand notation $c_\alpha = \cos(\alpha)$ and $s_\alpha = \sin(\alpha)$, and where the index $j$ on the right hand side indicates the $j$-th component of $[\ldots]$. Equating eqs.\ (\ref{EQN560}) and (\ref{EQN561}) relates the flux densities $\Delta F_{j,\Lambda_\eta^\epsilon}^2(r;\mathbf{x})$ and $\Delta F_{j,\Lambda_\eta^\epsilon}^2(r;\mathbf{x}_{-\alpha})$, i.e.\ yields their transformation law with respect to rotations around the $z$ axis \footnote{Eqs.\ (\ref{EQN560}) and (\ref{EQN561}) simplify for cubic rotations and, thus, are very helpful to improve statistical precision by symmetrizing the lattice results accordingly (see section~\ref{SEC773}).}. Clearly, one cannot expect that the flux densities $\Delta F_{j,\Lambda_\eta^+}^2(r;\mathbf{x})$ and $\Delta F_{j,\Lambda_\eta^-}^2(r;\mathbf{x})$ are invariant under rotations, nor that they appear to be identical, in particular not for $\Lambda \geq 1$, even though the corresponding potentials are degenerate. Numerical computations confirm that these flux densities are not invariant under rotations and that they are different from each other (see the discussion in section~\ref{SEC773} and the example plots in Figure~\ref{FIG201} and Figure~\ref{FIG202}).

Instead of quantum numbers $\Lambda_\eta^\epsilon$ one can also use quantum numbers $\lambda_\eta$ to label hybrid static potential states with $\Lambda \geq 1$, where $\lambda = \ldots,-2,-1,+1,+2,\ldots$ is the total angular momentum with respect to the $z$ axis, i.e.\ $\Lambda = |\lambda|$. Of course, there are again the same pairs of degenerate potentials, i.e.\ $V_{+\lambda_\eta}(r) = V_{-\lambda_\eta}(r) = V_{\Lambda_\eta^+}(r) = V_{\Lambda_\eta^-}(r)$. In this case the behavior of the corresponding states and flux densities under rotations is different,
\begin{eqnarray}
R_z(\alpha) | 0_{\lambda_\eta}(r) \rangle \ \ = \ \ e^{i \alpha \lambda} | 0_{\lambda_\eta}(r) \rangle ,
\end{eqnarray}
and eq.\ (\ref{EQN560}) simplifies,
\begin{eqnarray}
\label{EQN587} \langle 0_{\lambda_\eta}(r) | R_z^\dagger(\alpha) F_j^2(\mathbf{x}) R_z(\alpha) | 0_{\lambda_\eta}(r) \rangle - \langle \Omega | F_j^2 | \Omega \rangle \ \ = \ \ \Delta F_{j,\lambda_\eta}^2(r;\mathbf{x}) ,
\end{eqnarray}
while eq.\ (\ref{EQN561}) remains essentially unchanged (one just has to replace $\Lambda_\eta^\epsilon$ by $\lambda_\eta$). Consequently, the transformation law with respect to rotations around the $z$ axis and the angular dependence of $\Delta F_{j,\Lambda_\eta^\epsilon}^2(r;\mathbf{x})$ and $\Delta F_{j,\lambda_\eta}^2(r;\mathbf{x})$ is different, even though the corresponding hybrid static potentials are identical.

To eliminate this somewhat arbitrary angular dependence, which is a consequence of $\epsilon$ (when using quantum numbers $\Lambda_\eta^\epsilon$) or the sign of $\lambda$ (when using quantum numbers $\lambda_\eta$), but not related to $\Lambda = |\lambda|$ or $\eta$ ($\Lambda$ and $\eta$ fully characterize hybrid static potentials $V_{\Lambda_\eta}(r)$ for $\Lambda \geq 1$), we define for $\Lambda \geq 1$
\begin{eqnarray}
\label{EQN589} \Delta F_{j,\Lambda_\eta}^2(r;\mathbf{x}) \ \ = \ \ \frac{1}{2} \Big(\Delta F_{j,\Lambda_\eta^+}^2(r;\mathbf{x}) + \Delta F_{j,\Lambda_\eta^-}^2(r;\mathbf{x})\Big) \ \ = \ \ \frac{1}{2} \textrm{Tr}\Big(\mathcal{P}_{\Lambda_\eta} \Big(F_j^2(\mathbf{x}) - \langle \Omega | F_j^2 | \Omega \rangle\Big)\Big) .
\end{eqnarray}
This quantity represents the average over an ensemble of states with fixed $\Lambda$ and $\eta$, but arbitrary $\epsilon$. After the last equality the projector
\begin{eqnarray}
\nonumber & & \hspace{-0.7cm} \mathcal{P}_{\Lambda_\eta} \ \ = \ \ | 0_{\Lambda_\eta^+}(r) \rangle \langle 0_{\Lambda_\eta^+}(r) | + | 0_{\Lambda_\eta^-}(r) \rangle \langle 0_{\Lambda_\eta^-}(r) | \ \ = \ \ | 0_{+\lambda_\eta}(r) \rangle \langle 0_{+\lambda_\eta}(r) | + | 0_{-\lambda_\eta}(r) \rangle \langle 0_{-\lambda_\eta}(r) | \\
\label{EQN588} & &
\end{eqnarray}
to the corresponding 2-dimensional space of states has been introduced. This projector shows explicitly that $\Delta F_{j,\Lambda_\eta}^2(r;\mathbf{x})$ is independent of the basis used for that 2-dimensional space, i.e.\ independent of whether we use use $\epsilon$ or the sign of $\lambda$.

The transformation law with respect to rotations around the $z$ axis for $\Delta F_{j,\Lambda_\eta}^2(r;\mathbf{x})$ is
\begin{eqnarray}
\nonumber & & \hspace{-0.7cm} \Delta F_{j,\Lambda_\eta}^2(r;\mathbf{x}) \ \ = \\
\nonumber & & = \ \ \left[\begin{array}{c}
c_\alpha^2 \Delta F_{x,\Lambda_\eta}^2(r;\mathbf{x}_{-\alpha}) + s_\alpha^2 \Delta F_{y,\Lambda_\eta}^2(r;\mathbf{x}_{-\alpha}) + c_\alpha s_\alpha \sum_\epsilon \langle 0_{\Lambda_\eta^\epsilon}(r) | F_x(\mathbf{x}_{-\alpha}) F_y(\mathbf{x}_{-\alpha}) | 0_{\Lambda_\eta^\epsilon}(r) \rangle \\
c_\alpha^2 \Delta F_{y,\Lambda_\eta}^2(r;\mathbf{x}_{-\alpha}) + s_\alpha^2 \Delta F_{x,\Lambda_\eta}^2(r;\mathbf{x}_{-\alpha}) - c_\alpha s_\alpha \sum_\epsilon \langle 0_{\Lambda_\eta^\epsilon}(r) | F_x(\mathbf{x}_{-\alpha}) F_y(\mathbf{x}_{-\alpha}) | 0_{\Lambda_\eta^\epsilon}(r) \rangle \\
\Delta F_{z,\Lambda_\eta}^2(r;\mathbf{x}_{-\alpha})
\end{array}\right]_j , \\
\label{EQN691} & &
\end{eqnarray}
where the left hand side can be obtained by combining eqs.\ (\ref{EQN587}), (\ref{EQN589}) and (\ref{EQN588}) and the right hand side is essentially eq.\ (\ref{EQN561}). To simplify this even further it is convenient to define
\begin{eqnarray}
\label{EQN690} \Delta F_{\perp,\Lambda_\eta}^2(r;\mathbf{x}) \ \ = \ \ \frac{1}{2} \Big(\Delta F_{x,\Lambda_\eta}^2(r;\mathbf{x}) + \Delta F_{y,\Lambda_\eta}^2(r;\mathbf{x})\Big) ,
\end{eqnarray}
as e.g.\ also done in a similar way in \cite{Bicudo:2018jbb}. This quantity as well as $\Delta F_{z,\Lambda_\eta}^2(r;\mathbf{x})$ are invariant under rotations around the $z$ axis, i.e.\
\begin{eqnarray}
\Delta F_{\perp,\Lambda_\eta}^2(r;\mathbf{x}) \ \ = \ \ \Delta F_{\perp,\Lambda_\eta}^2(r;\mathbf{x}_{-\alpha}) \quad , \quad \Delta F_{z,\Lambda_\eta}^2(r;\mathbf{x}) \ \ = \ \ \Delta F_{z,\Lambda_\eta}^2(r;\mathbf{x}_{-\alpha}) .
\end{eqnarray}
Similarly, for $\Lambda = 0$
\begin{eqnarray}
\Delta F_{\perp,\Lambda_\eta^\epsilon}^2(r;\mathbf{x}) \ \ = \ \ \Delta F_{\perp,\Lambda_\eta^\epsilon}^2(r;\mathbf{x}_{-\alpha}) \quad , \quad \Delta F_{z,\Lambda_\eta^\epsilon}^2(r;\mathbf{x}) \ \ = \ \ \Delta F_{z,\Lambda_\eta^\epsilon}^2(r;\mathbf{x}_{-\alpha}) .
\end{eqnarray}
as can be read off from eqs.\ (\ref{EQN560}) and (\ref{EQN561}).

% ********************
% ********************
% ********************
% ********************
% ********************

\newpage

\section{\label{SEC549}Lattice setup}

The computations presented in this work have been performed using SU(2) and SU(3) lattice gauge theory. The corresponding gauge link configurations have been generated with the standard Wilson gauge action (see textbooks on lattice field theory, e.g.\ ref.\ \cite{Rothe:1992nt}). Since we are considering purely gluonic observables, we expect that there is little difference between our pure gauge theory results and corresponding results in full QCD. This expectation is supported by ref.\ \cite{Bali:2000vr}, where hybrid static potentials were computed both in pure gauge theory and QCD and no statistically significant differences were observed.

For the SU(2) simulations we have used a standard heatbath algorithm. To eliminate correlations in Monte Carlo time, the gauge link configurations are separated by 100 heatbath sweeps. For the SU(3) simulations we have used the Chroma QCD library \cite{Edwards:2004sx}. There, the gauge link configurations are separated by 20 update sweeps, where each update sweep comprises a heatbath and four over-relaxation steps. Details of our simulated ensembles are collected in Table~\ref{TAB101}, including the gauge coupling $\beta$, the lattice extent $(L/a)^3 \times T/a$ and the number of gauge link configurations used for the flux tube computations. We also list the lattice spacing $a$ in fm, which is obtained by identifying $r_0$ with $0.5 \, \textrm{fm}$ (see refs.\ \cite{Philipsen:2013ysa,Capitani:2018rox}). For the majority of computations for gauge group SU(2) we use the ensemble with $(L/a)^3 \times T/a = 24^4$. The ensemble with $(L/a)^3 \times T/a = 18^4$ is only used for exploring and excluding finite volume effects in section~\ref{SEC499}.

\begin{table}[htb]
\begin{center}

\def\arraystretch{1.2}

\begin{tabular}{cccccc}
\hline
gauge group & $\beta$ & $a$ in fm & $(L/a)^3 \times T/a$ & number of configurations \\
\hline
SU(2) & $2.5$ & $0.079$ & $18^4$           &           $13 \, 000$ \\
      &       &         & $24^4$           &           $48 \, 000$ \\
\hline
SU(3) & $6.0$ & $0.093$ & $24^3 \times 48$ & $\phantom{0}4 \, 500$ \\
\hline
\end{tabular}

\end{center}
\caption{\label{TAB101}SU(2) and SU(3) gauge link ensembles.}
\end{table}

To improve the signal quality, standard smearing techniques are applied, when sampling $\tilde{W}$ appearing in eqs.\ (\ref{EQN188}) and (\ref{EQN189}) and defined in eq.\ (\ref{EQN764}):
\begin{itemize}
\item Spatial gauge links, i.e.\ links in $a_{S;\Lambda_\eta^\epsilon}(-r/2,+r/2;t_0)$ and $a_{S;\Lambda_\eta^\epsilon}(-r/2,+r/2;t_2)$ (defined in eq.\ (\ref{EQN011})), are APE smeared gauge links (for detailed equations see e.g.\ ref.\ \cite{Jansen:2008si}), where the parameters $\alpha_\textrm{APE} = 0.5$ and $N_\text{APE} = 20$ have been optimized in ref.\ \cite{Capitani:2018rox} to generate large overlaps with the ground states $| 0_{\Lambda_\eta^\epsilon}(r) \rangle$ \footnote{The optimzation of APE smearing parameters in ref.\ \cite{Capitani:2018rox} was done for SU(3) gauge theory. We use the same parameters for our computations in SU(2) gauge theory and get similar ground state overlaps, which is indicated by effective mass plateaus of approximately the same quality.}. This allows to identify plateaus in $\Delta F_{\textrm{eff};j,\Lambda_\eta^\epsilon}^2(r,t_2,t_0;\mathbf{x},t_1)$ at smaller temporal separations $t_2 - t_1$ and $t_1 - t_0$ (see eq.\ (\ref{EQN532})).

\item For certain computations temporal gauge links, i.e.\ links in $U(+r/2;t_0,t_2)$ and \\ $U(-r/2;t_2,t_0)$, are HYP2-smeared gauge links \cite{Hasenfratz:2001hp,DellaMorte:2003mn,Della Morte:2005yc}, which lead to a reduced self energy of the static quarks and, consequently, to smaller statistical errors. This, however, introduces larger discretization errors for small $r$ as well as for $\mathbf{x}$ close to either $\mathbf{r}_Q=(0,0,+r/2)$ or $\mathbf{r}_{\bar{Q}}=(0,0,-r/2)$. Therefore, we use HYP2-smearing only, when computing field strengths $\Delta F_{j,\Lambda_\eta^\epsilon}^2(r;\mathbf{x} = (x,y,0))$ (see eq.\ (\ref{EQN532})), i.e.\ in the mediator plane $z = 0$. For a more detailed discussion see section~\ref{SEC575}.
\end{itemize}

All statistical errors shown and quoted throughout this paper are determined via jackknife. We perform a suitable binning of gauge link configurations to exclude statistical correlations in Monte Carlo time.

% ********************
% ********************
% ********************
% ********************
% ********************

\newpage

\section{\label{SEC698}Numerical results}

% ********************
% ********************
% ********************

\subsection{Investigation of systematic errors}

% ********************

\subsubsection{Plateaus of $\Delta F_{\textrm{eff};j,\Lambda_\eta^\epsilon}^2$ and contamination by excited states}

\begin{figure}[p]
\begin{center}
\includegraphics[width=15.8cm]{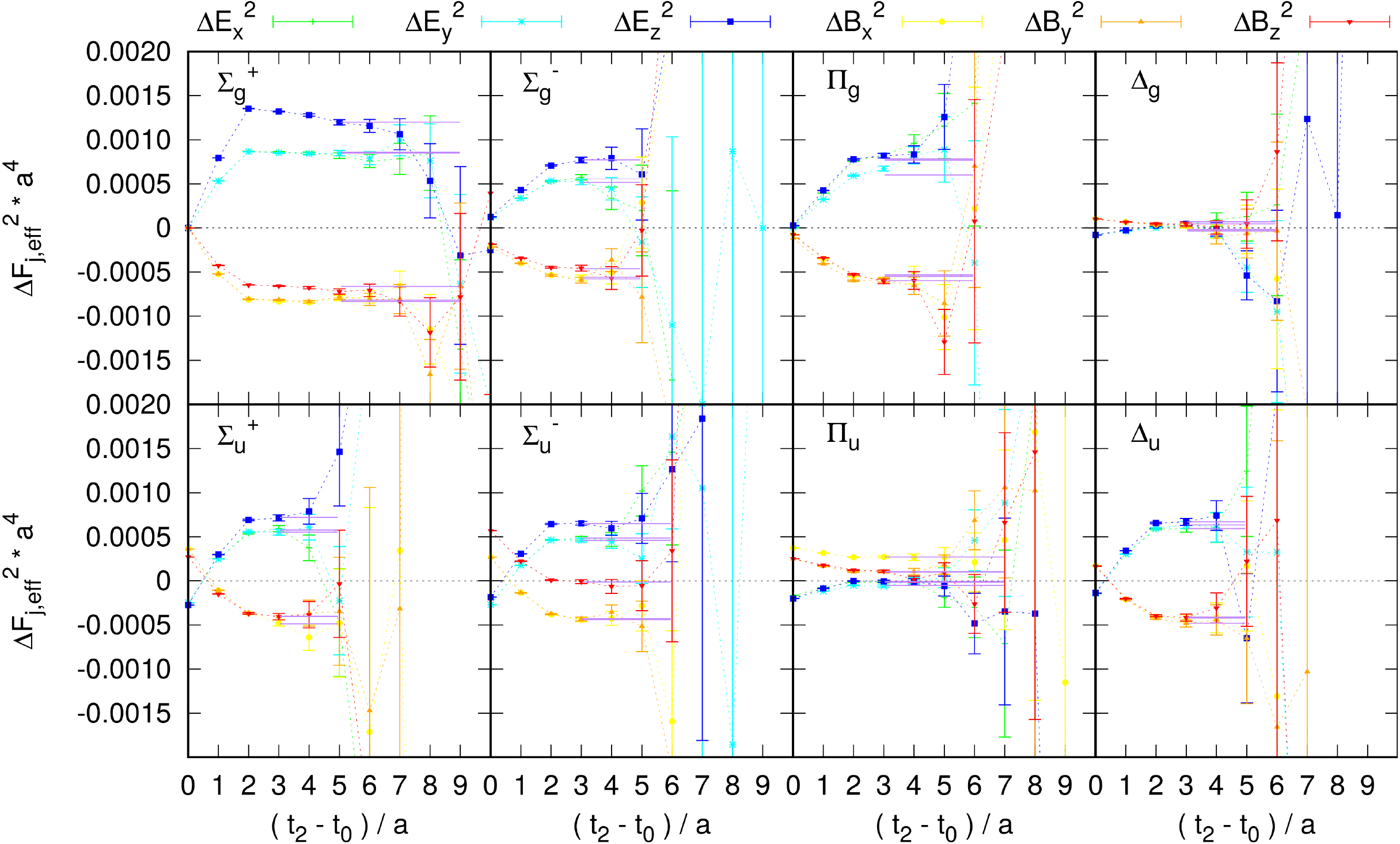} \\
\vspace{0.2cm}
\includegraphics[width=15.8cm]{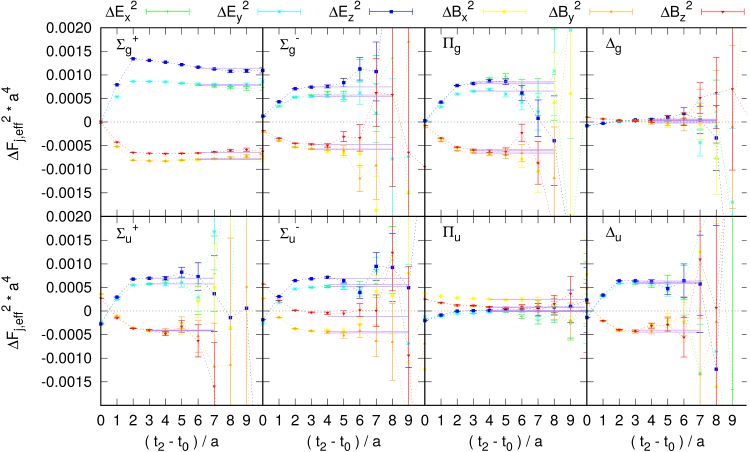}
\end{center}
\caption{\label{FIG060}$\Delta F_{\textrm{eff};j,\Lambda_\eta^\epsilon}^2(r,t_2,t_0;\mathbf{x} = \vec{0},t_1)$ as a function of $t_2 - t_0$ for gauge group SU(2), $\Lambda_\eta^\epsilon = \Sigma_g^+, \Sigma_u^+, \Sigma_g^-, \Sigma_u^-, \Pi_g,\Pi_u, \Delta_g, \Delta_u$ and $Q \bar{Q}$ separation $r = 10 \, a$. Plateau fits and fitting ranges $[t_\textrm{min},t_\textrm{max}]$ are indicated by horizontal straight lines. \textbf{(top)}~Temporal links in $\tilde{W}$ are unsmeared. \textbf{(bottom)}~Temporal links in $\tilde{W}$ are HYP2 smeared.}
\end{figure}

We have determined $\Delta F_{j,\Lambda_\eta^\epsilon}^2(r;\mathbf{x})$ by fitting a constant to the lattice result for \\ $\Delta F_{\textrm{eff};j,\Lambda_\eta^\epsilon}^2(r,t_2,t_0;\mathbf{x},t_1)$ at sufficiently large $t_2 - t_1$ and $t_1 -t_0$, where the data points are consistent with a plateau (see eqs.\ (\ref{EQN532}), (\ref{EQN188}) and (\ref{EQN189})). For even $(t_2 - t_0) / a$ we use $t_1 = (t_0 + t_2) / 2$, while for odd $(t_2 - t_0) / a$ we use $t_1 = (t_0 + t_2 + a) / 2$, i.e.\ equal or similar values for $t_2 - t_1$ and $t_1 -t_0$. Example plots of $\Delta F_{\textrm{eff};j,\Lambda_\eta^\epsilon}^2(r,t_2,t_0;\mathbf{x},t_1)$ as a function of $t_2 - t_0$ for gauge group SU(2), all investigated $\Lambda_\eta^\epsilon$ sectors, quark-antiquark separation $r = 10 \, a$ and $\mathbf{x} = \vec{0}$ are shown in Figure~\ref{FIG060}. We have performed an uncorrelated $\chi^2$-minimizing fit of a constant corresponding to $\Delta F_{j,\Lambda_\eta^\epsilon}^2(r;\mathbf{x})$ in the region $t_\textrm{min} \leq t_2 - t_0 \leq t_\textrm{max}$. Since the statistical errors of $\Delta F_{\textrm{eff};j,\Lambda_\eta^\epsilon}^2(r,t_2,t_0;\mathbf{x},t_1)$ rapidly increase with increasing $t_2 - t_0$, the results are almost independent of $t_\textrm{max}$. We have taken the largest $t_\textrm{max}$, where the signal is not lost in noise. $t_\textrm{min}$ has been chosen such that $\chi^2 / \textrm{dof} \ltapprox 1$ for the majority of fits. This results in $t_\textrm{min} \approx 3 \, a \ldots 4 \, a$ and $t_\textrm{max} \approx 5 \, a \ldots 8 \, a$ for hybrid static potentials with $\Lambda_\eta^\epsilon = \Sigma_u^+,\Sigma_g^-,\Sigma_u^-,\Pi_g,\Pi_u,\Delta_g,\Delta_u$, while $t_\textrm{min} \approx 5 \, a \ldots 6 \, a$ and $t_\textrm{max} = 10 \, a$ for the ordinary static potential with $\Lambda_\eta^\epsilon = \Sigma_g^+$.

As an additional check that $t_\textrm{min}$ is chosen sufficiently large, i.e.\ that excited states are strongly suppressed, we have repeated the computation of $\Delta F_{j,\Lambda_\eta^\epsilon}^2(r;\mathbf{x})$ for gauge group SU(2), $\Lambda_\eta^\epsilon = \Pi_u$, $r = 6 \, a$ and $\mathbf{x} = (x,0,0)$ using a creation operator $S$ (see eq.\ (\ref{EQN011})), which has a structure significantly different from that shown in Table~\ref{TAB001}, namely $S_{\textrm{I},1}$ as defined in ref.\ \cite{Capitani:2018rox}, Figure~2. Within statistical errors we find identical flux densities $\Delta F_{j,\Lambda_\eta^\epsilon}^2(r;\mathbf{x})$, which we interpret as confirmation, that we indeed measure the flux densities of the ground states in the $\Lambda_\eta^\epsilon$ sectors and not flux densities, which depend on the creation operators and trial states we are using.

% ********************

\FloatBarrier

\subsubsection{\label{SEC575}Discretization errors and smearing}

Until now we have performed computations only at a single value of the lattice spacing $a$. Therefore, we are not yet able to study the continuum limit. Strong discretization errors are expected, when either $r = |\mathbf{r}_Q - \mathbf{r}_{\bar{Q}}|$, $|\mathbf{x} - \mathbf{r}_Q|$ or $|\mathbf{x} - \mathbf{r}_{\bar{Q}}|$ is small, where $\mathbf{r}_Q = (0,0,+r/2)$ and $\mathbf{r}_{\bar{Q}} = (0,0,-r/2)$ are the positions of the static charges and $\mathbf{x}$ is the spatial argument of the flux density $\Delta F_{j,\Lambda_\eta^\epsilon}^2(r;\mathbf{x})$. These discretization errors are expected to be even more pronounced, when using HYP2 smeared temporal links in $\tilde{W}$ (see eq.\ (\ref{EQN764})), which can be interpreted as increasing the radii of the static charges. We, therefore, compare results for $\Delta F_{j,\Lambda_\eta^\epsilon}^2(r;\mathbf{x})$ obtained with and without HYP2-smeared temporal links.

\begin{figure}[htb]
\begin{center}
\includegraphics[width=12.0cm]{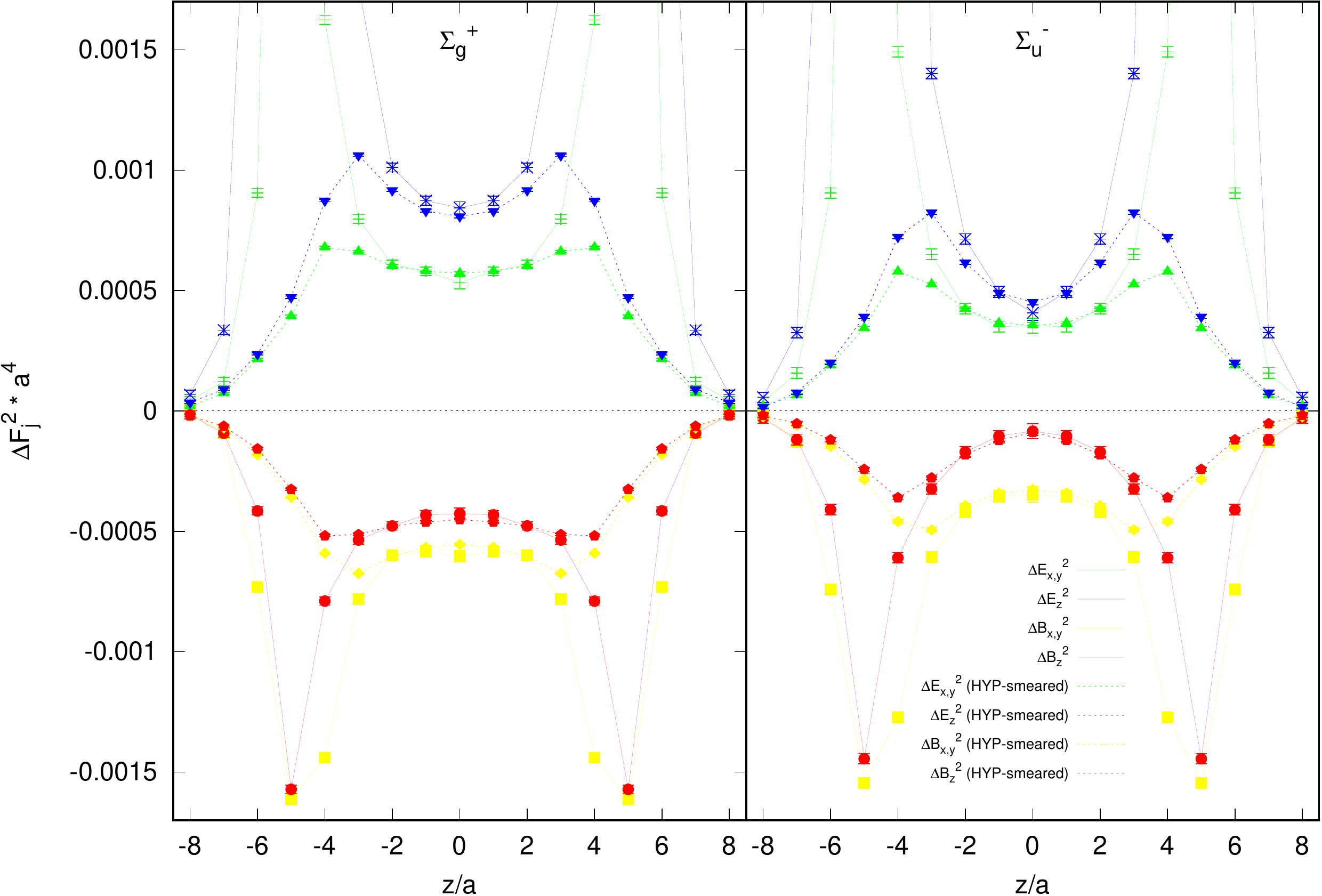}
\end{center}
\caption{\label{FIG021}Flux densities $\Delta F_{j,\Lambda_\eta^\epsilon}^2(r;\mathbf{x} = (0,0,z))$  as a function of $z$ for gauge group SU(3), $\Lambda_\eta^\epsilon = \Sigma_g^+ , \Sigma_u^-$ and $Q \bar{Q}$ separation $r = 10 \, a$.}
\end{figure}

In Figure~\ref{FIG021} we show results for $\Lambda_\eta^\epsilon = \Sigma_g^+ , \Sigma_u^-$ and $Q \bar{Q}$ separation $r = 10 \, a$ on the separation axis, $\mathbf{x} = (0,0,z)$. For $|\mathbf{x} - \mathbf{r}_Q| \leq a$ or $|\mathbf{x} - \mathbf{r}_{\bar{Q}}| \leq a$ drastic discrepancies between unsmeared and HYP2-smeared results can be observed, while for $|\mathbf{x} - \mathbf{r}_Q| , |\mathbf{x} - \mathbf{r}_{\bar{Q}}| \geq 4 \, a$ and $\Delta F_{j,\Lambda_\eta^\epsilon}^2(r;\mathbf{x}) = \Delta E_{z,\Lambda_\eta^\epsilon}^2(r;\mathbf{x})$ as well as for $|\mathbf{x} - \mathbf{r}_Q| , |\mathbf{x} - \mathbf{r}_{\bar{Q}}| \geq 3 \, a$ and all other field strength components there is agreement within statistical errors. When using HYP2-smeared temporal links, the pronounced peaks at the positions of the charges, which are present in the unsmeared results, are essentially gone. This is expected and can be observed in a qualitatively similar way also in much simpler theories, for example in classical electrodynamics, when smearing the charge density of a point charge. Analogous plots for other $\Lambda_\eta^\epsilon$ sectors look very similar and are not shown. Therefore, for the computations of $\Delta F_{j,\Lambda_\eta^\epsilon}^2(r;\mathbf{x})$ in a plane containing the separation axis (see section~\ref{SEC386}) we do not use HYP2-smearing. Note, however, that even unsmeared results within a radius of $\approx 2 \, a$ around either of the two static charges will exhibit sizable discretization errors and should be considered as crude estimates only. In other words, instead of the poles related to the infinite self energy of the static charges, $\Delta F_{j,\Lambda_\eta^\epsilon}^2(r;\mathbf{x})$ will exhibit pronounced but finite peaks.

\begin{figure}[htb]
\begin{center}
\includegraphics[width=12.0cm]{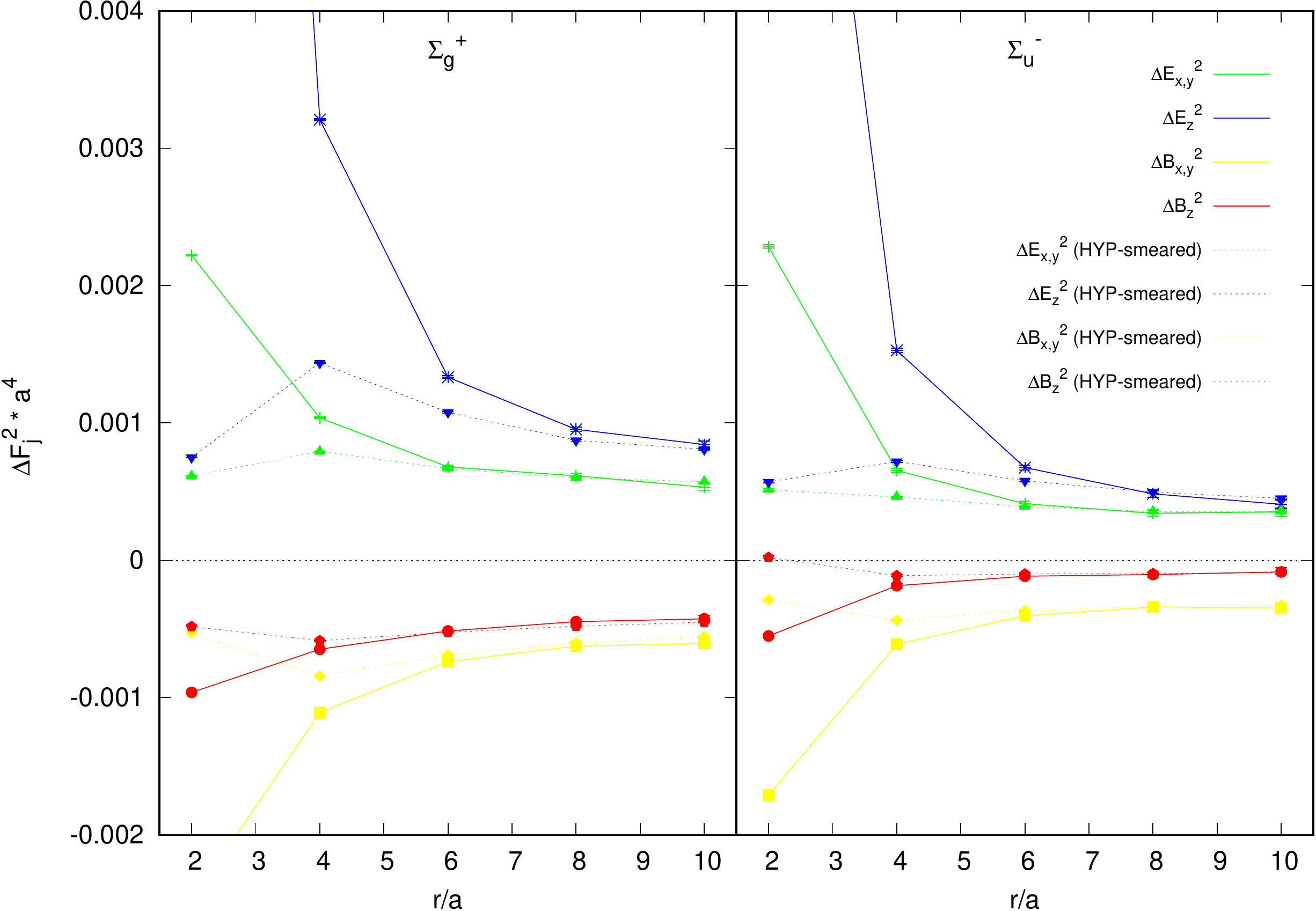}
\end{center}
\caption{\label{FIG022}Flux densities $\Delta F_{j,\Lambda_\eta^\epsilon}^2(r;\mathbf{x} = \vec{0})$ as a function of the $Q \bar{Q}$ separation $r$ for gauge group SU(3) and $\Lambda_\eta^\epsilon = \Sigma_g^+ , \Sigma_u^-$.}
\end{figure}

We also study the effect of HYP2-smearing on $\Delta F_{j,\Lambda_\eta^\epsilon}^2(r;\mathbf{x})$ in the mediator plane defined by $z = 0$ for various $Q \bar{Q}$ separations $r$. For $r \geq 6 \, a$ we find agreement within statistical errors for all $\Lambda_\eta^\epsilon$ sectors and all field strength components with exception of $\Delta E_{z,\Lambda_\eta^\epsilon}^2(r;\mathbf{x}= (x,y,0))$. For $\Delta E_{z,\Lambda_\eta^\epsilon}^2(r;\mathbf{x}= (x,y,0))$ there is agreement for $r \geq 10 \, a$ for $\Lambda_\eta^\epsilon = \Sigma_g^+$ and for $r \geq 8 \, a$ for all other $\Lambda_\eta^\epsilon$ sectors. Example plots for $\Lambda_\eta^\epsilon = \Sigma_g^+ , \Sigma_u^-$ and $\mathbf{x} = (0,0,0)$ are shown in Figure~\ref{FIG022}. Therefore, for computations of $\Delta F_{j,\Lambda_\eta^\epsilon}^2$  in the mediator plane, which we show for $r = 10 \, a$ in section~\ref{SEC386}, we use HYP2-smearing, which reduces statistical errors significantly.

% ********************

\subsubsection{\label{SEC499}Finite volume corrections}

Finite volume corrections are rather mild for static potentials, when $r < L/2$, where $L$ is the spatial lattice extent. In particular for pure gauge theory, where the lightest particle (the $J^{\mathcal{P} \mathcal{C}} = 0^{+ +}$ glueball) is very heavy, finite volume corrections should be almost negligible. A comparison of flux densities $\Delta F_{j,\Lambda_\eta^\epsilon}^2(r;\mathbf{x})$ for gauge group SU(2) on the two gauge link ensembles with $(L/a)^3 \times T/a = 18^4$ and $(L/a)^3 \times T/a = 24^4$ (see Table~\ref{TAB101}) supports this expectation.

% ********************
% ********************
% ********************

\FloatBarrier

\subsection{\label{SEC773}Angular dependence and symmetrization of hybrid static potential flux densities}

In section~\ref{SEC478} we have discussed, how hybrid static potential flux densities transform under rotations around the $z$ axis. On a hypercubic lattice the relevant eqs.\ (\ref{EQN560}) and (\ref{EQN561}) are exact only for cubic rotations, i.e.\ for rotations with angle $\alpha$, which is a multiple of $\pi/2$. For $\alpha = \pm \pi/2$ they become for even $\Lambda$, i.e.\ $\Lambda = \Sigma$ and $\Lambda = \Delta$,
\begin{eqnarray}
 & & \hspace{-0.7cm} \Delta F_{x,\Lambda_\eta^\pm}^2(r;\mathbf{x}) \ \ = \ \ \Delta F_{y,\Lambda_\eta^\pm}^2(r;\mathbf{x}_{\pm \pi/2}) \\
 & & \hspace{-0.7cm} \Delta F_{y,\Lambda_\eta^\pm}^2(r;\mathbf{x}) \ \ = \ \ \Delta F_{x,\Lambda_\eta^\pm}^2(r;\mathbf{x}_{\pm \pi/2}) \\
 & & \hspace{-0.7cm} \Delta F_{z,\Lambda_\eta^\pm}^2(r;\mathbf{x}) \ \ = \ \ \Delta F_{z,\Lambda_\eta^\pm}^2(r;\mathbf{x}_{\pm \pi/2})
\end{eqnarray}
and for odd $\Lambda$, i.e.\ $\Lambda = \Pi$,
\begin{eqnarray}
 & & \hspace{-0.7cm} \Delta F_{x,\Lambda_\eta^\pm}^2(r;\mathbf{x}) \ \ = \ \ \Delta F_{y,\Lambda_\eta^\mp}^2(r;\mathbf{x}_{\pm \pi/2}) \\
 & & \hspace{-0.7cm} \Delta F_{y,\Lambda_\eta^\pm}^2(r;\mathbf{x}) \ \ = \ \ \Delta F_{x,\Lambda_\eta^\mp}^2(r;\mathbf{x}_{\pm \pi/2}) \\
 & & \hspace{-0.7cm} \Delta F_{z,\Lambda_\eta^\pm}^2(r;\mathbf{x}) \ \ = \ \ \Delta F_{z,\Lambda_\eta^\mp}^2(r;\mathbf{x}_{\pm \pi/2}) .
\end{eqnarray}
We have verified our numerical computation of flux densities using these equations, i.e.\ we have checked that all our results are consistent with these equations within statistical errors. In a second step we have used these equations to reduce the statistical errors of our results by averaging related flux densities.

\begin{figure}[htb]
\begin{center}
\includegraphics[width=7.0cm]{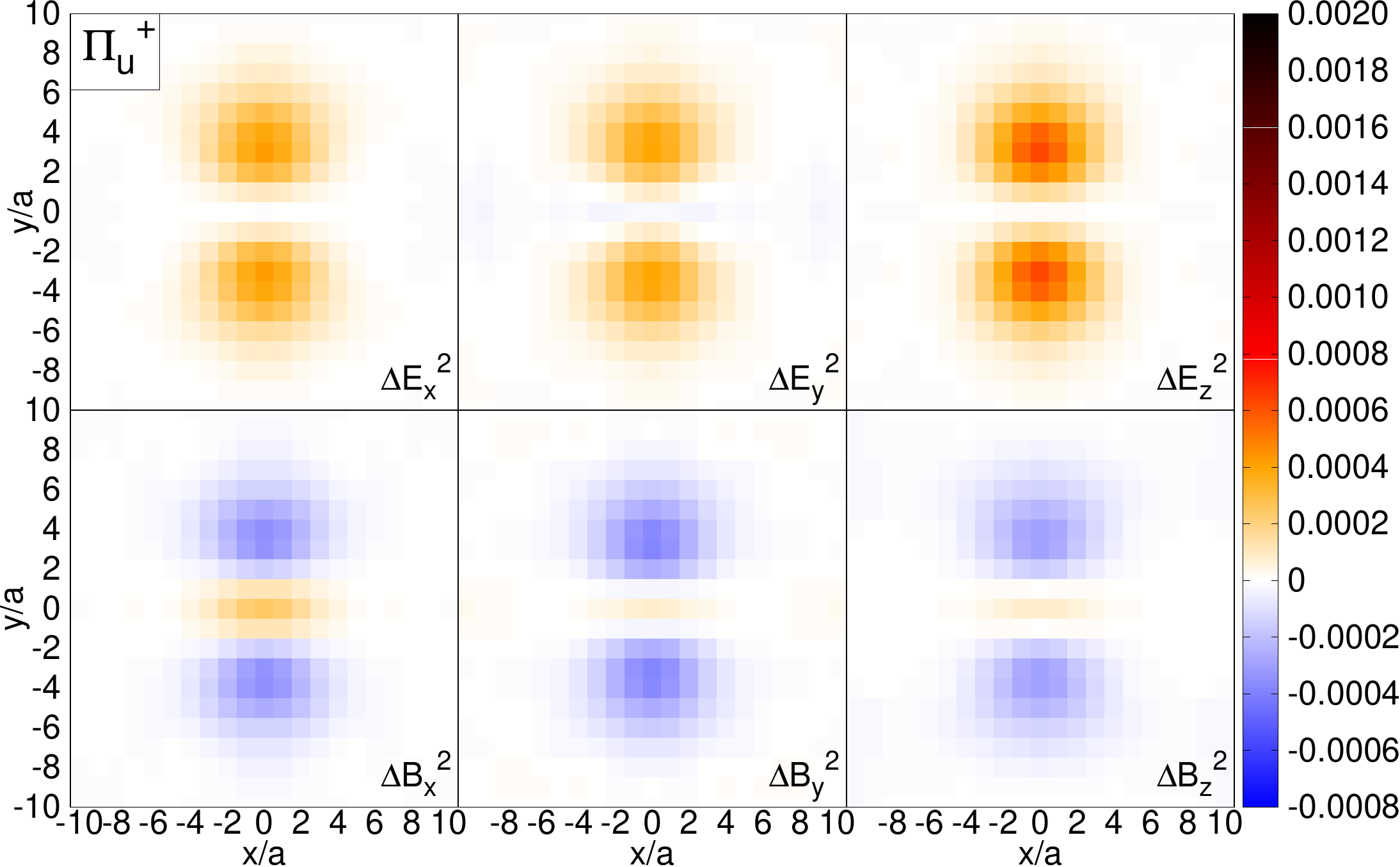}
\includegraphics[width=7.0cm]{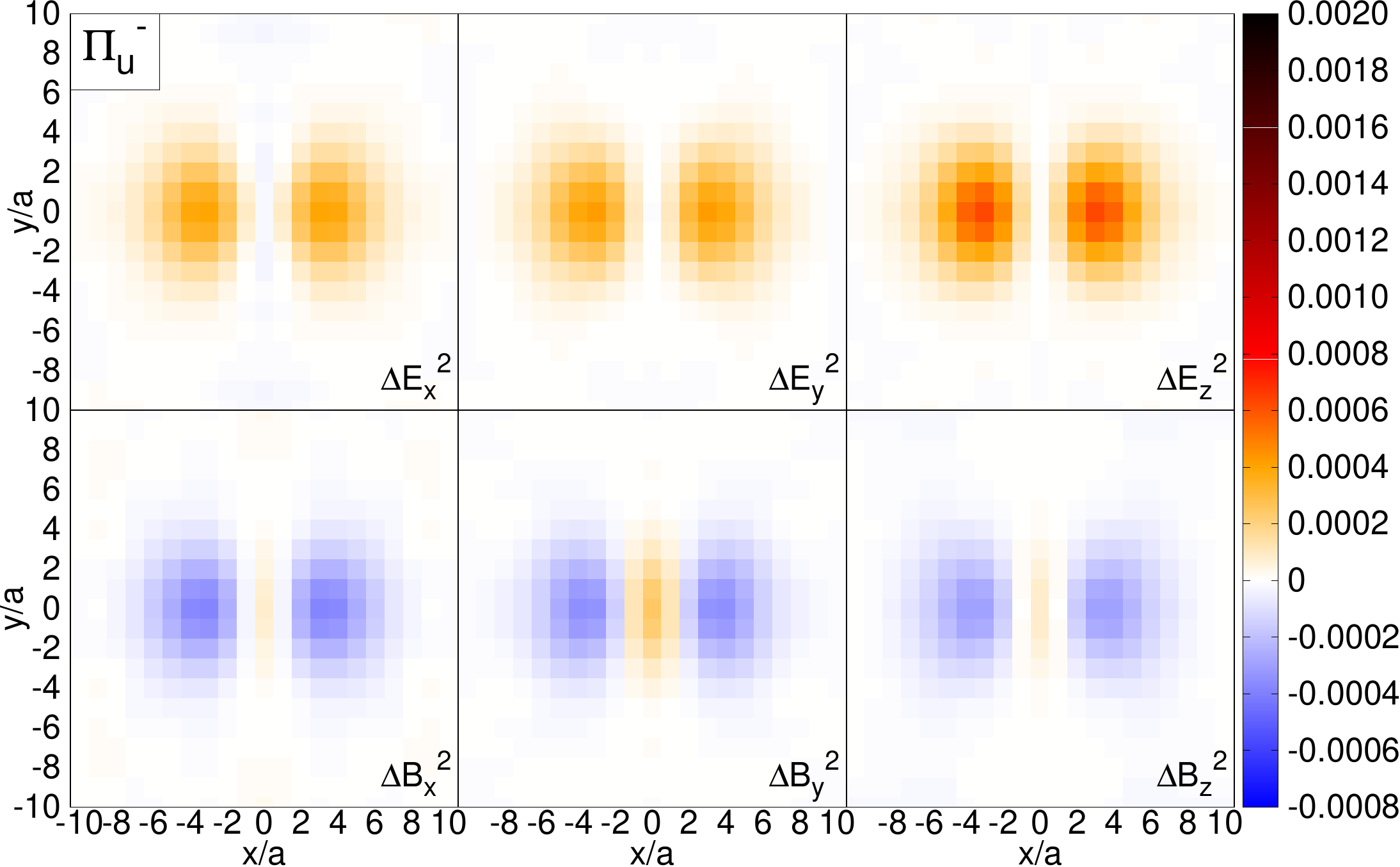}
\includegraphics[width=7.0cm]{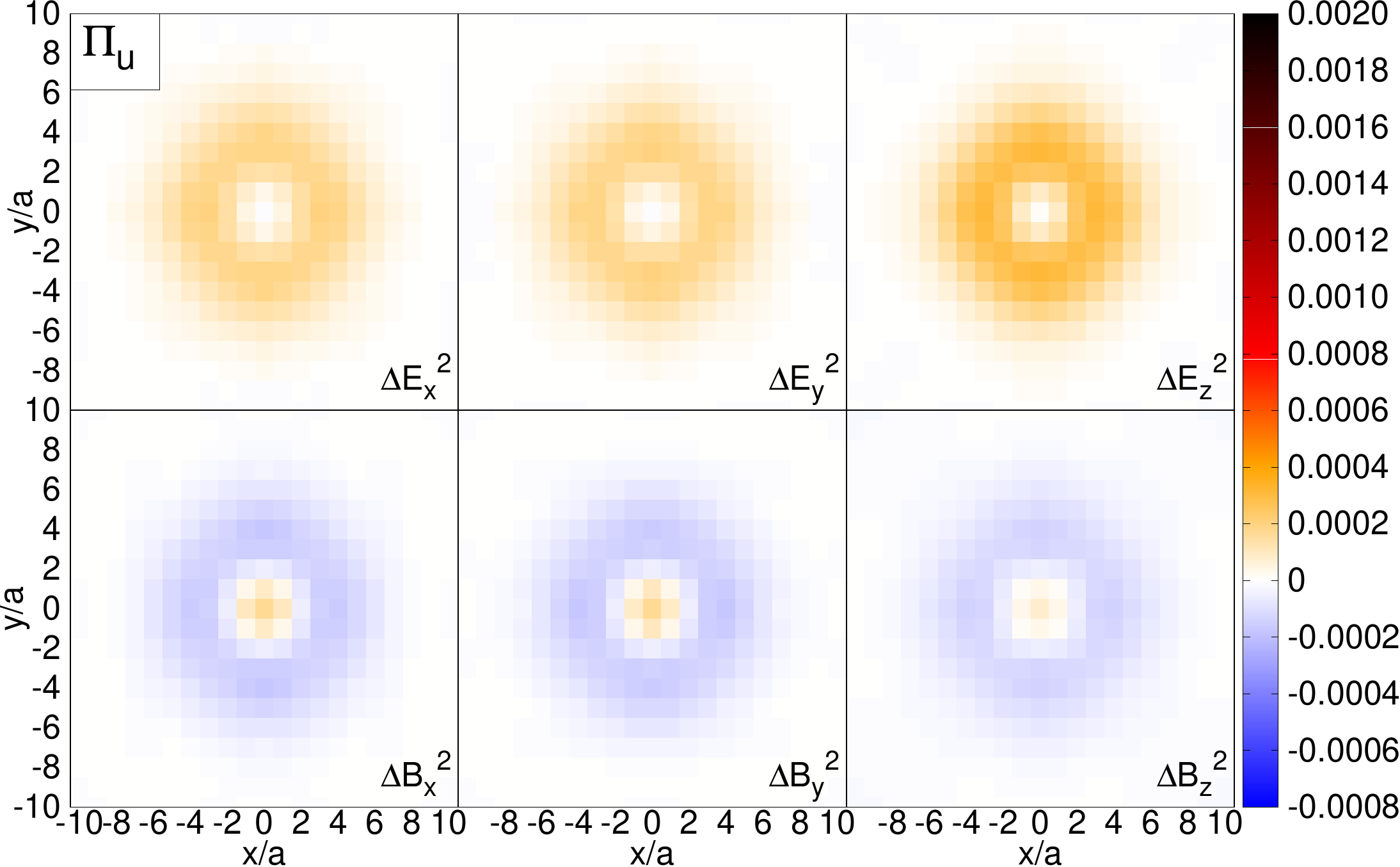}
\end{center}
\caption{\label{FIG201}Flux densities $\Delta F_{j,\Pi_u^+}^2(r;\mathbf{x})$, $\Delta F_{j,\Pi_u^-}^2(r;\mathbf{x})$ and $\Delta F_{j,\Pi_u}^2(r;\mathbf{x})$ in the mediator plane ($z = 0$) for gauge group SU(2) and $Q \bar{Q}$ separation $r = 10 \, a$.}
\end{figure}

\begin{figure}[htb]
\begin{center}
\includegraphics[width=7.0cm]{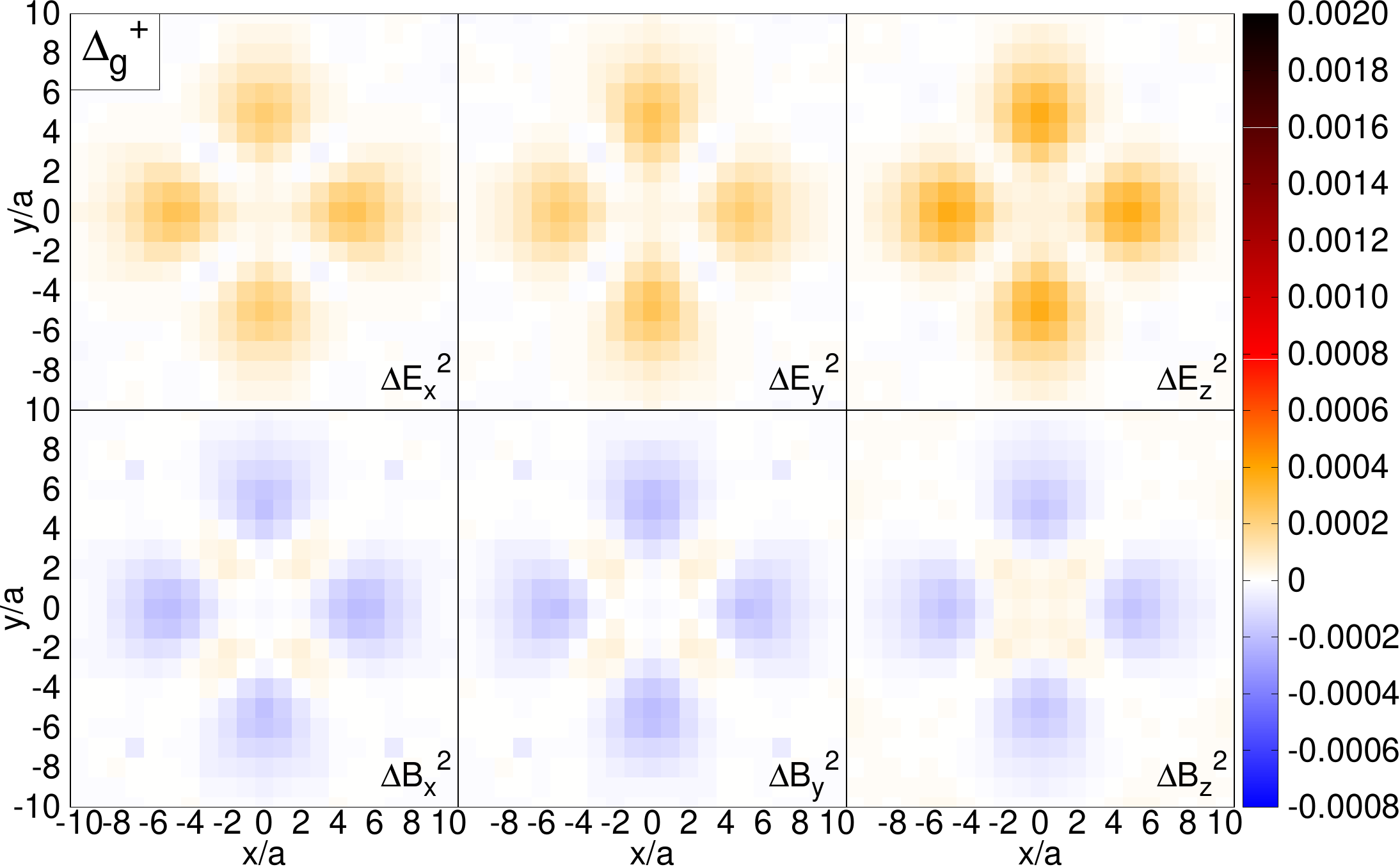}
\includegraphics[width=7.0cm]{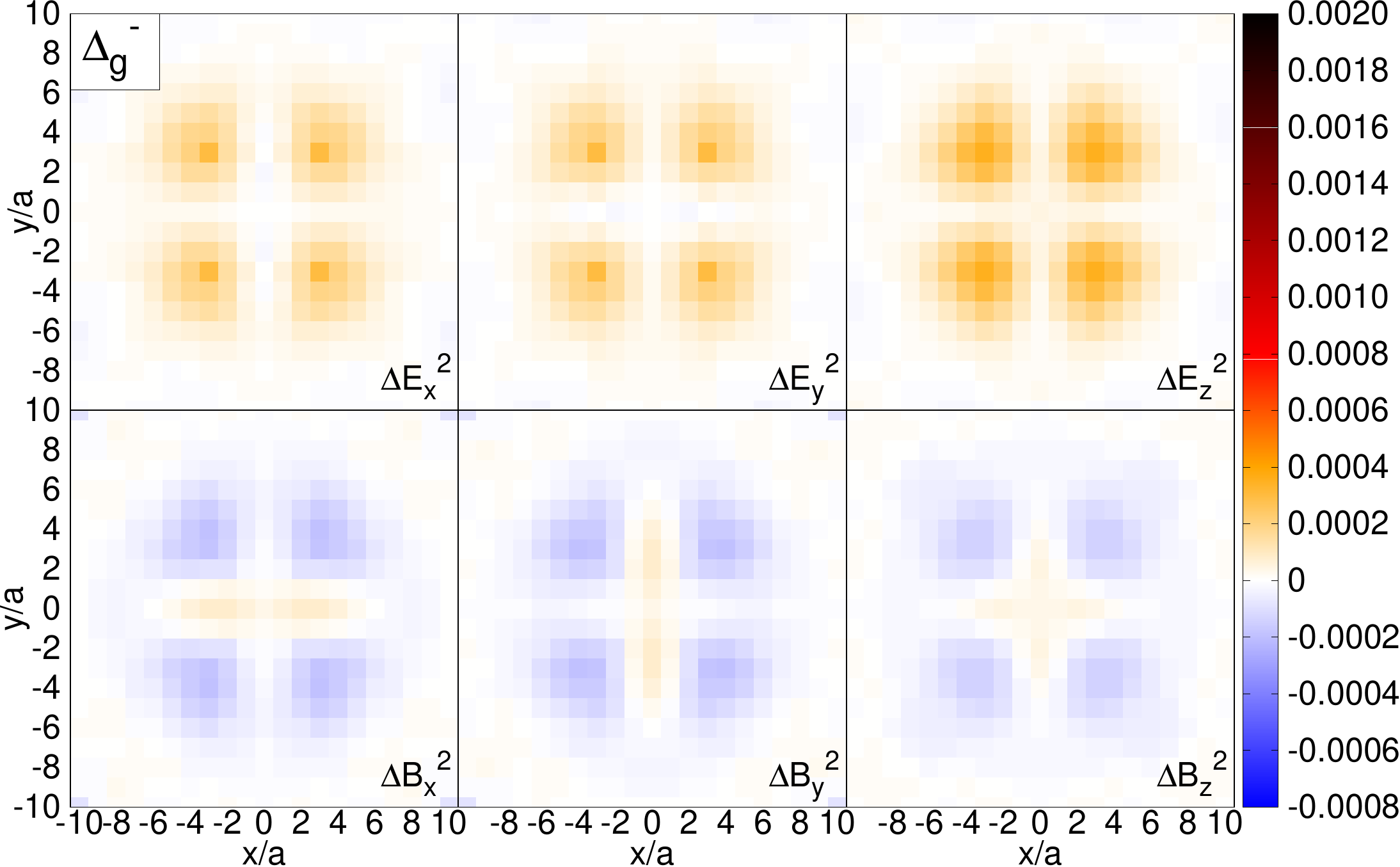}
\includegraphics[width=7.0cm]{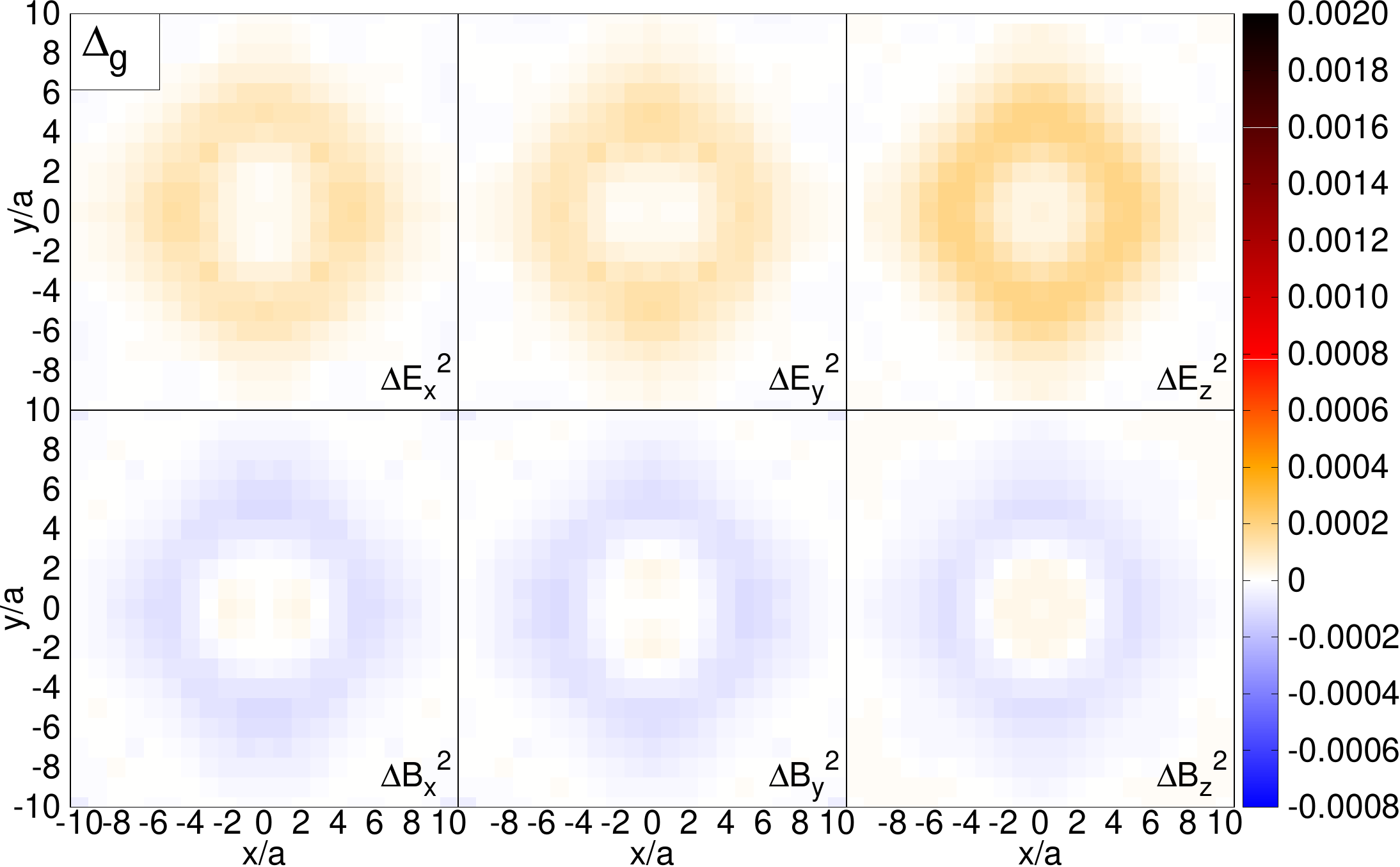}
\end{center}
\caption{\label{FIG202}Flux densities $\Delta F_{j,\Delta_g^+}^2(r;\mathbf{x})$, $\Delta F_{j,\Delta_g^-}^2(r;\mathbf{x})$ and $\Delta F_{j,\Delta_g}^2(r;\mathbf{x})$ in the mediator plane ($z = 0$) for gauge group SU(2) and $Q \bar{Q}$ separation $r = 10 \, a$.}
\end{figure}

In section~\ref{SEC478} we have also discussed that hybrid static potential flux densities $\Delta F_{j,\Lambda_\eta^+}^2(r;\mathbf{x})$ and $\Delta F_{j,\Lambda_\eta^-}^2(r;\mathbf{x})$ with $\Lambda \geq 1$ are not expected to be identical, even though the corresponding potentials are degenerate (see eqs.\ (\ref{EQN560}) and (\ref{EQN561})). This expectation is confirmed by the plots in the upper row of Figure~\ref{FIG201} and Figure~\ref{FIG202}, where we show two examples, the flux densities $\Delta F_{j,\Lambda_\eta^\epsilon}^2(r;\mathbf{x})$ for $\Lambda_\eta^\epsilon = \Pi_u^+ , \Pi_u^-$ and for $\Lambda_\eta^\epsilon = \Delta_g^+ , \Delta_g^-$ in the mediator plane $z = 0$.

In the plots at the bottom of Figure~\ref{FIG201} and Figure~\ref{FIG202} we show the flux densities $\Delta F_{j,\Lambda_\eta}^2(r;\mathbf{x})$ defined in eq.\ (\ref{EQN589}), again for $\Lambda_\eta = \Pi_u$ and for $\Lambda_\eta = \Delta_g$. As discussed in section~\ref{SEC478} these are ensemble averages over states with fixed $\Lambda$ and $\eta$, but indefinite $\epsilon$. Note that $\Delta F_{z,\Lambda_\eta}^2(r;\mathbf{x})$ is invariant under cubic rotations, while $\Delta F_{x,\Lambda_\eta}^2(r;\mathbf{x})$ and $\Delta F_{y,\Lambda_\eta}^2(r;\mathbf{x})$, even though quite similar, are related by rotations with angle $\alpha = \pm \pi/2$ (see eq.\ (\ref{EQN691})). Averaging $\Delta F_{x,\Lambda_\eta}^2(r;\mathbf{x})$ and $\Delta F_{y,\Lambda_\eta}^2(r;\mathbf{x})$ according to eq.\ (\ref{EQN690}) would lead to another quantity invariant under cubic rotations. From now on we always show the flux densities $\Delta F_{j,\Lambda_\eta}^2(r;\mathbf{x})$ for $\Lambda \geq 1$, i.e.\ not anymore $\Delta F_{j,\Lambda_\eta^\epsilon}^2(r;\mathbf{x})$.

% ********************
% ********************
% ********************

\FloatBarrier

\subsection{\label{SEC386}Hybrid static potential flux densities for all $\Lambda_\eta^\epsilon$ sectors}

In this section we show and discuss the main numerical results of this work, the flux densities $\Delta F_{j,\Lambda_\eta^{(\epsilon)}}^2(r;\mathbf{x})$, $j = x,y,z,\perp$ for the eight sectors $\Lambda_\eta^{(\epsilon)} = \Sigma_g^+, \Sigma_g^-, \Sigma_u^+, \Sigma_u^-, \Pi_g,\Pi_u, \Delta_g, \Delta_u$, both in the mediator plane $z = 0$ and in the separation plane $y = 0$. All plots in this section are for SU(2) gauge theory. Corresponding plots for SU(3) gauge theory are very similar and collected in appendix~\ref{APP001}.

We decided to perform computations for two $Q \bar{Q}$ separations, $r = 6 \, a$ and $r = 10 \, a$. This allows to compare results for two significantly different $r$, i.e.\ to see, how the shapes of the hybrid static potential flux tubes change, when the quark and the antiquark are pulled apart. We did not study separations $r < 6 \, a$, because for such small separations flux densities exhibit sizable discretization errors in the region between the two charges (see the discussion in section~\ref{SEC575}). Since the signal for a Wilson loop decays exponentially with its area, we also refrained from performing computations for $r > 10 \, a$, which are very costly in terms of CPU time.

\begin{figure}[p]
\begin{center}
\includegraphics[width=7.0cm]{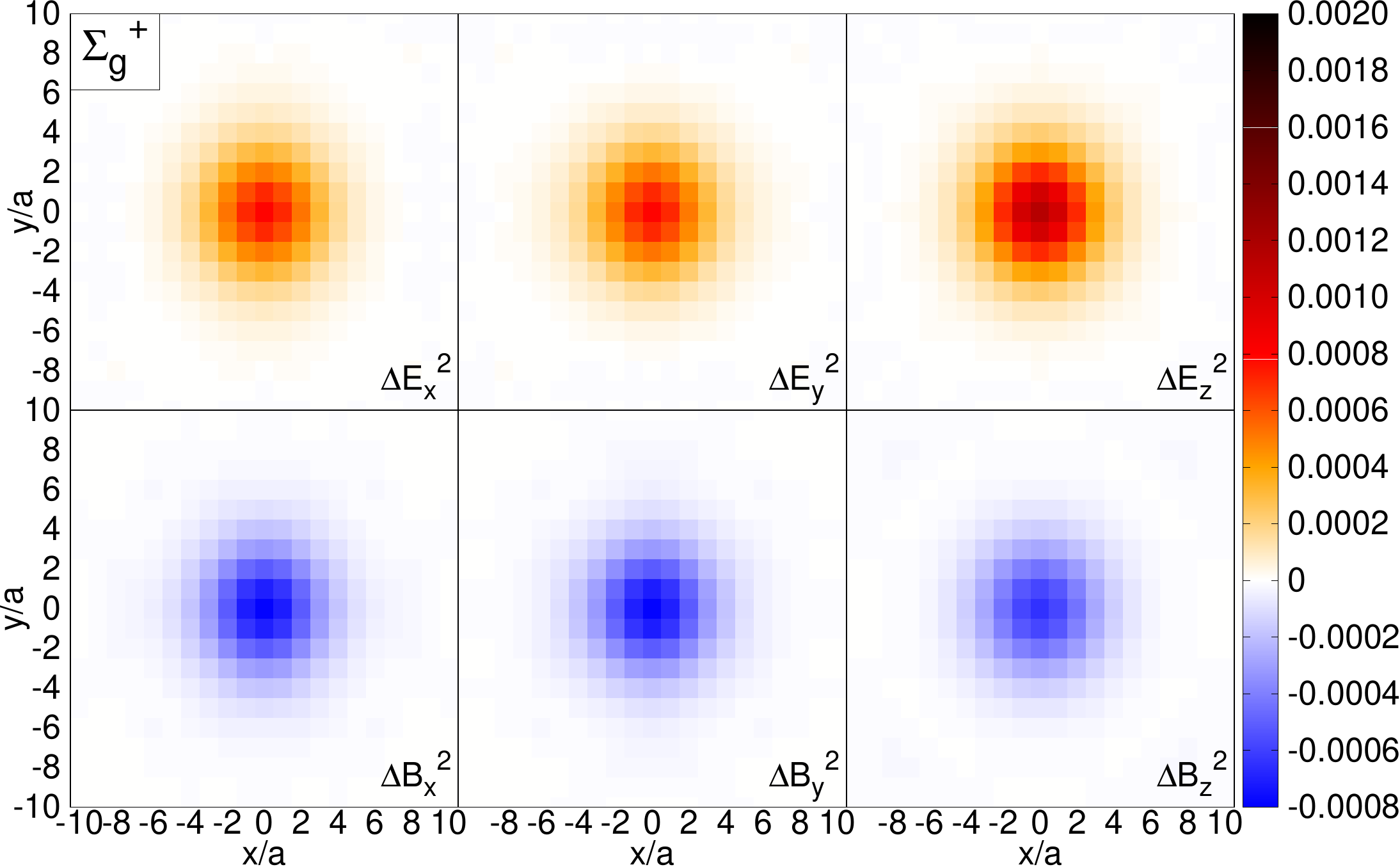}
\includegraphics[width=7.0cm]{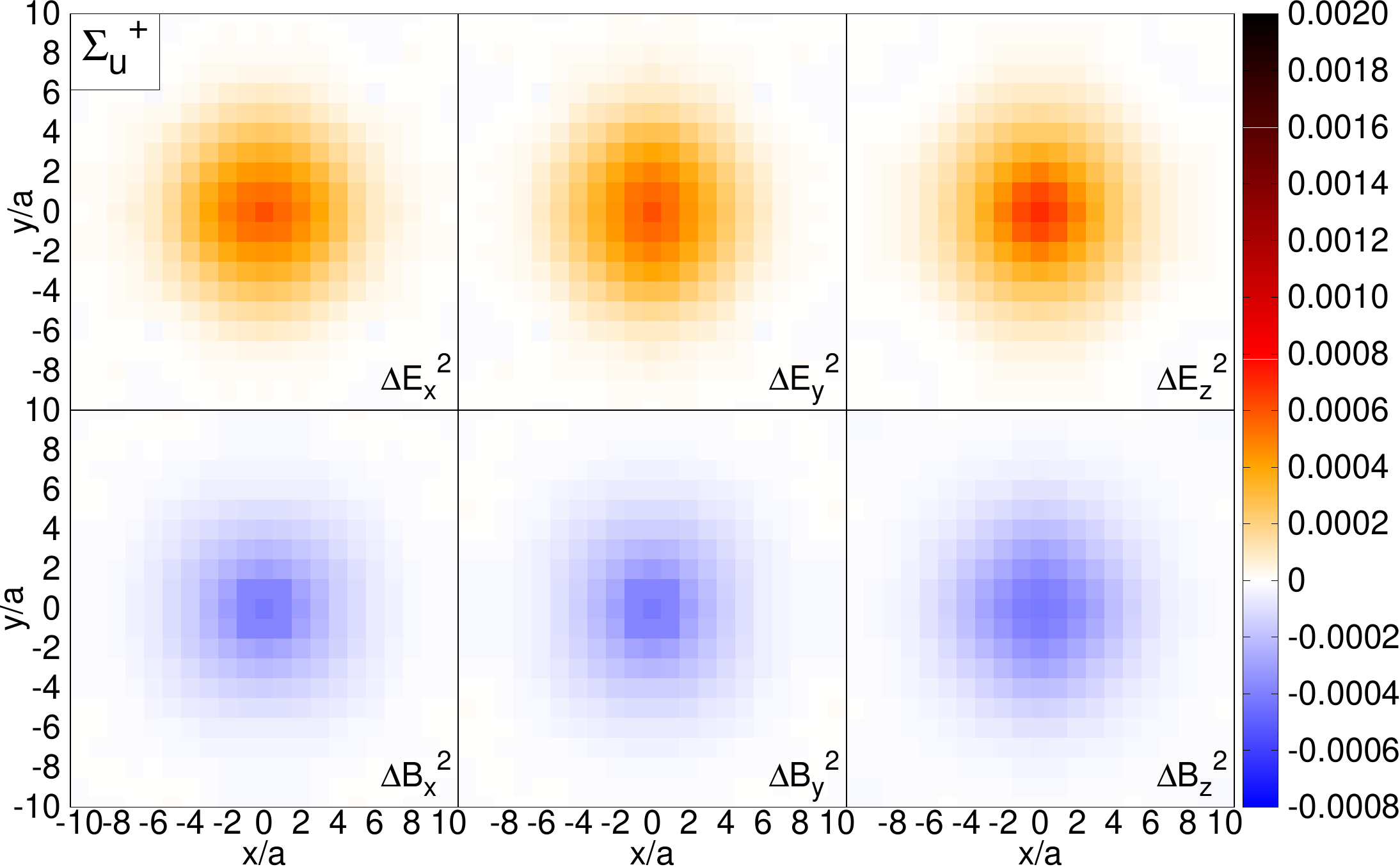}
\includegraphics[width=7.0cm]{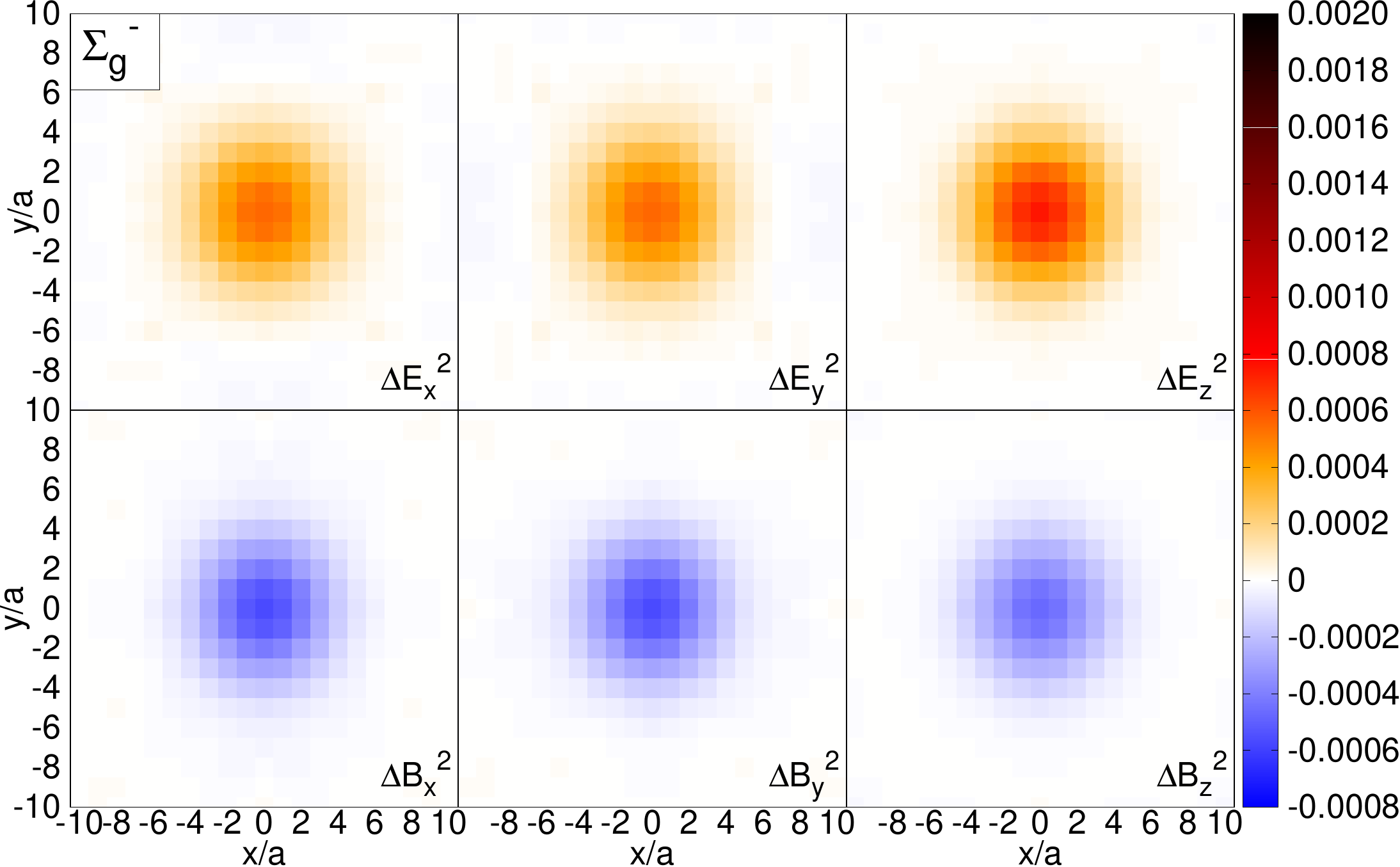}
\includegraphics[width=7.0cm]{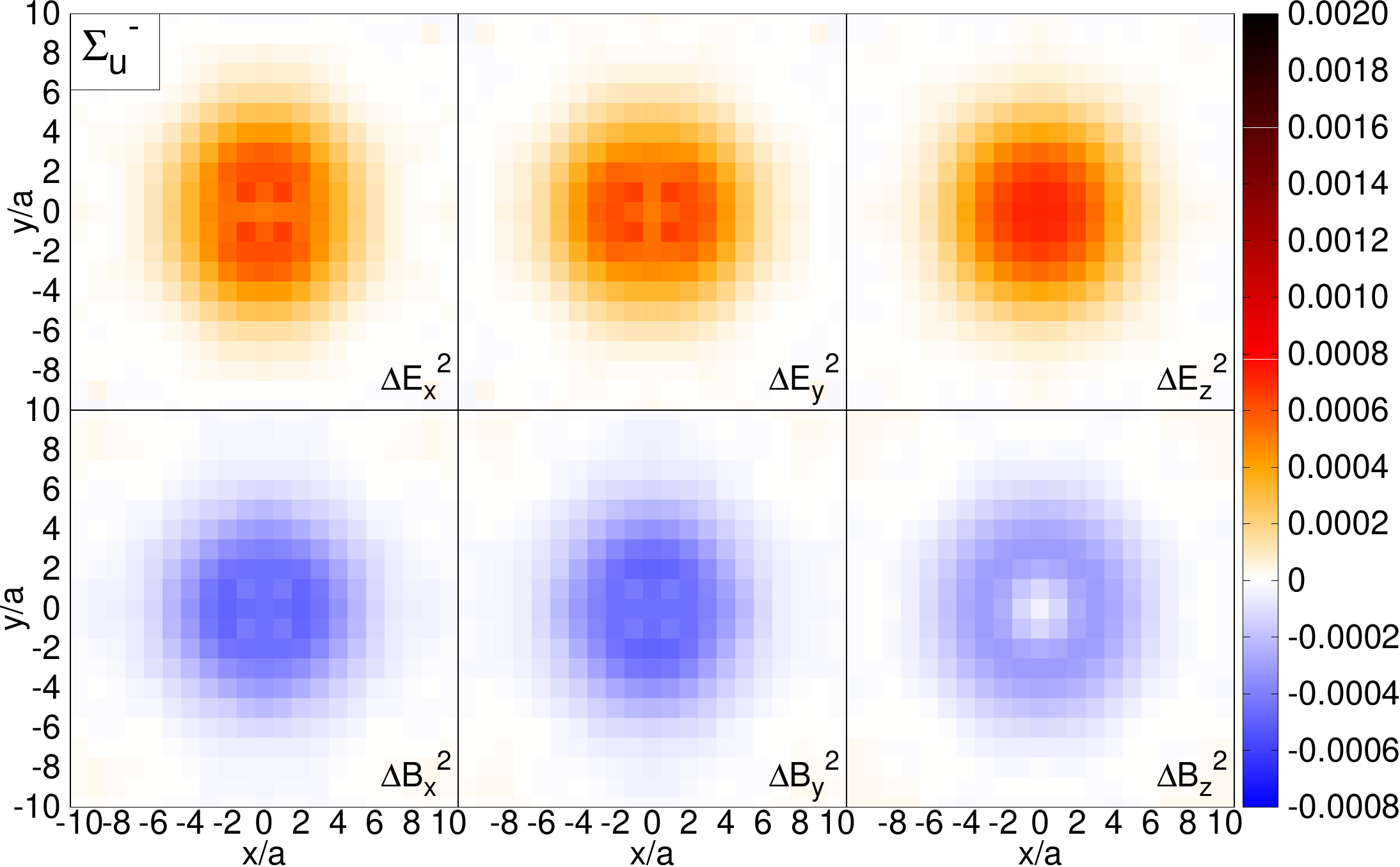}
\includegraphics[width=7.0cm]{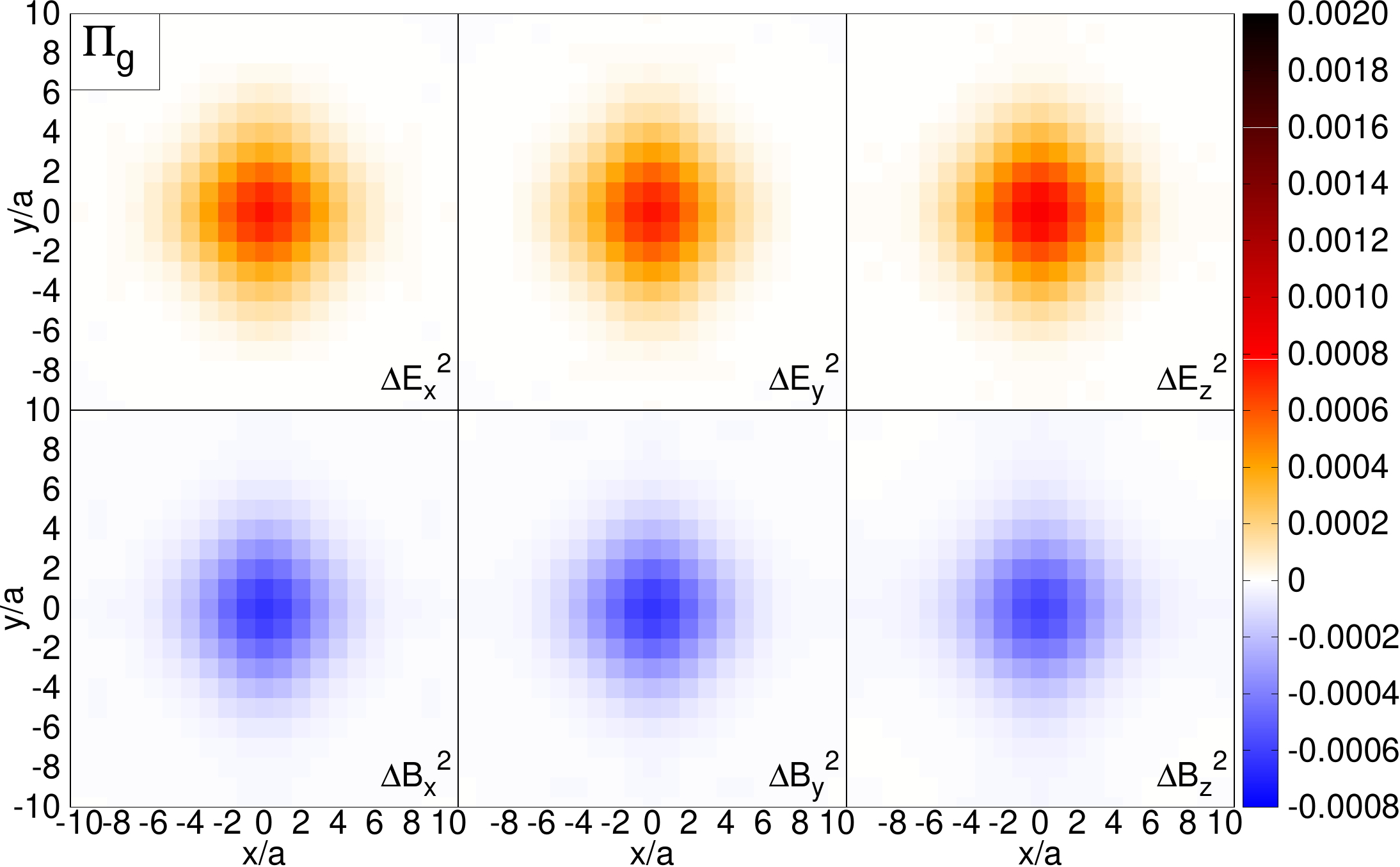}
\includegraphics[width=7.0cm]{R10_Pi_u_heatmap_med_symm.pdf}
\includegraphics[width=7.0cm]{R10_Delta_g_heatmap_med_symm.pdf}
\includegraphics[width=7.0cm]{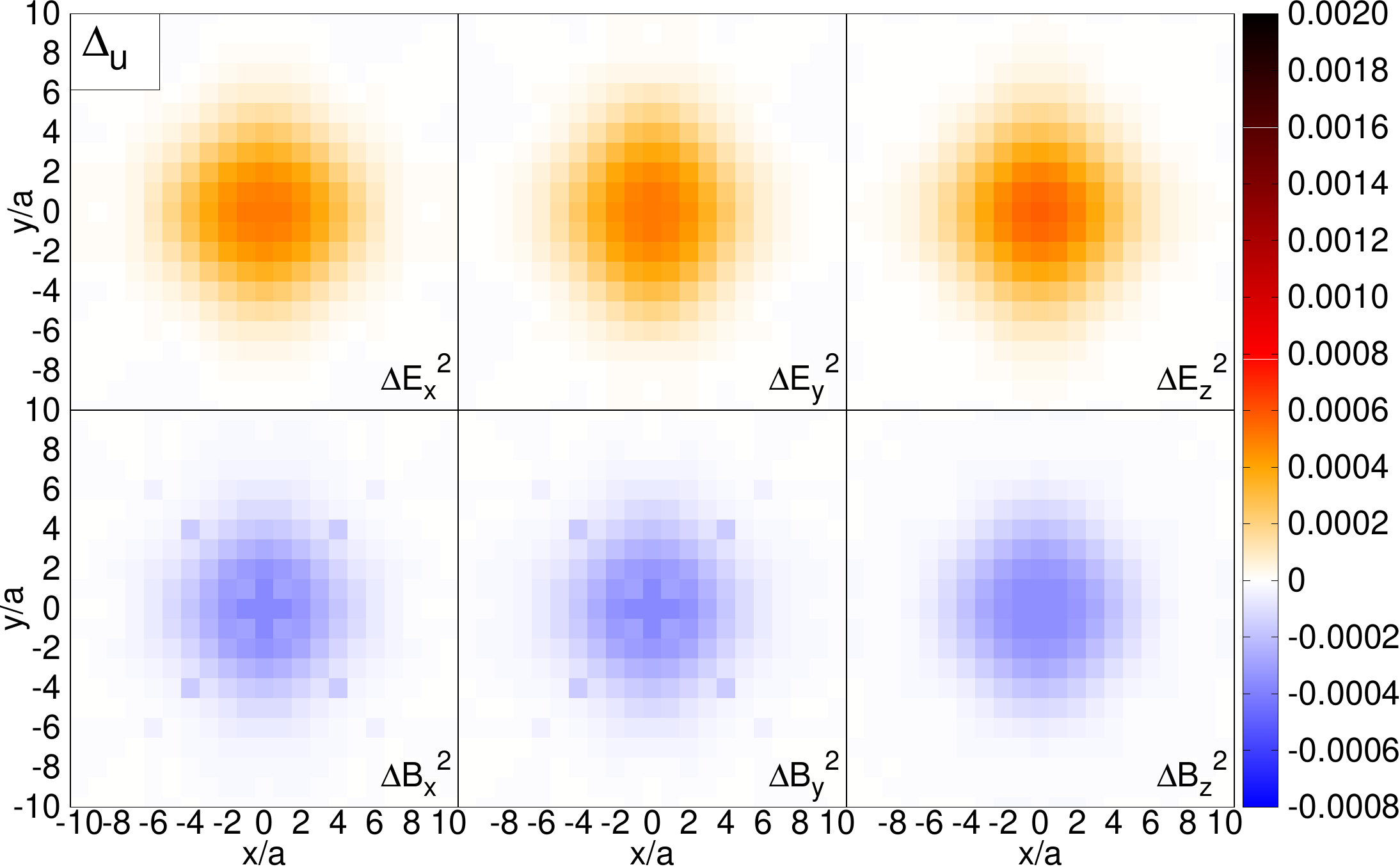}
\end{center}
\caption{\label{FIG_med_10}Flux densities $\Delta F_{j,\Lambda_\eta^{(\epsilon)}}^2(r;\mathbf{x} = (x,y,0))$, $j = x,y,z$ in the mediator plane for gauge group SU(2), all investigated sectors $\Lambda_\eta^{(\epsilon)} = \Sigma_g^+, \Sigma_u^+, \Sigma_g^-, \Sigma_u^-, \Pi_g, \Pi_u, \Delta_g, \Delta_u$ and $Q \bar{Q}$ separation $r = 10 \, a$.}
\end{figure}

\begin{figure}[p]
\begin{center}
\includegraphics[width=15.0cm]{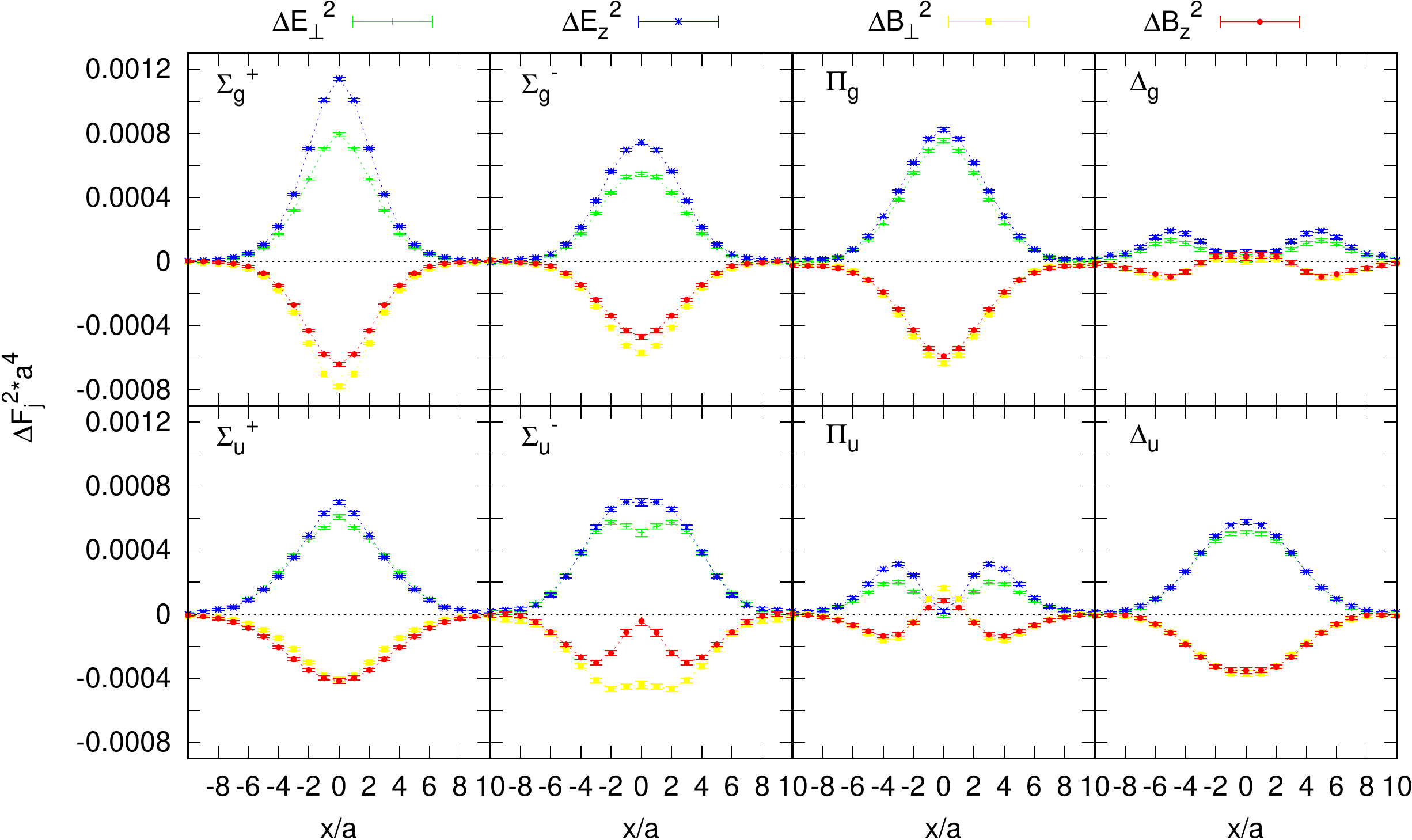}

\vspace{+0.5cm}
\includegraphics[width=15.0cm]{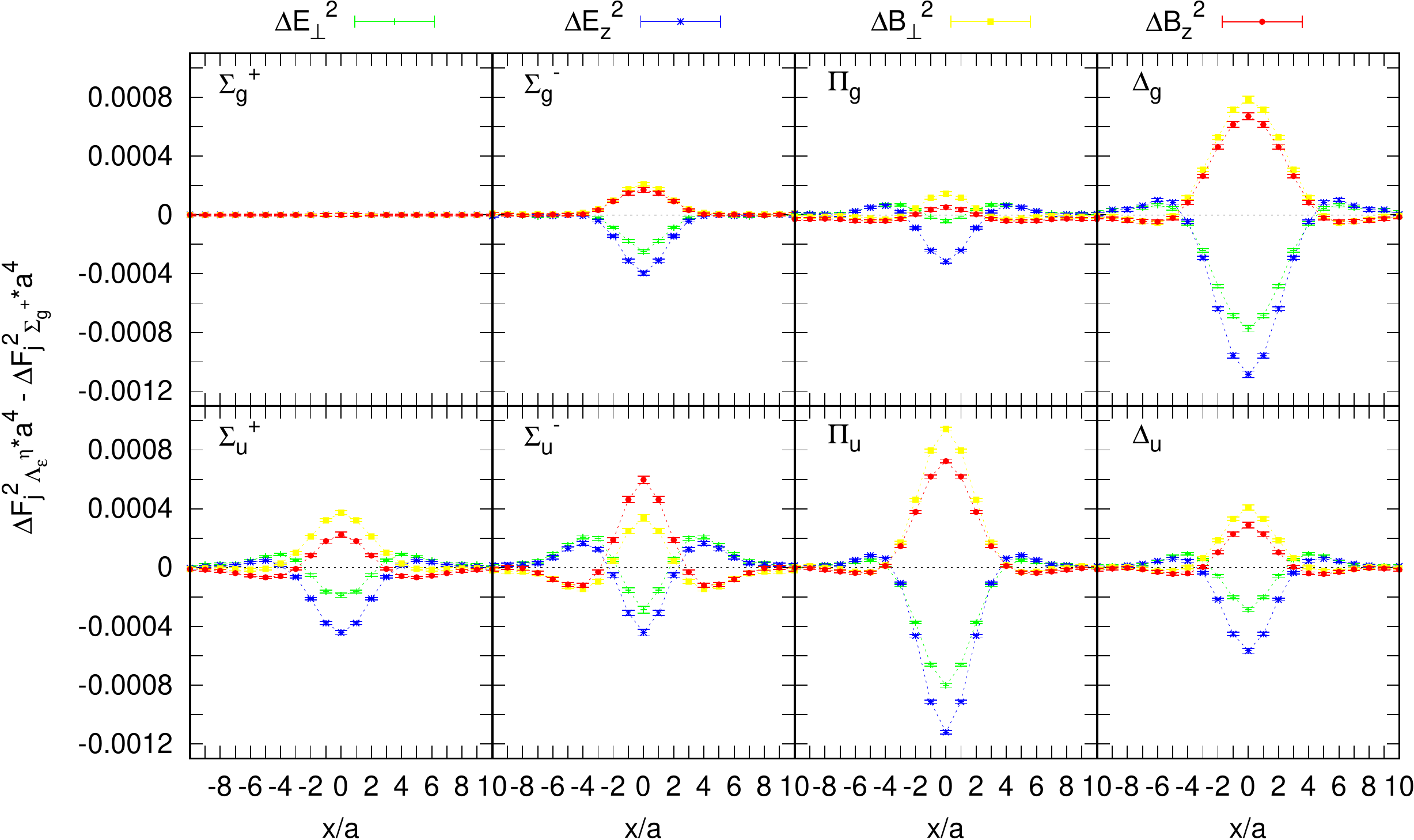}
\end{center}
\caption{\label{FIG_med_10_x_axis}Flux densities on the $x$ axis for gauge group SU(2), all investigated sectors \\ $\Lambda_\eta^{(\epsilon)} = \Sigma_g^+, \Sigma_u^+, \Sigma_g^-, \Sigma_u^-, \Pi_g, \Pi_u, \Delta_g, \Delta_u$ and $Q \bar{Q}$ separation $r = 10 \, a$. \\ \textbf{(top)} $\Delta F_{j,\Lambda_\eta^{(\epsilon)}}^2(r;\mathbf{x} = (x,0,0))$, $j = \perp,z$. \\ \textbf{(bottom)}~$\Delta F_{j,\Lambda_\eta^{(\epsilon)}}^2(r;\mathbf{x} = (x,0,0)) - \Delta F_{j,\Sigma_g^+}^2(r;\mathbf{x} = (x,0,0))$, $j = \perp,z$.}
\end{figure}

\begin{figure}[p]
\begin{center}
\includegraphics[width=7.0cm]{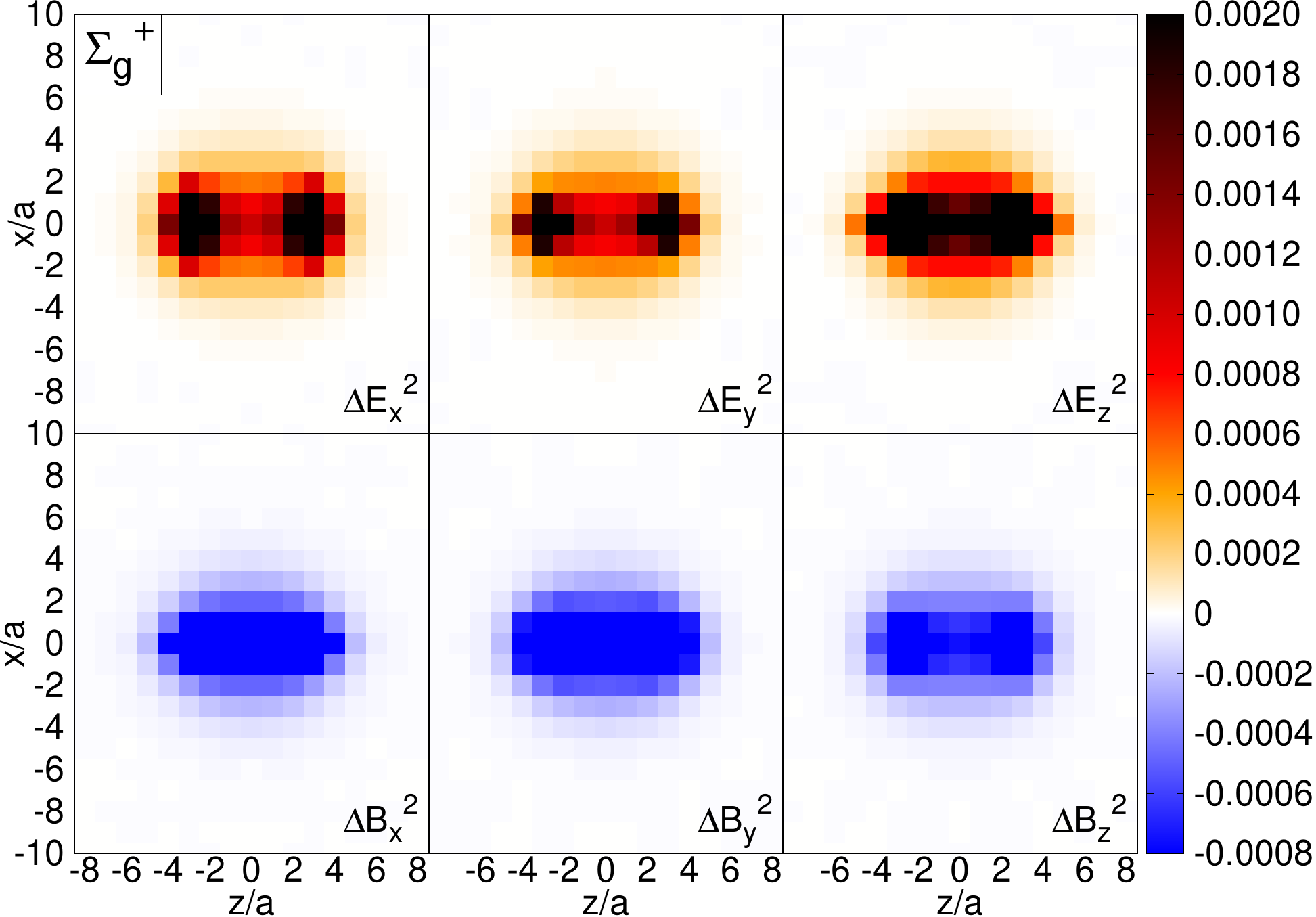}
\includegraphics[width=7.0cm]{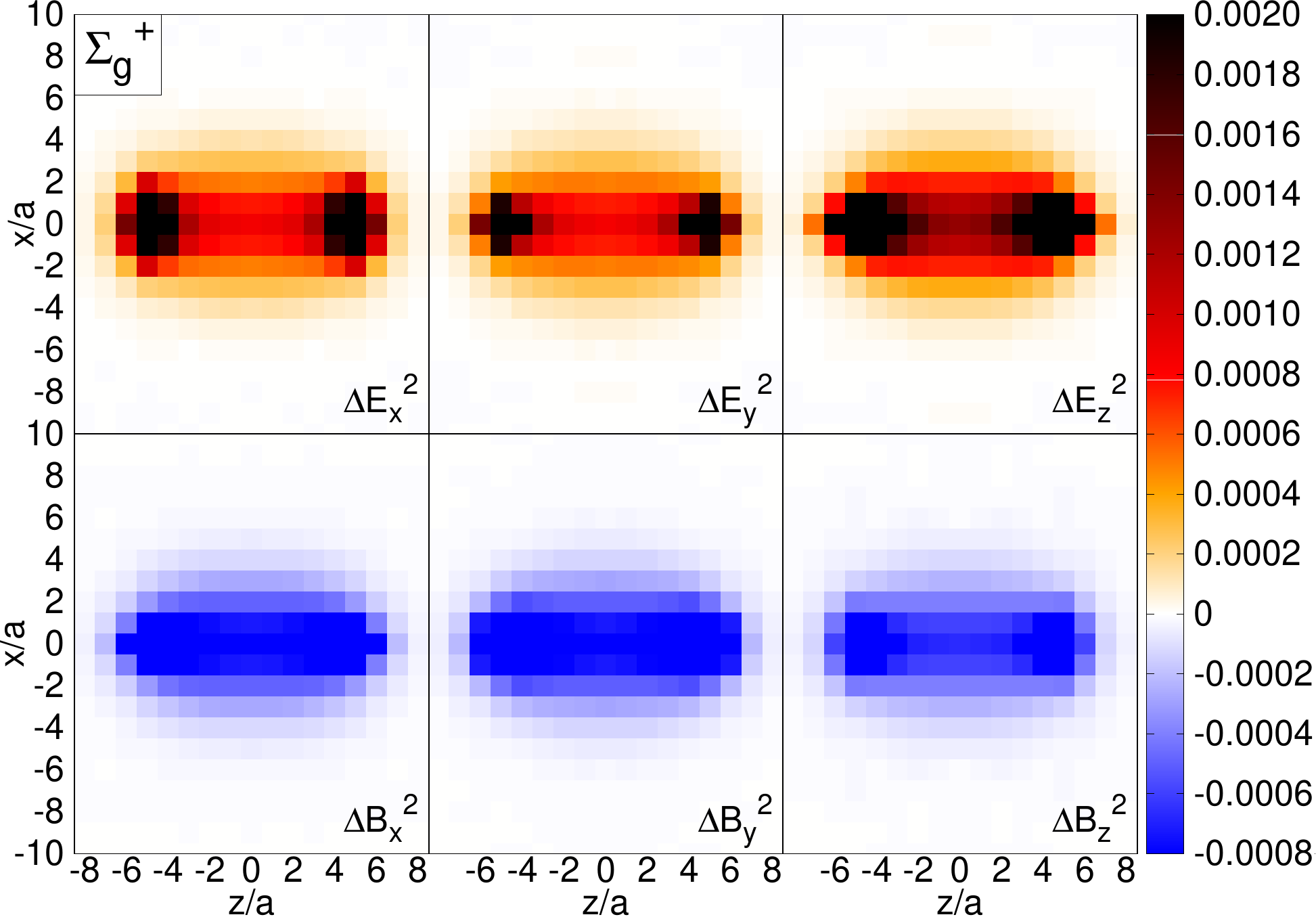}
\includegraphics[width=7.0cm]{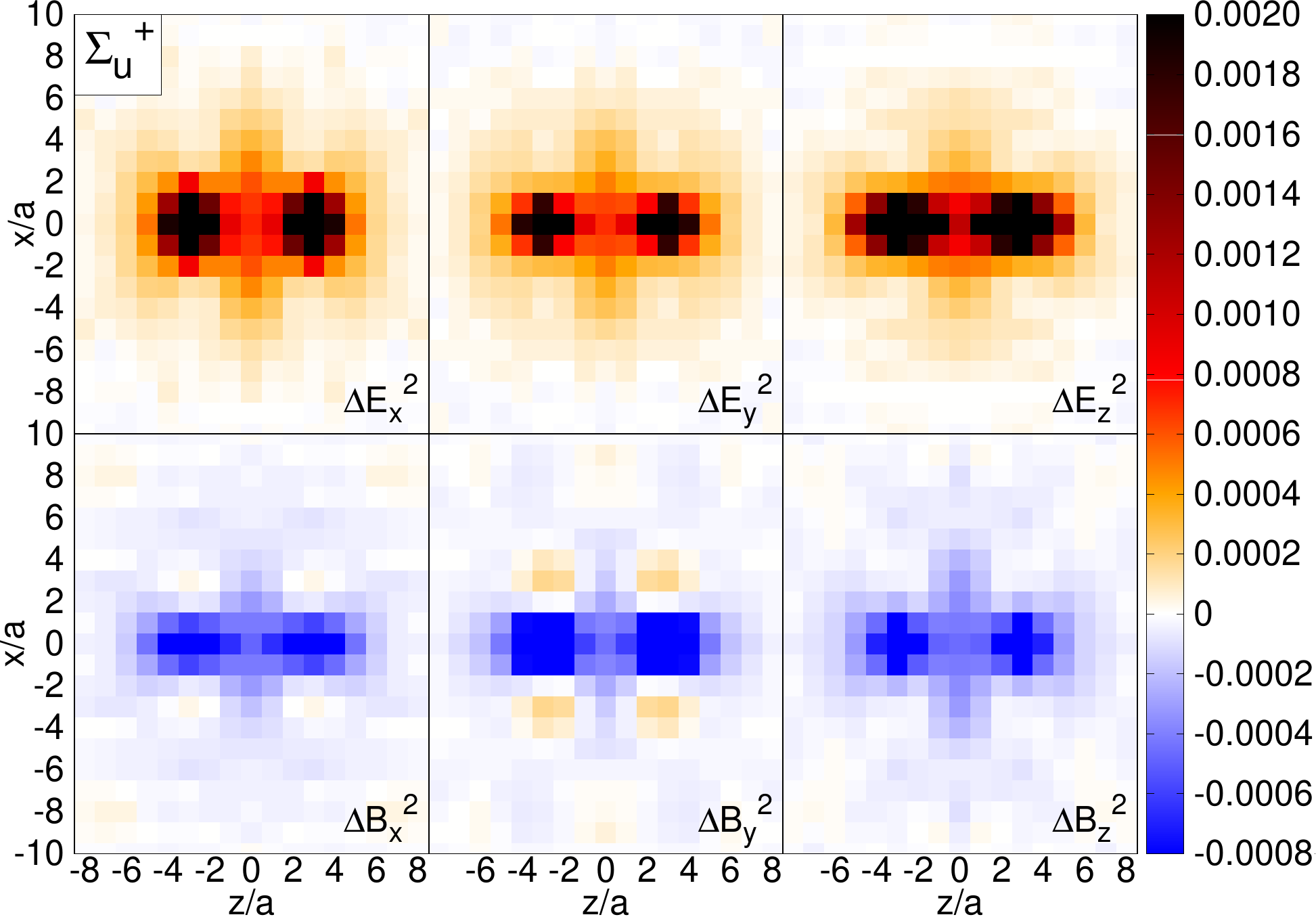}
\includegraphics[width=7.0cm]{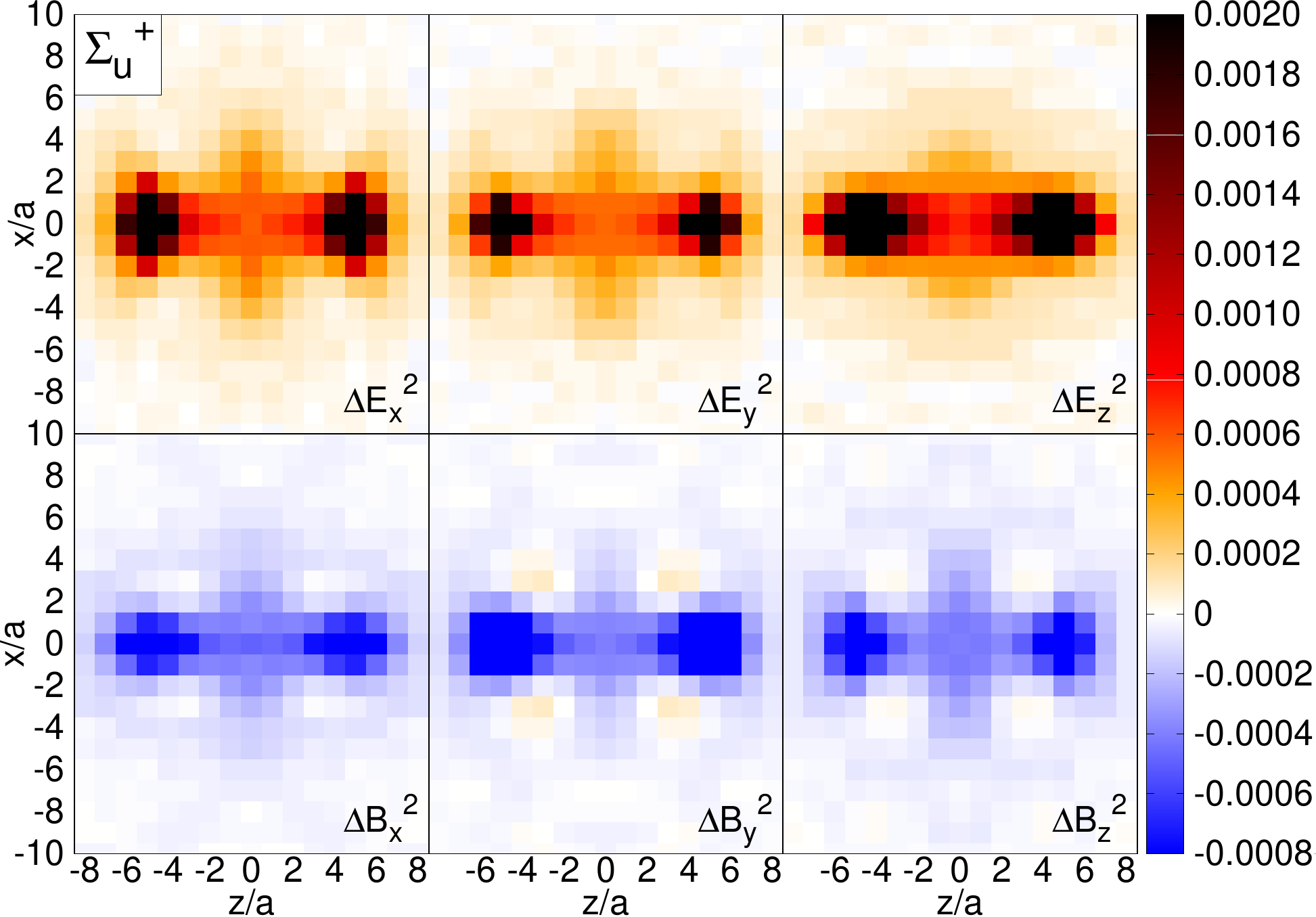}
\includegraphics[width=7.0cm]{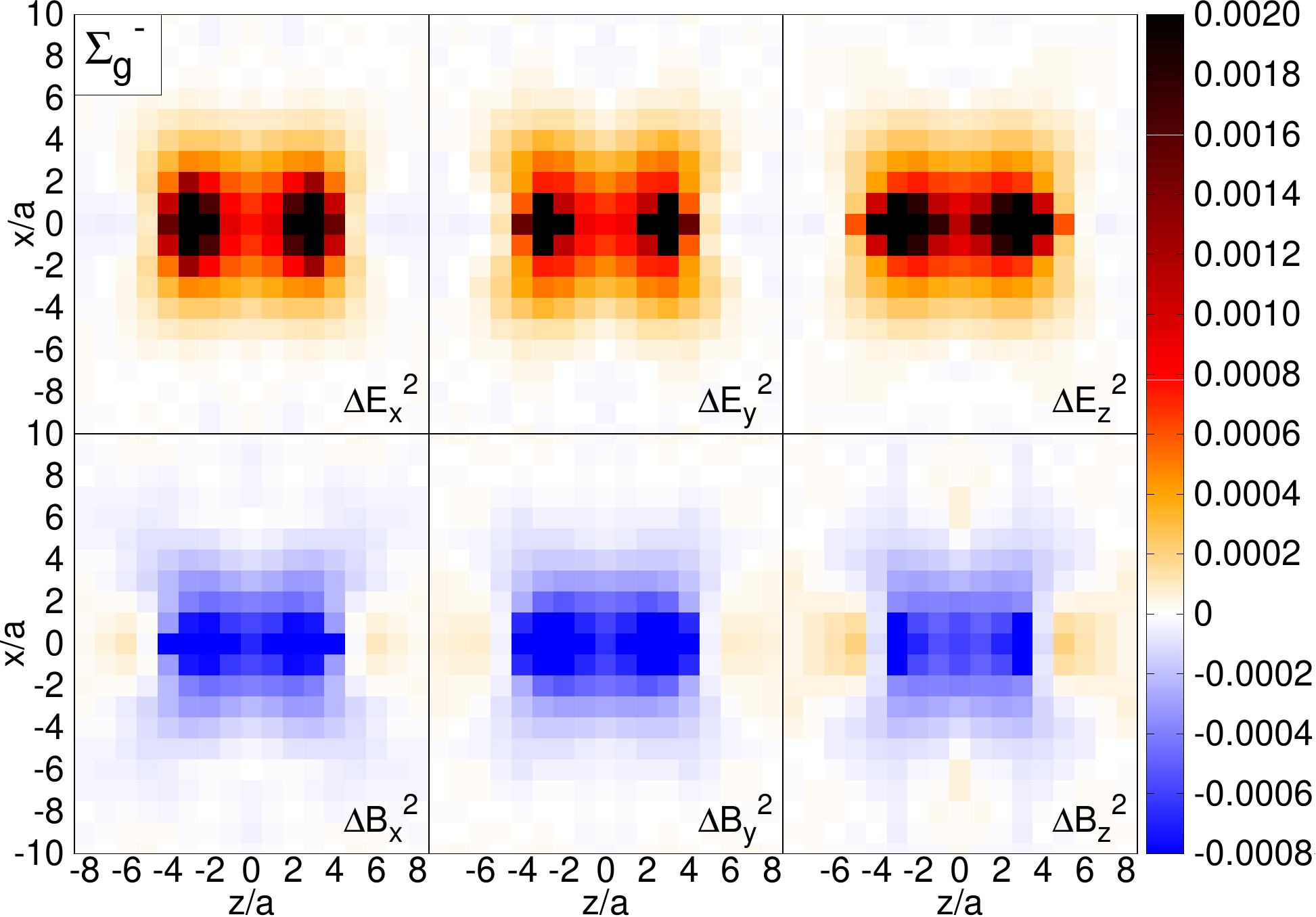}
\includegraphics[width=7.0cm]{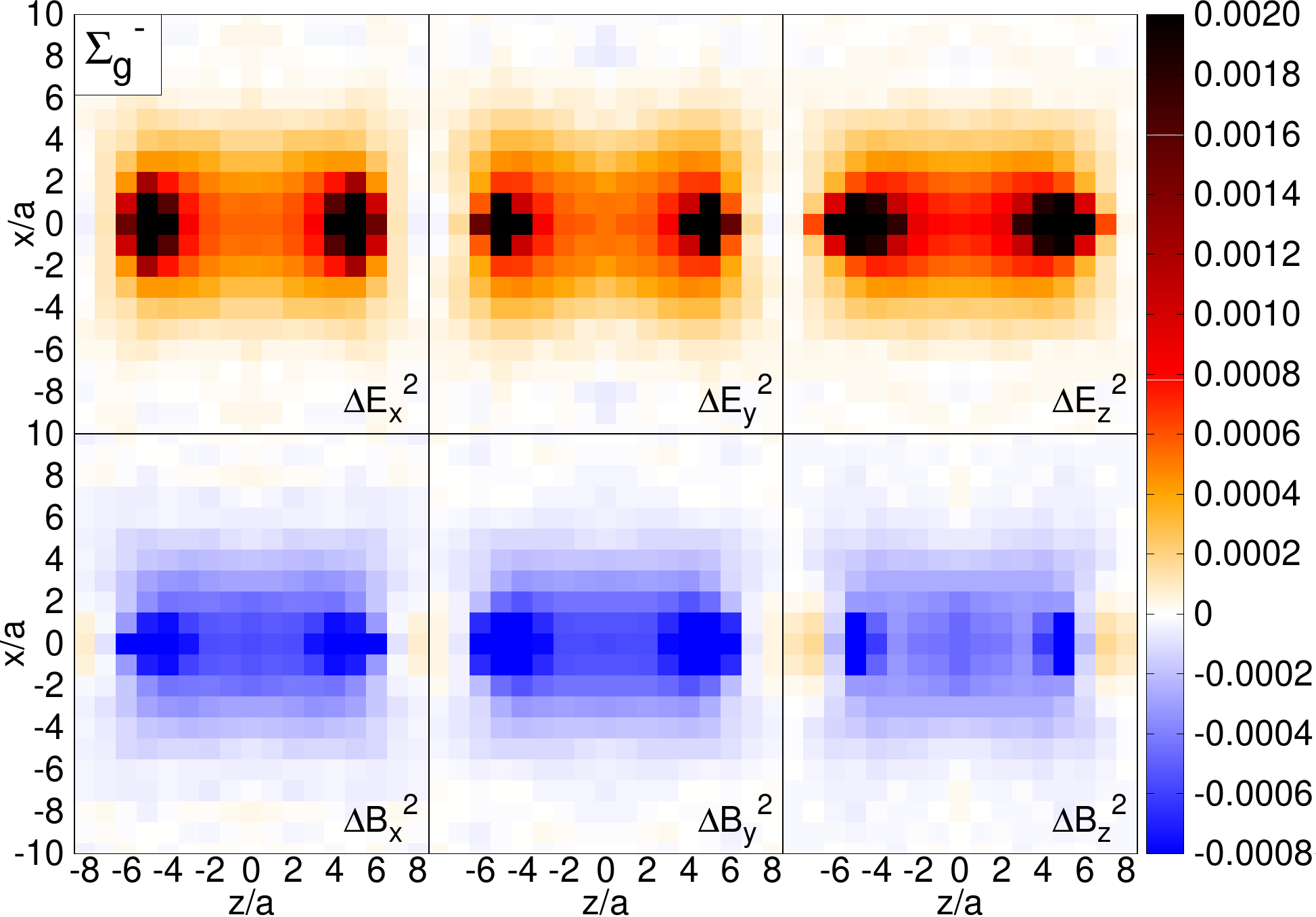}
\includegraphics[width=7.0cm]{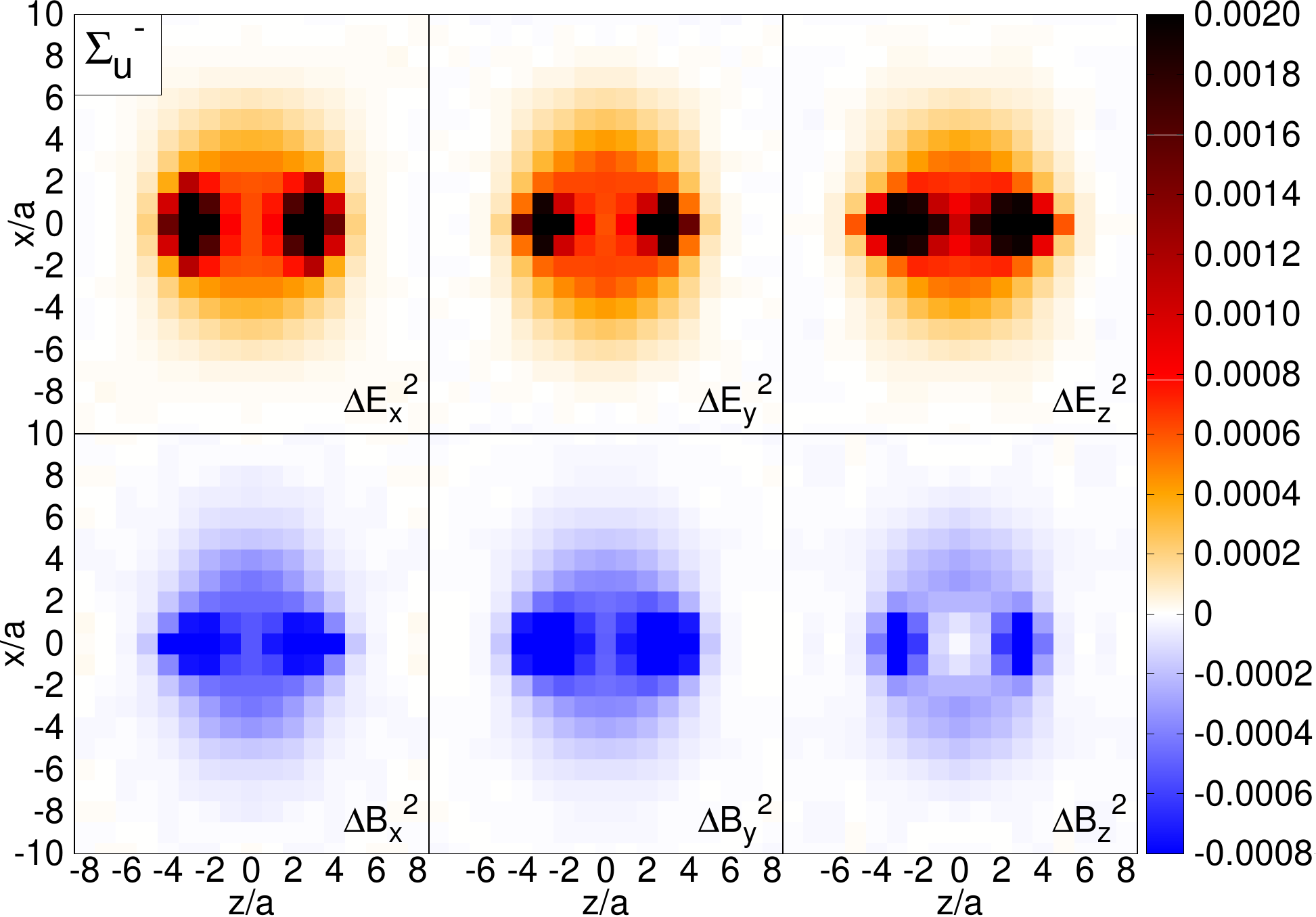}
\includegraphics[width=7.0cm]{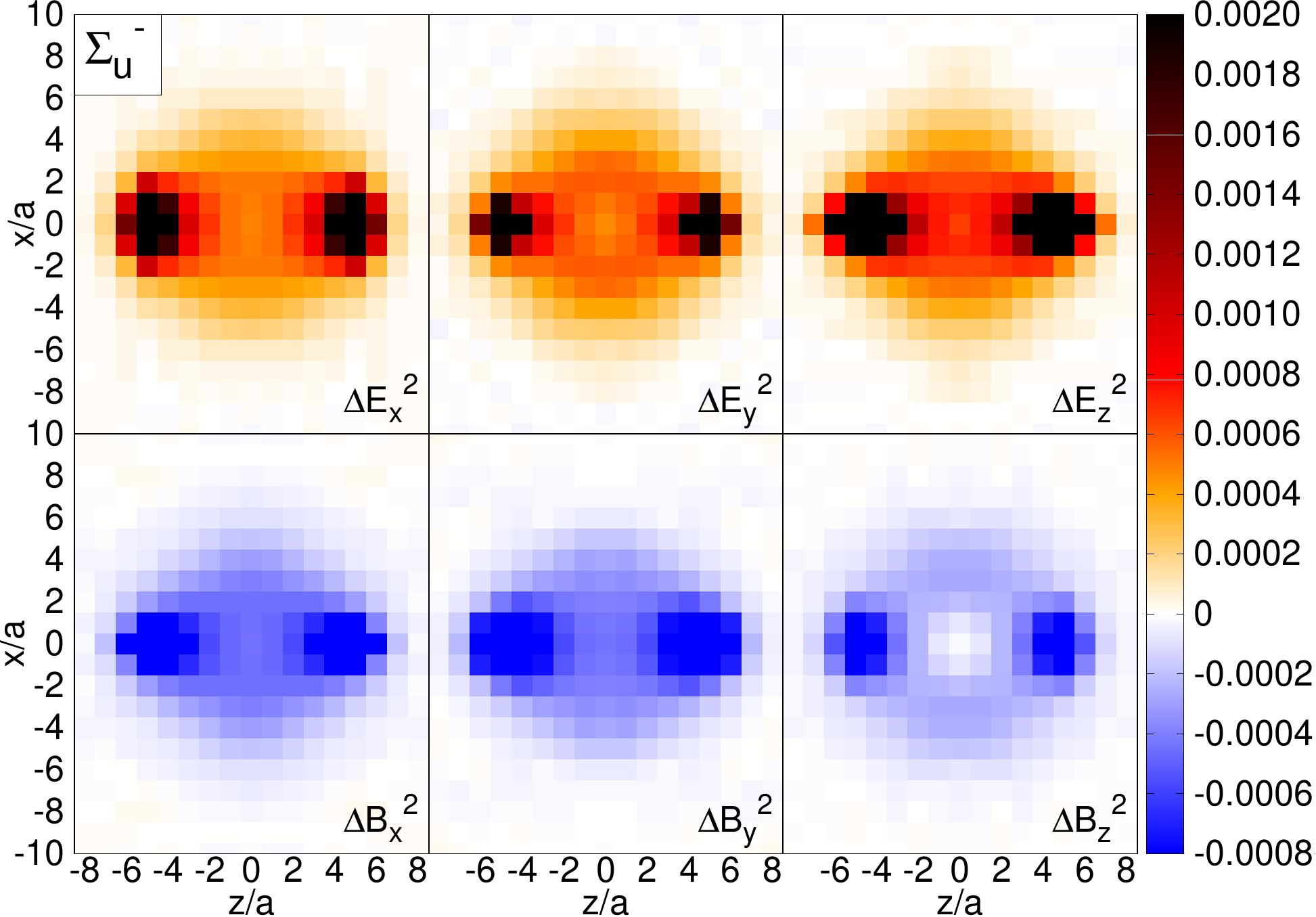}
\end{center}
\caption{\label{FIG_sep_Sigma}Flux densities $\Delta F_{j,\Lambda_\eta^\epsilon}^2(r;\mathbf{x} = (x,0,z))$, $j = x,y,z$ in the separation plane for gauge group SU(2) and sectors $\Lambda_\eta^\epsilon = \Sigma_g^+, \Sigma_u^+, \Sigma_g^-, \Sigma_u^-$. \textbf{(left)} $Q \bar{Q}$ separation $r = 6 \, a$. \textbf{(right)} $Q \bar{Q}$ separation $r = 10 \, a$.}
\end{figure}

\begin{figure}[p]
\begin{center}
\includegraphics[width=7.0cm]{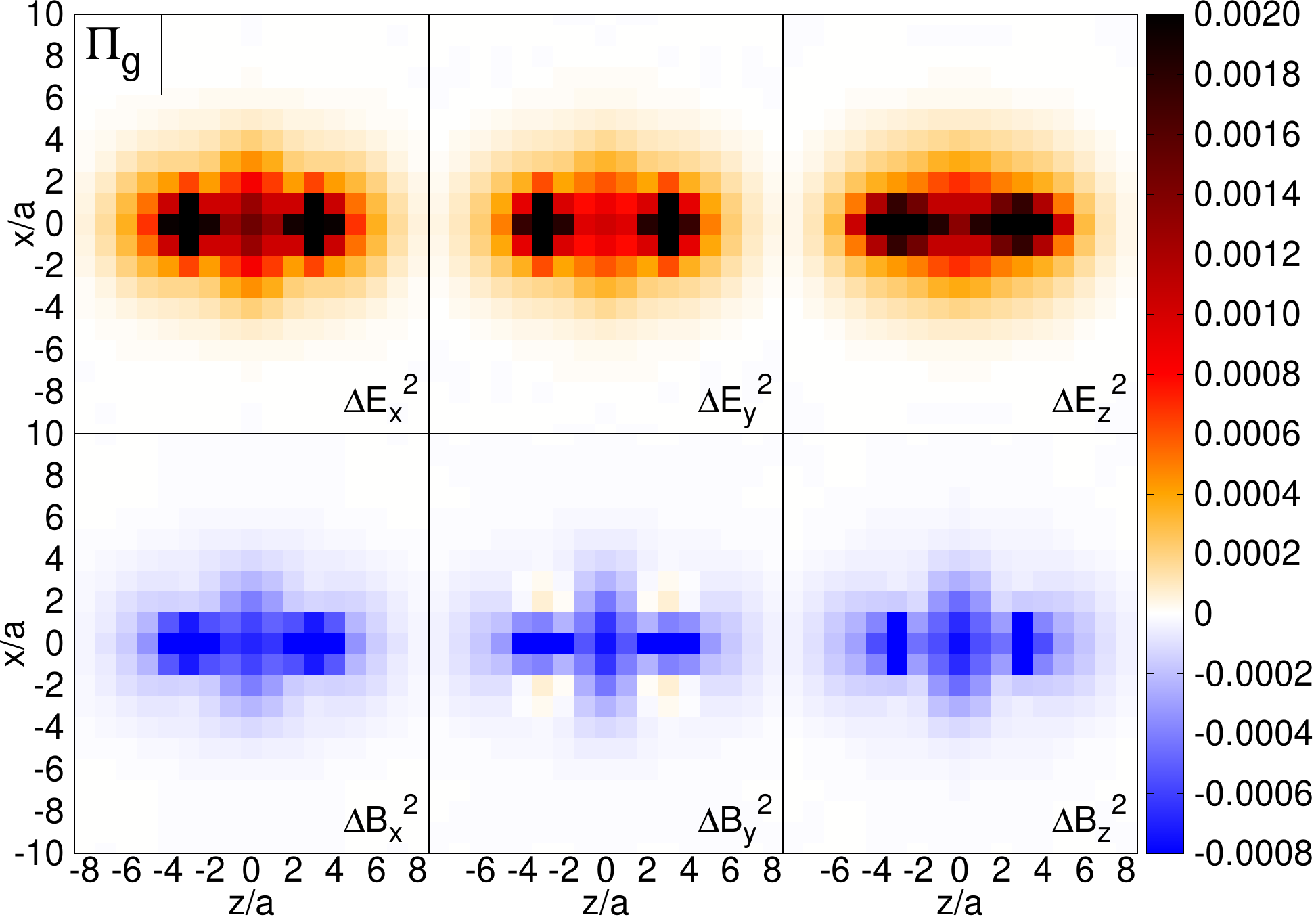}
\includegraphics[width=7.0cm]{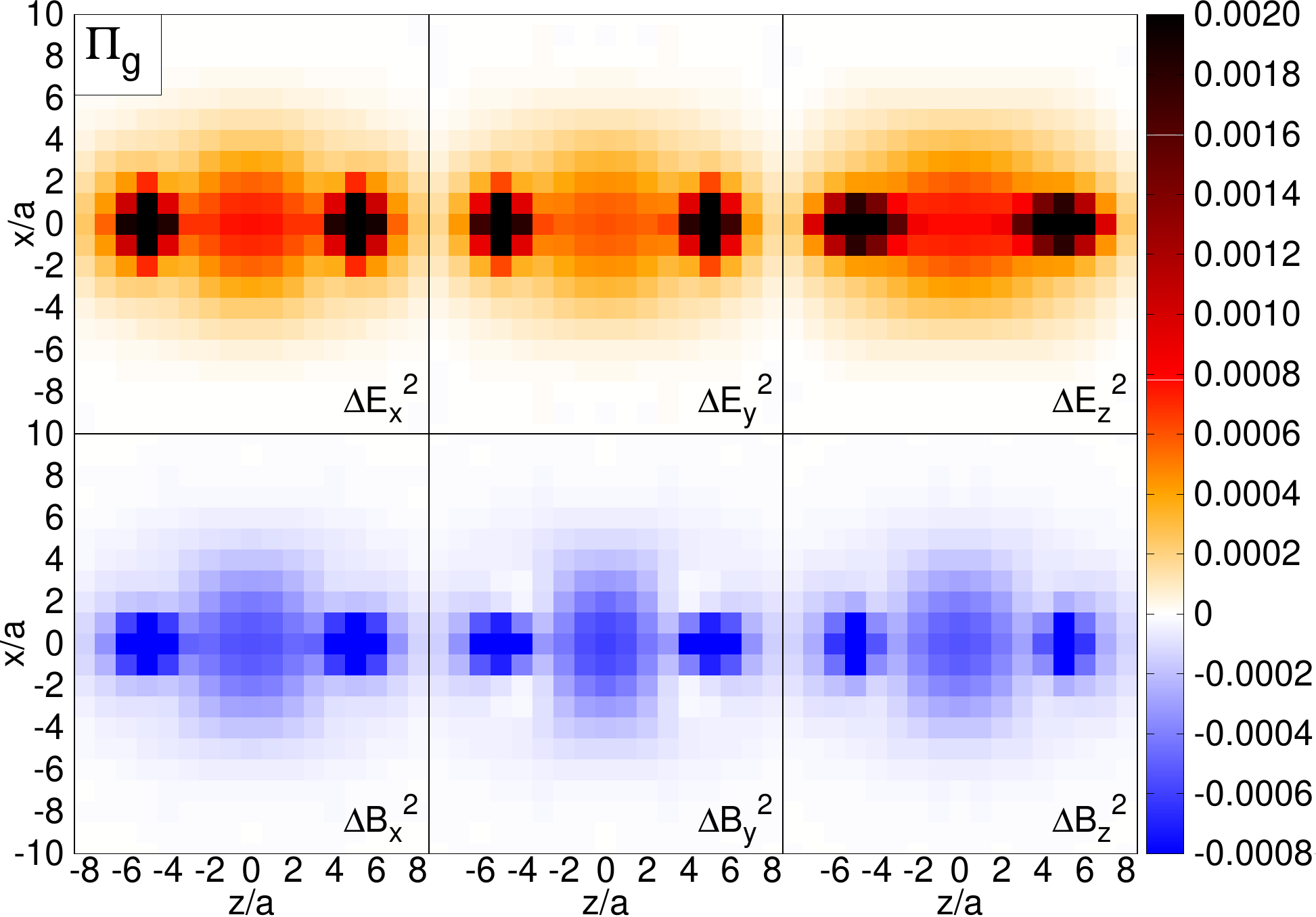}
\includegraphics[width=7.0cm]{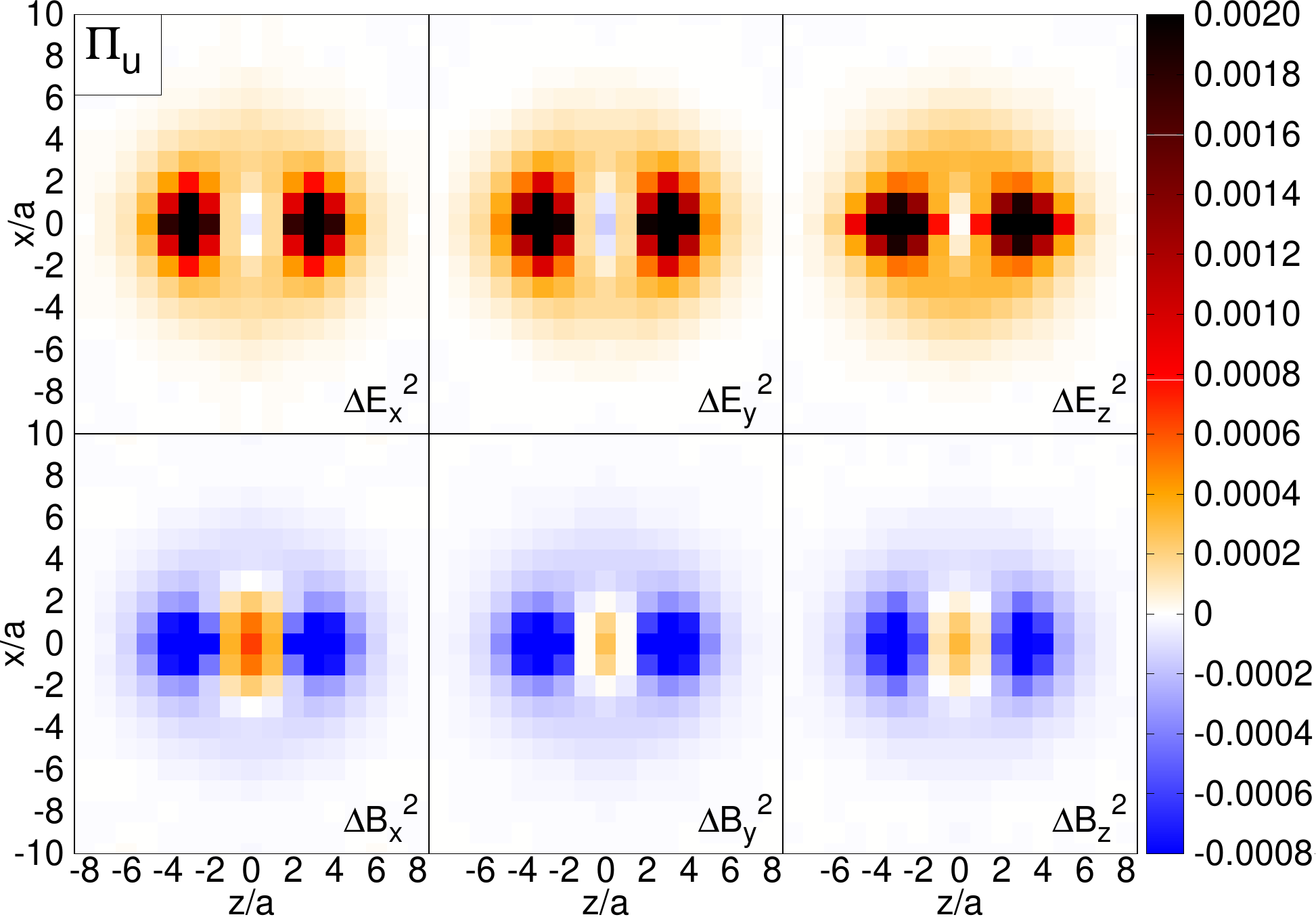}
\includegraphics[width=7.0cm]{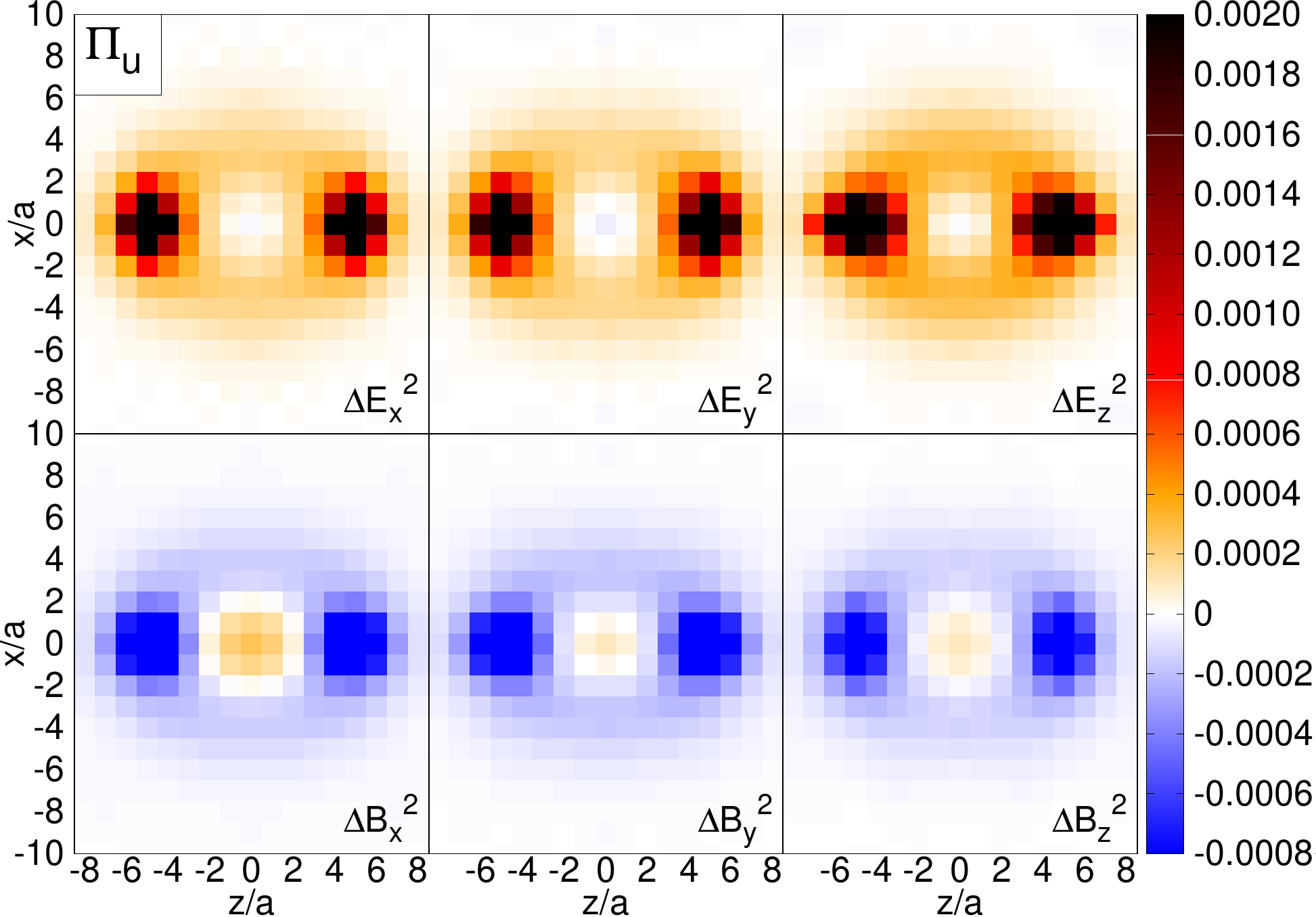}
\includegraphics[width=7.0cm]{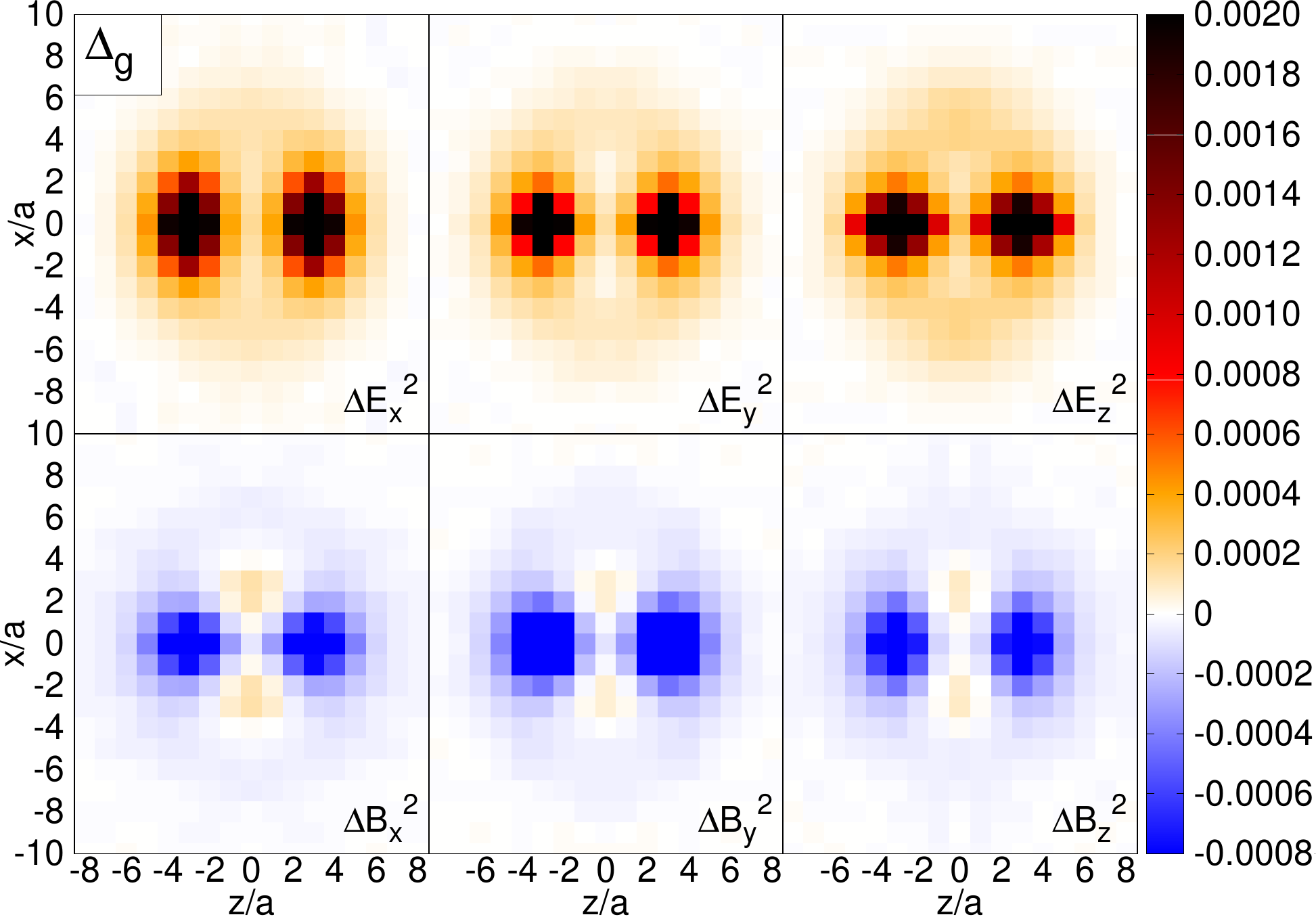}
\includegraphics[width=7.0cm]{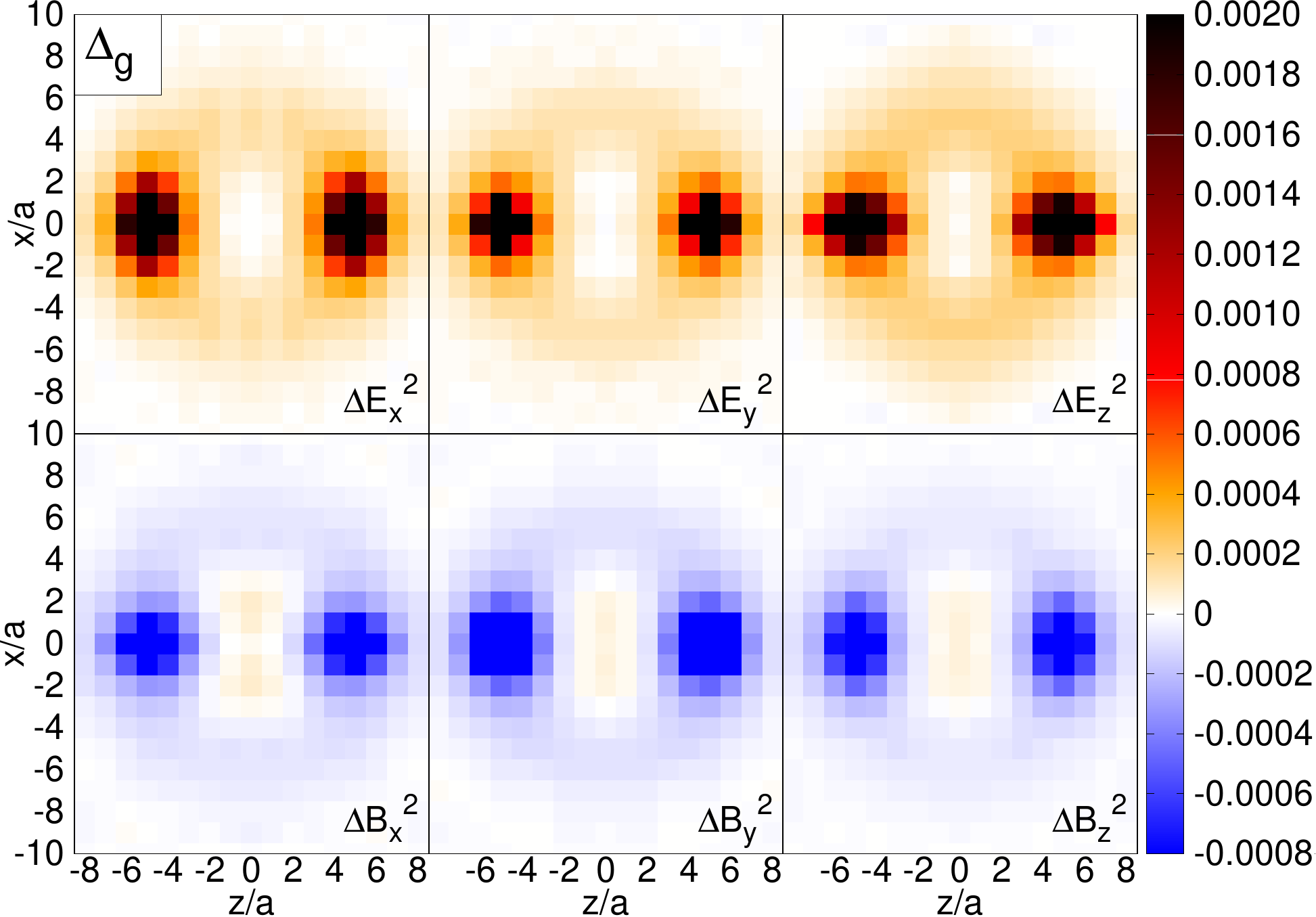}
\includegraphics[width=7.0cm]{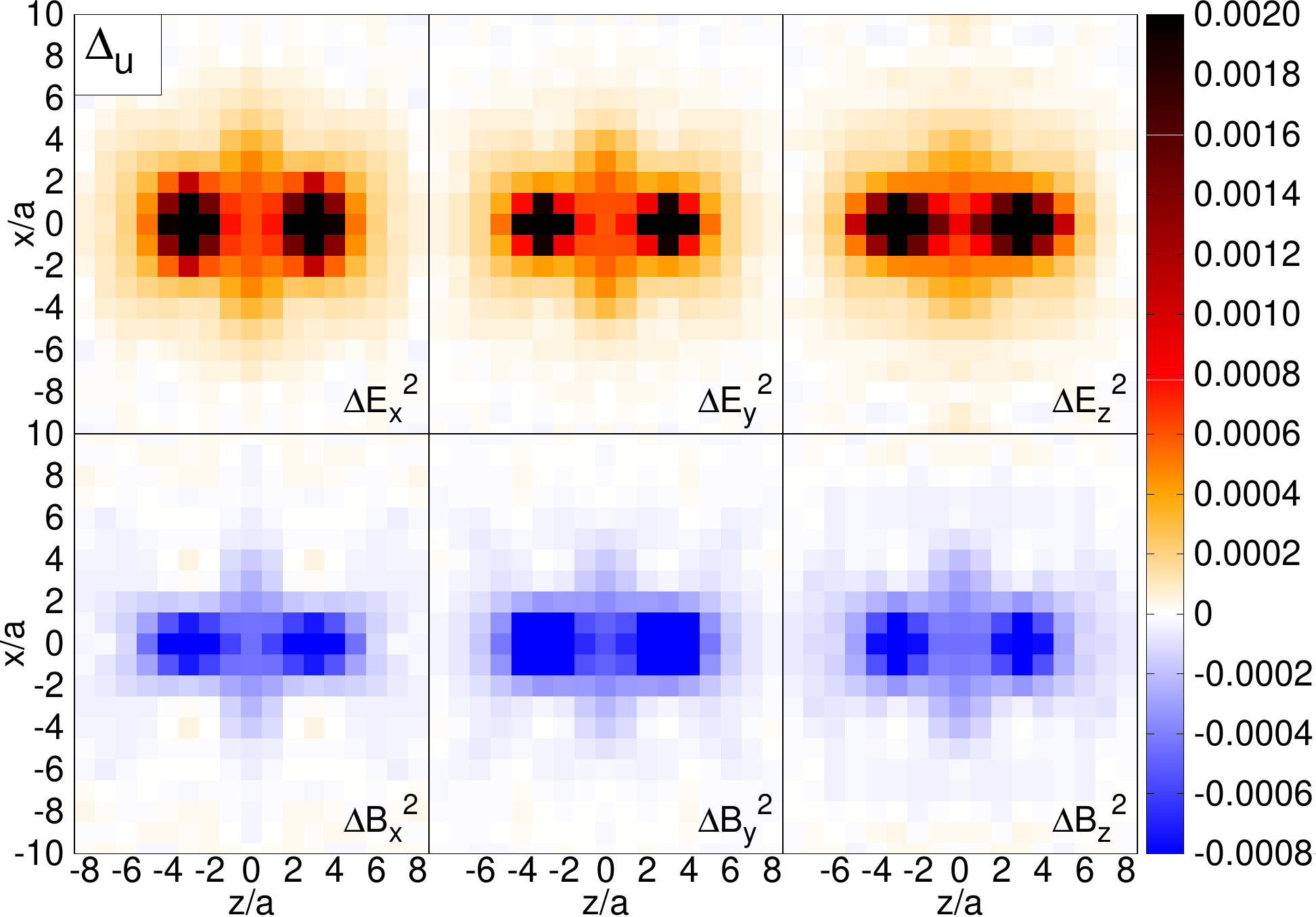}
\includegraphics[width=7.0cm]{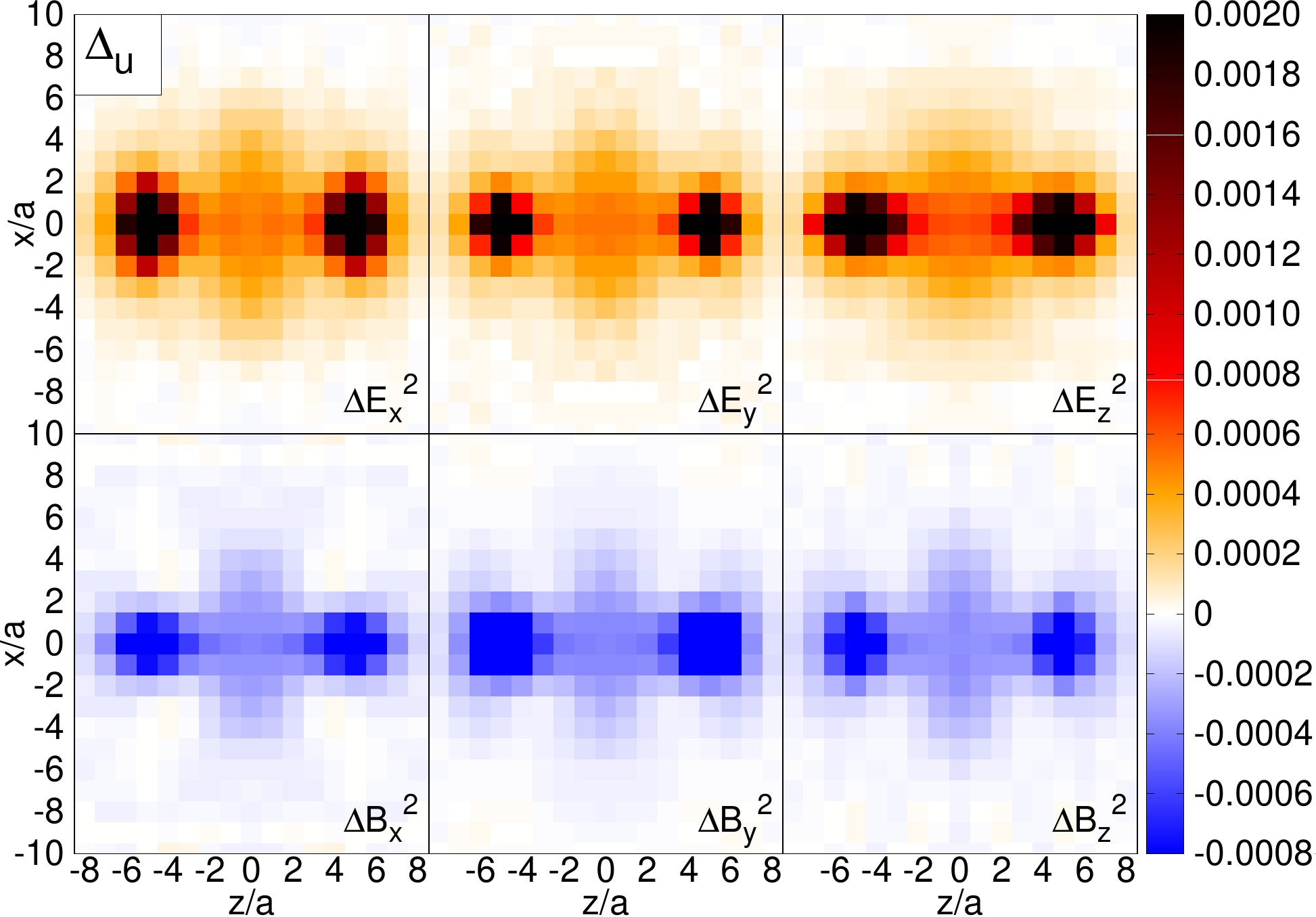}
\end{center}
\caption{\label{FIG_sep_Pi_Delta}Flux densities $\Delta F_{j,\Lambda_\eta}^2(r;\mathbf{x} = (x,0,z))$, $j = x,y,z$ in the separation plane for gauge group SU(2) and sectors $\Lambda_\eta = \Pi_g, \Pi_u, \Delta_g, \Delta_u$. \textbf{(left)} $Q \bar{Q}$ separation $r = 6 \, a$. \textbf{(right)} $Q \bar{Q}$ separation $r = 10 \, a$.}
\end{figure}

Since the resulting flux densities in the mediator plane for $r = 6 \, a$ and $r = 10 \, a$ are very similar, we only present them for $r = 10 \, a$. In Figure~\ref{FIG_med_10} these flux densities $\Delta F_{j,\Lambda_\eta^{(\epsilon)}}^2(r;\mathbf{x}=(x,y,0))$, $j = x,y,z$ are shown as 2D color maps. In the upper panel of Figure~\ref{FIG_med_10_x_axis} we present similar results, the rotationally invariant $\Delta F_{j,\Lambda_\eta^{(\epsilon)}}^2(r;\mathbf{x}=(x,0,0))$, $j = \perp,z$ along the $x$ axis, i.e.\ in the mediator plane as a function of the radial coordinate. In contrast to the 2D color maps, these 1D curves allow to also show statistical errors and, thus, provide information about the precision of our numerical results. In the lower panel of Figure~\ref{FIG_med_10_x_axis} we present in the same style differences of hybrid static potential flux densities to those of the ordinary static potential, i.e.\ \\ $\Delta F_{j,\Lambda_\eta^{(\epsilon)}}^2(r;\mathbf{x} = (x,0,0)) - \Delta F_{j,\Sigma_g^+}^2(r;\mathbf{x} = (x,0,0))$, $j = \perp,z$. Flux densities \\ $\Delta F_{j,\Lambda_\eta^{(\epsilon)}}^2(r;\mathbf{x} = (x,0,z))$, $j = x,y,z$ in the separation plane are shown as 2D color maps in Figure~\ref{FIG_sep_Sigma} and Figure~\ref{FIG_sep_Pi_Delta} for both separations $r = 6 \, a$ and $r = 10 \, a$. Note that flux densities close to one of the static charges, in particular for $|\mathbf{x} - \mathbf{r}_{Q}| \leq a$ or $|\mathbf{x} - \mathbf{r}_{\bar{Q}}| \leq a$, exhibit sizable discretization errors (see the discussion in section~\ref{SEC575}).

The flux densities of the ordinary static potential form a cigar-shaped flux tube with strong positive contributions to the energy density from the chromoelectric and smaller negative contributions from the chromomagnetic field strength components. The maxima are on the $Q \bar{Q}$ separation axis, i.e.\ at $x = y = 0$. While this is known from previous lattice gauge theory investigations of the ordinary static potential (see e.g.\ ref.\ \cite{Bali:1994de}), the corresponding flux densities for hybrid static potentials show a variety of different and interesting structures. For example chromomagnetic flux densities of hybrid static potentials are typically larger close to the center of the flux tube than those of the ordinary static potential, as can be seen in Figure~\ref{FIG_med_10_x_axis}, lower panel. Hybrid static potential flux tubes are also wider, i.e.\ have a larger extension in $x$ and $y$ direction (see e.g.\ Figure~\ref{FIG_med_10_x_axis}, Figure~\ref{FIG_sep_Sigma} and Figure~\ref{FIG_sep_Pi_Delta}). Another interesting difference is that some hybrid static potentials show a clear reduction of the chromoelectric flux densities close to the center, while the chromomagnetic flux densities exhibit peaks (most prominently for $\Lambda_\eta = \Pi_u, \Delta_g$, but to some extent also for $\Lambda_\eta^\epsilon = \Sigma_u^-$). For other sectors, $\Lambda_\eta^{(\epsilon)} = \Sigma_u^+, \Pi_g, \Delta_u$, the opposite is the case, i.e.\ there is a positive localized peak at the center for the chromoelectric flux densities and a corresponding negative contribution of the chromomagnetic flux densities. These peaks in either the chromoelectric or chromomagnetic flux densities can be interpreted as ``valence gluons'' generating the hybrid quantum numbers, as discussed in models and phenomenological descriptions of hybrid mesons. The positive or negative peaks are surrounded by spherical shells, where flux densities are smaller or larger, respectively (see in particular the 2D color maps in Figure~\ref{FIG_sep_Sigma} and Figure~\ref{FIG_sep_Pi_Delta}, where these shells are visible as rings). These structures remind of and might indicate vibrating strings, which have either nodes or maxima at $z = 0$. Moreover, the transverse extent of the structures formed by the chromoelectric or chromomagnetic flux densities as almost the same for $Q \bar{Q}$ separation $r = 6 \, a$ and $r = 10 \, a$, which is consistent with a string interpretation.

It is also interesting to compare the resulting flux densities to the gluonic excitation operators for hybrid static potentials at leading order in the multipole expansion of pNRQCD (see e.g.\ refs.\ \cite{Brambilla:1999xf,Berwein:2015vca}). Similar operators were also used in lattice gauge theory computations of hybrid static potentials as local insertions in Wilson loops (see e.g.\ ref.\ \cite{Wolf:2014tta}), but numerically it turned out that they generate less ground state overlap than optimized non-local operators (like those discussed in ref.\ \cite{Capitani:2018rox} and in section~\ref{SEC477} of this work) and are, thus, less suited for computations in lattice gauge theory. The leading order gluonic excitation operators of pNRQCD are listed in Table~\ref{TAB501}, where the $Q \bar{Q}$ separation axis is again the $z$ axis. For certain $\Lambda_\eta^{(\epsilon)}$ sectors the flux densities we have obtained by our lattice computation closely resemble the pNRQCD operators. For example in the lower panel of Figure~\ref{FIG_med_10_x_axis} one can clearly see that the chromomagnetic flux densities for $\Pi_u$ and $\Delta_g$ are significantly larger than for the ordinary static potential $\Sigma_g^+$, in particular the $x$ and $y$ components. The corresponding pNRQCD operators include $B_x$ and $B_y$ as well as $D_x B_x - D_y B_y$ and $D_x B_y + D_y B_x$. Similarly, for $\Sigma_u^-$ the $z$ component of the chromomagnetic flux density is rather large, where one of the corresponding pNRQCD operators is $B_z$. Further interesting structures are the double peaks in the the chromoelectric flux densities for $\Sigma_u^-$ and $\Pi_u$ as shown in the upper panel of Figure~\ref{FIG_med_10_x_axis}. The pNRQCD operators for these sectors contain derivatives in $x$ direction of the corresponding chromoelectric field operators, $D_x E_y - D_y E_x$ and $D_x E_z - D_z E_x$, respectively. Again this is consistent, because from lattice gauge theory it is known that such derivative operators generate nodes in the corresponding wave functions.

\begin{table}[htb]
\begin{center}

\def\arraystretch{1.2}

\begin{tabular}{ll|ll}
\hline
$\Sigma_g^+$ & $1$ &   % 1st excitation: $E_z$ , $D_x B_y - D_y B_x$
$\Sigma_u^+$ & $D_z E_z$ \\
\hline
$\Sigma_g^-$ & $D_z B_z$ &
$\Sigma_u^-$ & $B_z$ \\
 &  &
 & $D_x E_y - D_y E_x$ \\
%
% What about $D_x E_x + D_y E_y$, $D_x B_x + D_y B_y$? These are \Sigma operators, but appear in \cite{Brambilla:1999xf,Berwein:2015vca} as \Delta operators ...
%
\hline
$\Pi_g$ & $E_x \quad , \quad E_y$ &   % 1st excitation: $D_x B_z + D_z B_x$, $D_y B_z + D_z B_y$
$\Pi_u$ & $B_x \quad , \quad B_y$ \\   % 1st excitation: $D_x E_z + D_z E_x$, $D_y E_z + D_z E_y$
 & $D_x B_z - D_z B_x \quad , \quad D_y B_z - D_z B_y$ &
 & $D_x E_z - D_z E_x \quad , \quad D_y E_z - D_z E_y$ \\
\hline
$\Delta_g$ & $D_x B_x - D_y B_y$ &
$\Delta_u$ & $D_x E_x - D_y E_y$ \\
 & $D_x B_y + D_y B_x$ &
 & $D_x E_y + D_y E_x$ \\
\hline
\end{tabular}

\end{center}
\caption{\label{TAB501}Gluonic excitation operators at leading order in the multipole expansion of pNRQCD (the $Q \bar{Q}$ separation axis is the $z$ axis; $D_j$ denotes the covariant derivative; see e.g.\ refs.\ \protect\cite{Brambilla:1999xf,Berwein:2015vca}).}
\end{table}

Finally we compare and discuss our results in the context of a recent and similar lattice computation of hybrid static potential flux densities \cite{Bicudo:2018jbb}. There the flux densities for two hybrid sectors $\Lambda_\eta^{(\epsilon)} = \Sigma_u^+, \Pi_u$ were computed for gauge group SU(3), for $\Lambda_\eta = \Pi_u$ not only for the ground state, but also for the first excitation. We have computed the flux densities for the ground states of the seven hybrid sectors $\Lambda_\eta^{(\epsilon)} = \Sigma_u^+, \Sigma_g^-, \Sigma_u^-, \Pi_g, \Pi_u, \Delta_g, \Delta_u$ for gauge groups SU(2) as well as SU(3). Lattice spacings, spacetime volumes and $Q \bar{Q}$ separations are similar in both works. Compared to ref.\ \cite{Bicudo:2018jbb} our presentation of results is different in the following aspects:
\begin{itemize}
\item We show flux densities $\Delta F_{j,\Lambda_\eta^{(\epsilon)}}^2(r;\mathbf{x})$ for $j = x, y$ separately, while in \cite{Bicudo:2018jbb} only the average of the $x$ and the $y$ component is shown, i.e.\ $\Delta F_{\perp,\Lambda_\eta^{(\epsilon)}}^2(r;\mathbf{x})$ (cf.\ also eq.\ (\ref{EQN690})).

\item In contrast to ref.\ \cite{Bicudo:2018jbb} we do not show the flux densities on the $Q \bar{Q}$ separation axis $x = y = 0$ as curves, because several of the hybrid static potentials have small flux densities on the separation axis, but large flux densities on spherical shells rather far away. Since the latter information is lost in such 1D curve plots, we prefer to show 2D color maps including the separation axis (see Figure~\ref{FIG_sep_Sigma} and Figure~\ref{FIG_sep_Pi_Delta} for SU(2) and Figure~\ref{FIG_sep_Sigma_} and Figure~\ref{FIG_sep_Pi_Delta_} for SU(3)).
\end{itemize}
Comparing the ground state flux densities for $\Lambda_\eta^{(\epsilon)} = \Sigma_u^+ , \Pi_u$ and gauge group SU(3), which were computed both in ref.\ \cite{Bicudo:2018jbb} (see Figure~7, Figure~10, Figure~11 and Figure~12 in ref.\ \cite{Bicudo:2018jbb}) and this work (see Figure~\ref{FIG_med_10_} to Figure~\ref{FIG_sep_Pi_Delta_}), we find fair agreement. A detailed comparison is, however, difficult, because of the different $Q \bar{Q}$ separations considered. Concerning statistical errors, our results are more precise by a factor of up to five. An obvious reasons for this is the larger number of gauge link configurations we have been using ($4 \, 500$ compared to $1 \, 199$). Moreover, we have improved our statistical precision by averaging data points, which are related by symmetries, e.g.\ rotational symmetry as discussed in detail in section~\ref{SEC455} and  section~\ref{SEC773}. Such a symmetrization was not done in \cite{Bicudo:2018jbb} as indicated by various plots presented in ref.\ \cite{Bicudo:2018jbb}.

\FloatBarrier

% ********************
% ********************
% ********************
% ********************
% ********************

\newpage

\section{\label{SEC697}Conclusions}

We have computed chromoelectric and chromomagnetic flux densities for hybrid static potential states for seven different sectors $\Lambda_\eta^{(\epsilon)} = \Sigma_u^+, \Sigma_g^-, \Sigma_u^-, \Pi_g, \Pi_u, \Delta_g, \Delta_u$ both in SU(2) and SU(3) lattice gauge theory. These flux densities can be interpreted as flux densities inside heavy hybrid mesons and, thus, provide insights into the structure of such mesons. Five of these sectors, $\Lambda_\eta^{(\epsilon)} = \Sigma_g^-, \Sigma_u^-, \Pi_g, \Delta_g, \Delta_u$, are investigated for the first time in this work, while our computation of the remaining two sectors, $\Lambda_\eta^{(\epsilon)} = \Sigma_u^+, \Pi_u$, confirm results recently published \cite{Bicudo:2018jbb}, now provided with improved precision. We find flux tubes with interesting structure, significantly different from that of the ordinary static potential with $\Lambda = \Sigma_g^+$ and reminiscent of different vibrational modes of a string. There are also localized peaks in the flux densities, which can be interpreted as valence gluons. Moreover, we compared the resulting flux densities to local operators typically used to study such states, e.g.\ in pNRQCD.

Concerning future work a straightforward direction would be to consider smaller lattice spacings and larger spatial volumes, i.e.\ to study the continuum and infinite volume limit. However, we do not expect significant changes in the numerical results presented here, since we already have partly investigated discretization errors (by comparing results obtained with unsmeared and with HYP2 smeared static propagators) and finite volume corrections (by comparing our SU(2) main results to an identical computation with a smaller volume and lower statistics). A more interesting direction would be to extend the computation by including also dynamical light quarks. In principle, one could then study not just heavy-heavy hybrid mesons, but, more generally, heavy-heavy exotic mesons and explore their gluon and light quark distribution at the same time. In practice, however, this might be very challenging, because a hybrid static potential state might decay into the ordinary static potential and one or more light mesons.

% ********************
% ********************
% ********************
% ********************
% ********************

\newpage

\appendix

\section{\label{APP600}Transformation of $| 0_{\Lambda_\eta^\epsilon}(r) \rangle$ under rotations}

In this appendix we derive eq.\ (\ref{EQN239}).

Static potential eigenstates $| 0_{\lambda_\eta}(r) \rangle$ (introduced in section~\ref{SEC455} for $|\lambda| \geq 1$) are also eigenstates of the $z$ component of the total angular momentum operator $J_z$ and, thus, transform under rotations around the $z$ axis with angle $\varphi$ according to
\begin{eqnarray}
\label{EQN244} R_z(\varphi) | 0_{\lambda_\eta}(r) \rangle \ \ = \ \ e^{i \varphi J_z} | 0_{\lambda_\eta}(r) \rangle \ \ = \ \ e^{i \varphi \lambda} | 0_{\lambda_\eta}(r) \rangle .
\end{eqnarray}
Consequently,
\begin{eqnarray}
R_z(\varphi) \Big(P_x | 0_{\lambda_\eta}(r) \rangle\Big) \ \ = \ \ e^{i \varphi J_z} P_x | 0_{\lambda_\eta}(r) \rangle \ \ = \ \ P_x e^{i (-\varphi) J_z} | 0_{\lambda_\eta}(r) \rangle \ \ = \ \ e^{i (-\varphi) \lambda}  \Big(P_x | 0_{\lambda_\eta}(r) \rangle\Big)
\end{eqnarray}
($J_z P_x = -P_x J_z$ has been used) and
\begin{eqnarray}
P_x | 0_{+\lambda_\eta}(r) \rangle \ \ = \ \ | 0_{-\lambda_\eta}(r) \rangle .
\end{eqnarray}
The last equation allows to express states $| 0_{\Lambda_\eta^\epsilon}(r) \rangle$ for $\Lambda \geq 1$ in terms of states $| 0_{\lambda_\eta}(r) \rangle$,
\begin{eqnarray}
| 0_{\Lambda_\eta^{\pm}}(r) \rangle \ \ = \ \frac{1}{\sqrt{2}} \Big(| 0_{+\lambda_\eta}(r) \rangle \pm | 0_{-\lambda_\eta}(0) \rangle\Big) ,
\end{eqnarray}
where $\lambda = \Lambda$. Using eq.\ (\ref{EQN244}) one can infer
\begin{eqnarray}
\label{EQN110} R_z(\varphi) | 0_{\Lambda_\eta^\pm}(r) \rangle \ \ = \ \ \cos(\varphi \Lambda) | 0_{\Lambda_\eta^\pm}(r) \rangle + i \sin(\varphi \Lambda) | 0_{\Lambda_\eta^\mp}(r) \rangle
\end{eqnarray}
for $\Lambda \geq 1$, which is eq.\ (\ref{EQN239}). Note that eq.\ (\ref{EQN110}) also holds for $\Lambda = 0$, because it reduces to $R_z(\varphi) | 0_{\Lambda_\eta^\pm}(r) \rangle = | 0_{\Lambda_\eta^\pm}(r) \rangle$, i.e.\ correctly indicates rotational invariance for states with total angular momentum $\Lambda = 0$.

% ********************
% ********************
% ********************
% ********************
% ********************

\newpage

\section{\label{APP001}Hybrid static potential flux densities for all $\Lambda_\eta^\epsilon$ sectors: plots for SU(3) gauge theory}

\begin{figure}[p]
\begin{center}
\includegraphics[width=7.0cm]{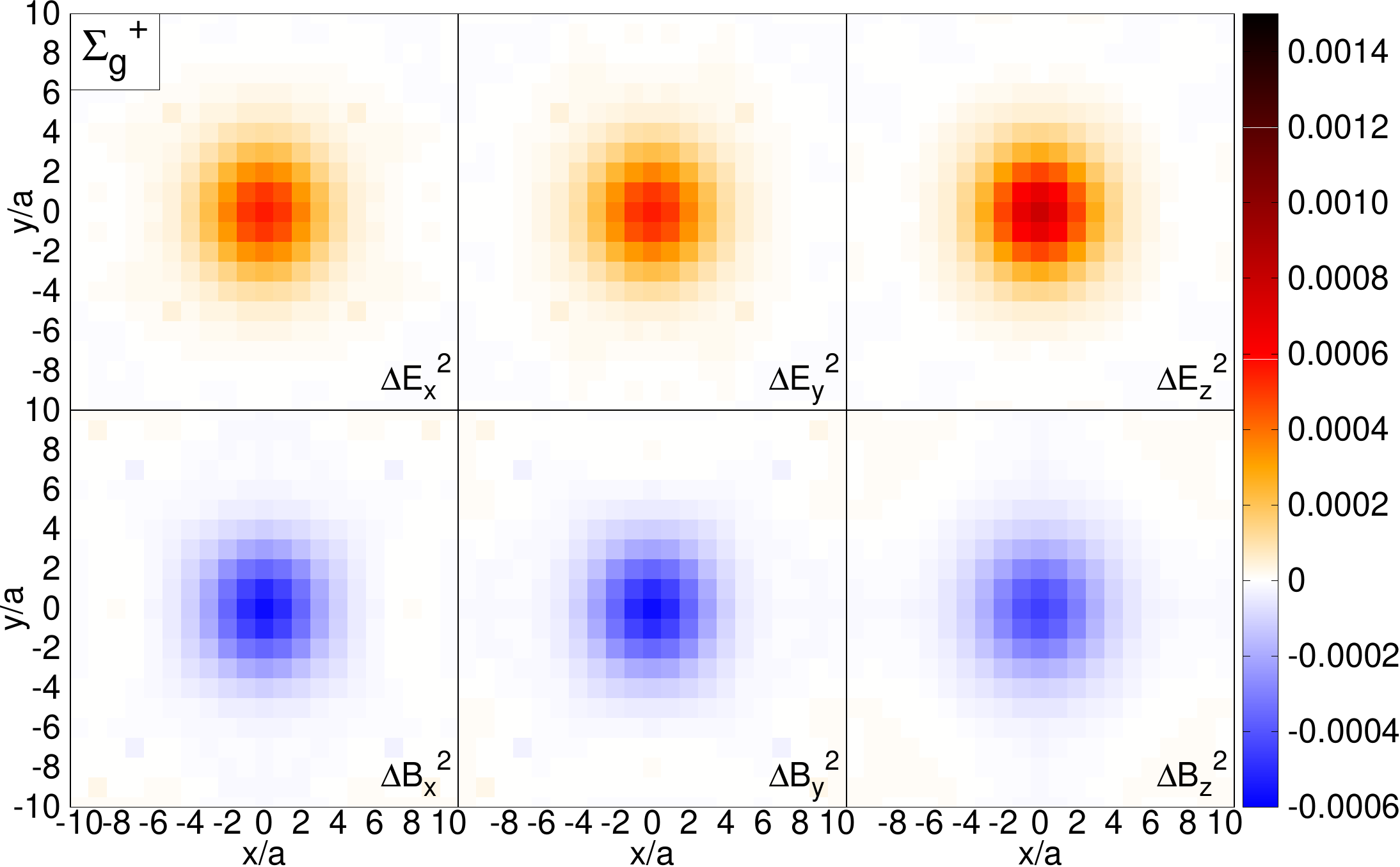}
\includegraphics[width=7.0cm]{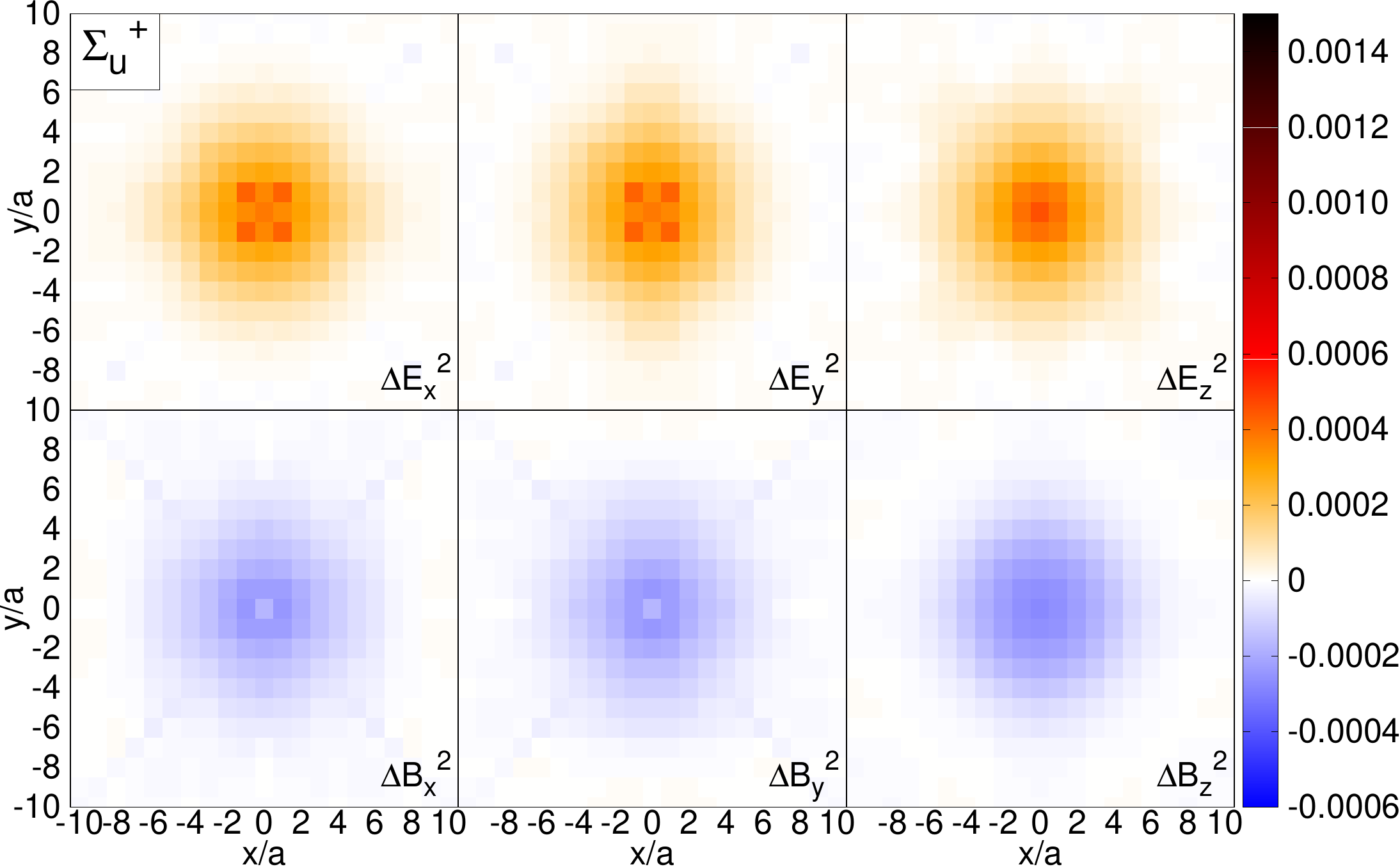}
\includegraphics[width=7.0cm]{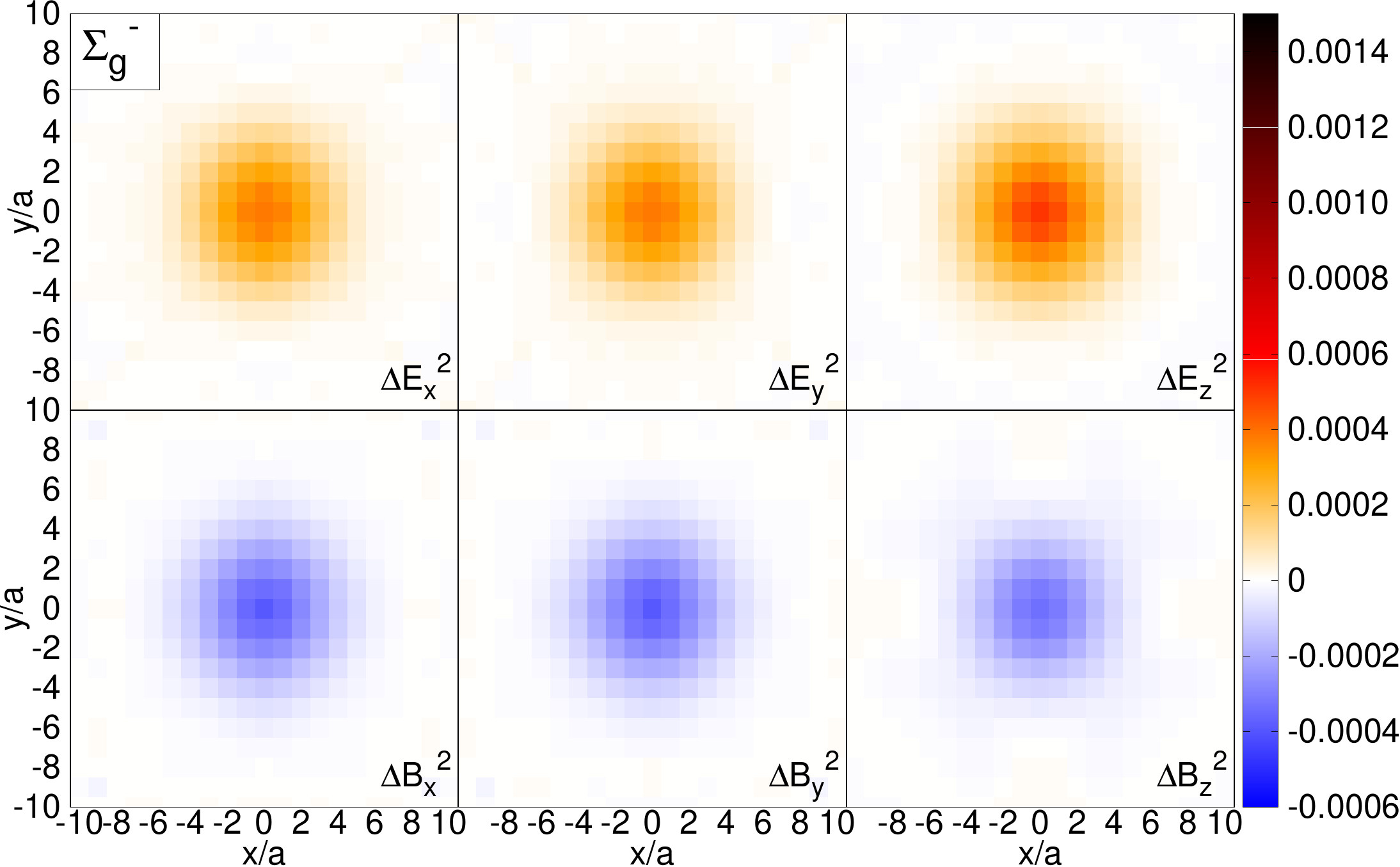}
\includegraphics[width=7.0cm]{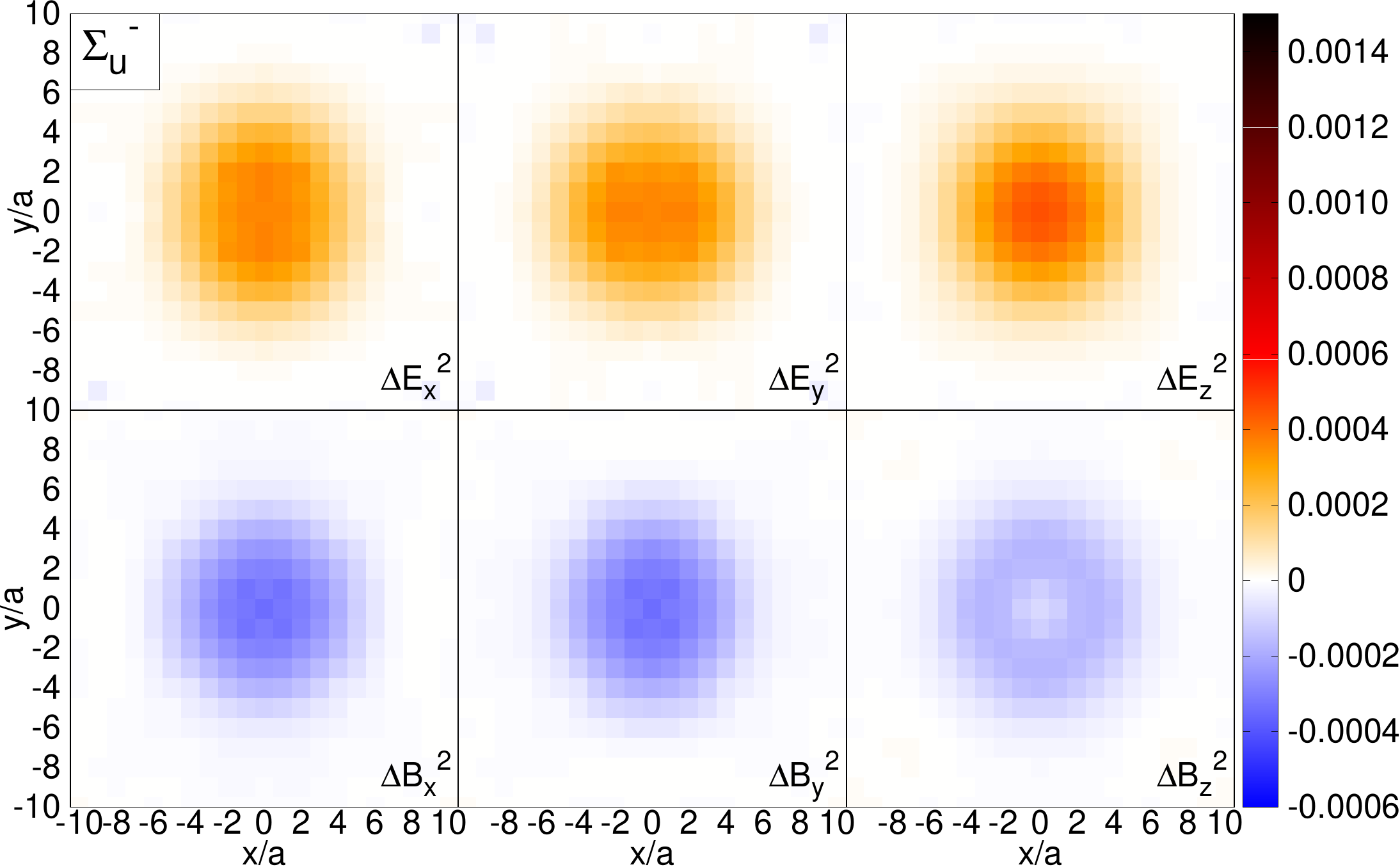}
\includegraphics[width=7.0cm]{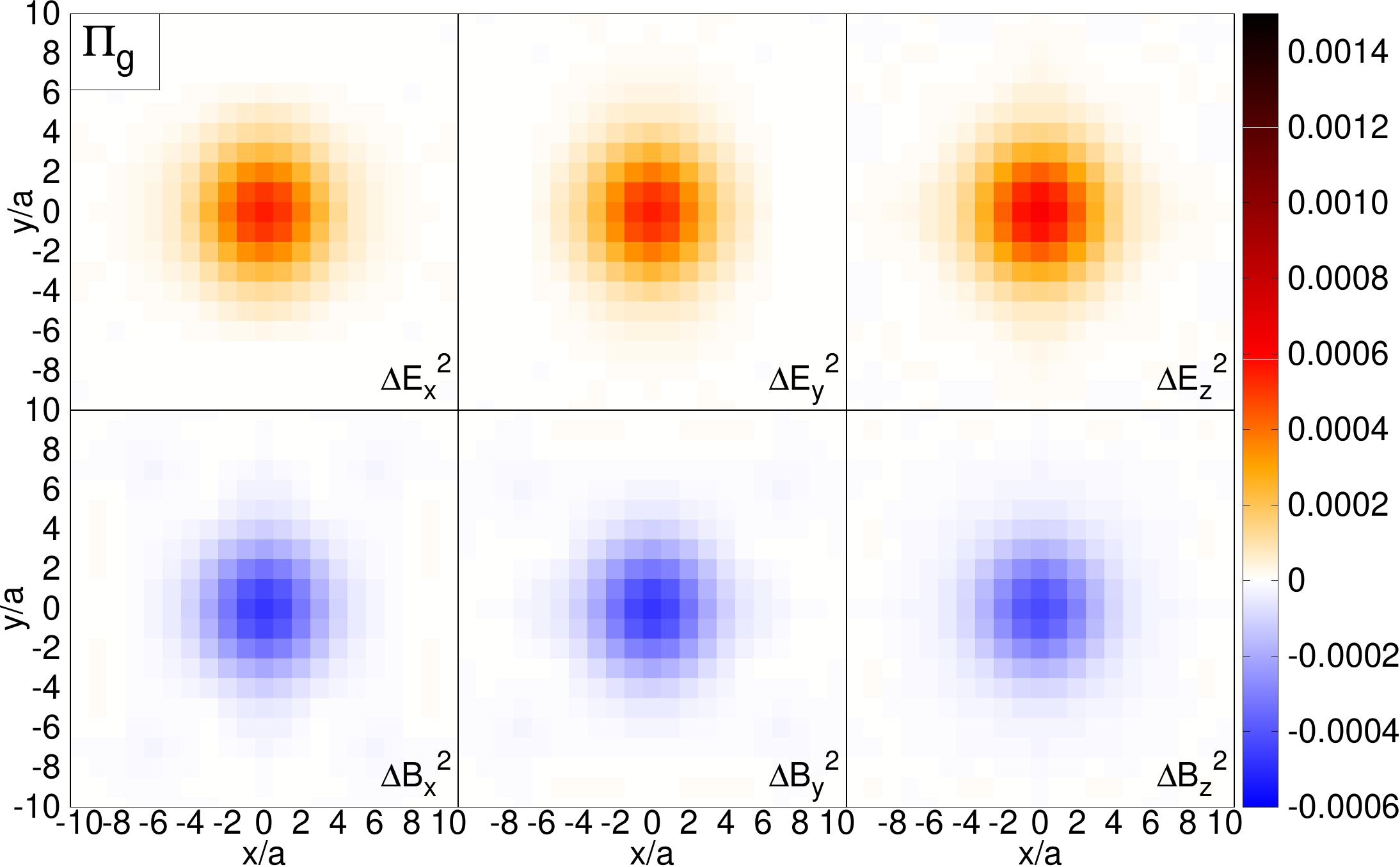}
\includegraphics[width=7.0cm]{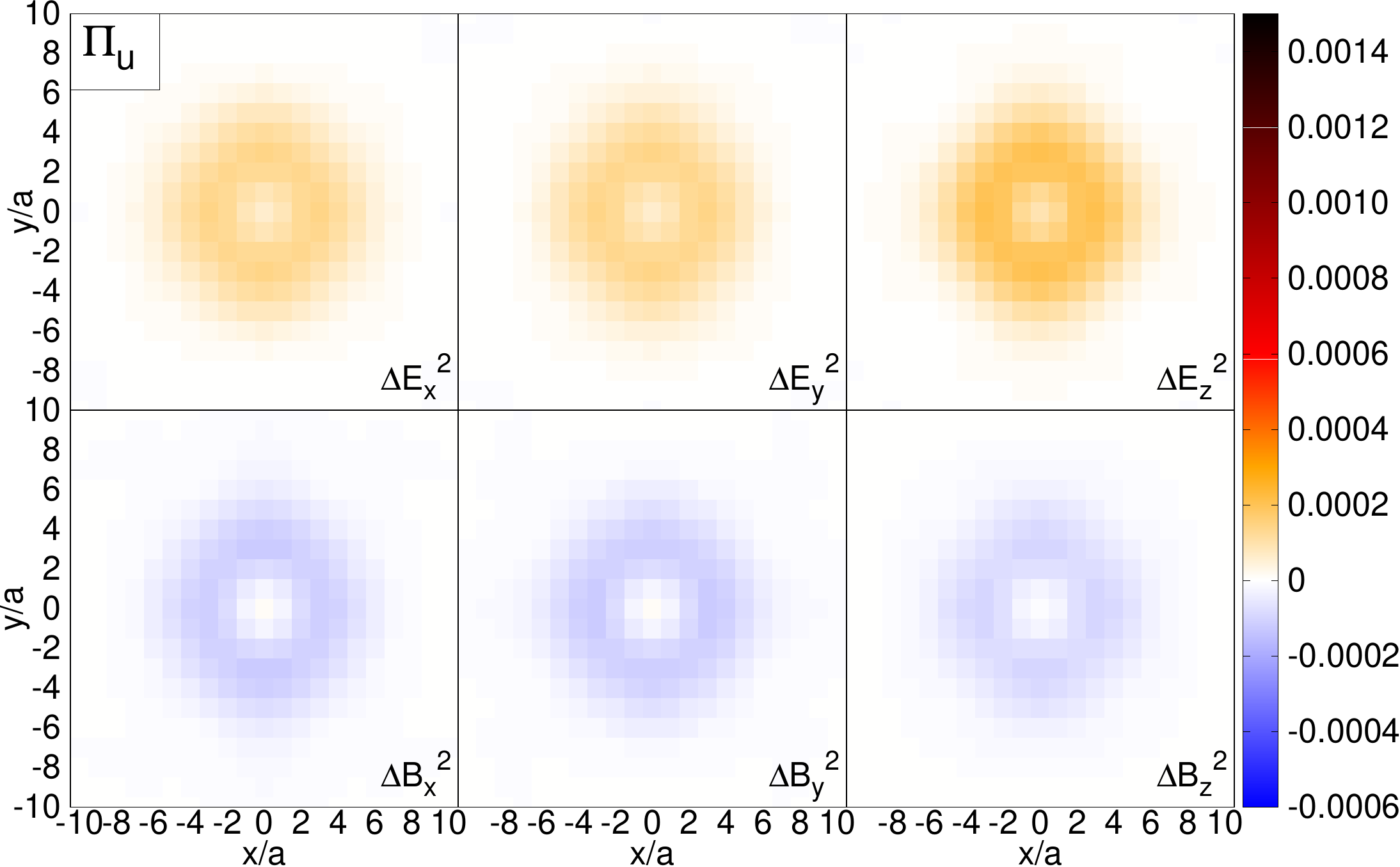}
\includegraphics[width=7.0cm]{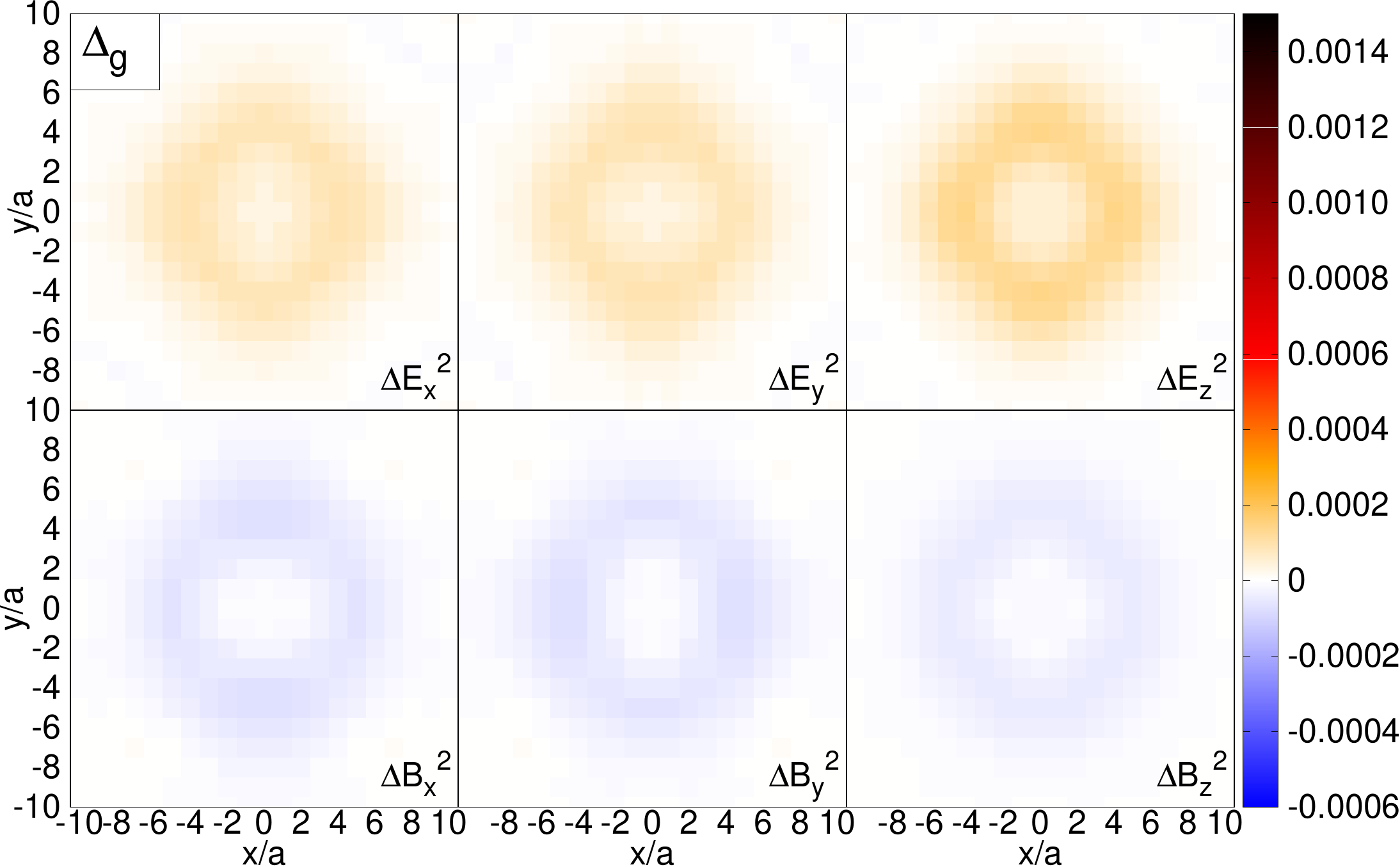}
\includegraphics[width=7.0cm]{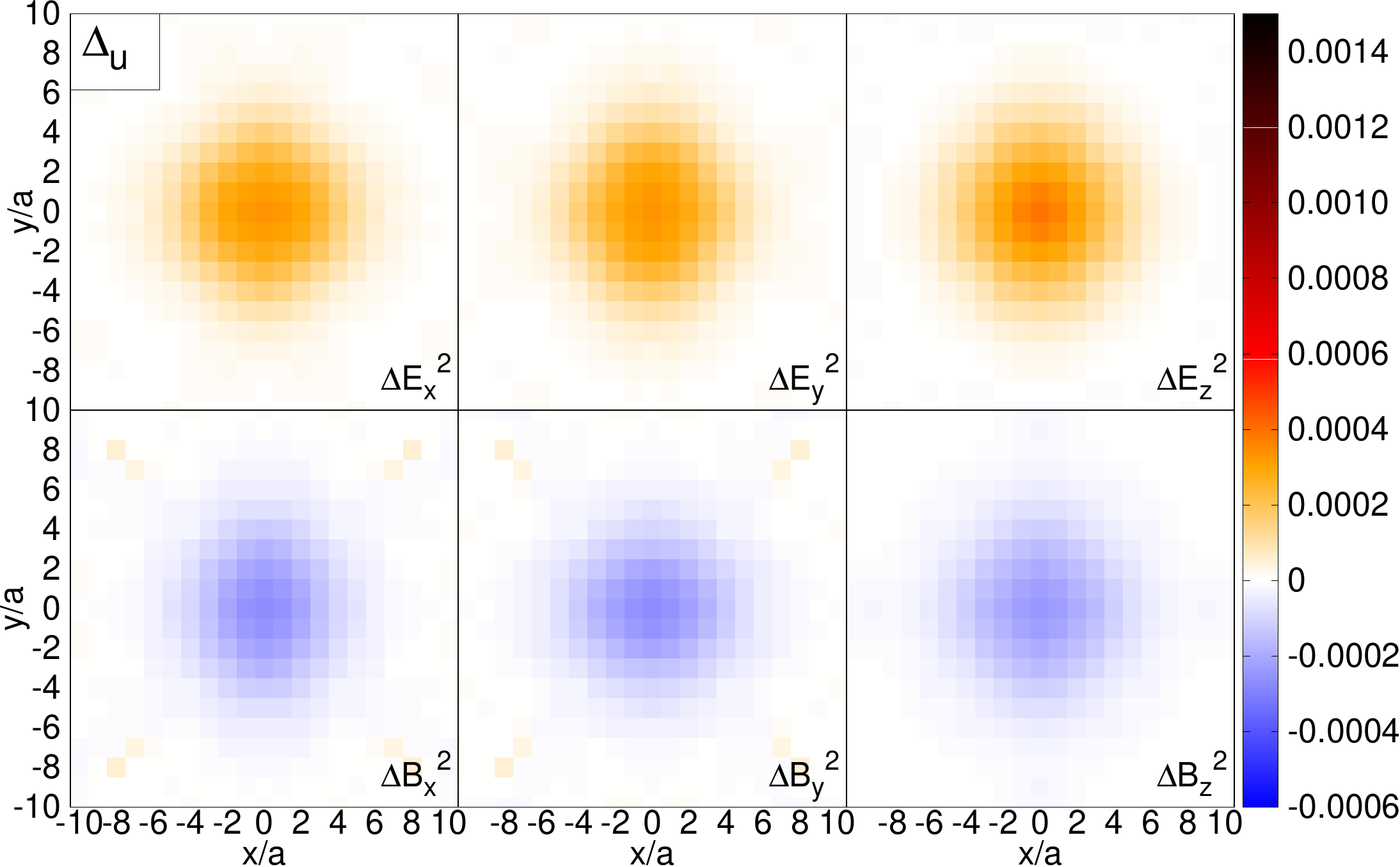}
\end{center}
\caption{\label{FIG_med_10_}Flux densities $\Delta F_{j,\Lambda_\eta^{(\epsilon)}}^2(r;\mathbf{x} = (x,y,0))$, $j = x,y,z$ in the mediator plane for gauge group SU(3), all investigated sectors $\Lambda_\eta^{(\epsilon)} = \Sigma_g^+, \Sigma_u^+, \Sigma_g^-, \Sigma_u^-, \Pi_g, \Pi_u, \Delta_g, \Delta_u$ and $Q \bar{Q}$ separation $r = 10 \, a$.}
\end{figure}

\begin{figure}[p]
\begin{center}
\includegraphics[width=15.0cm]{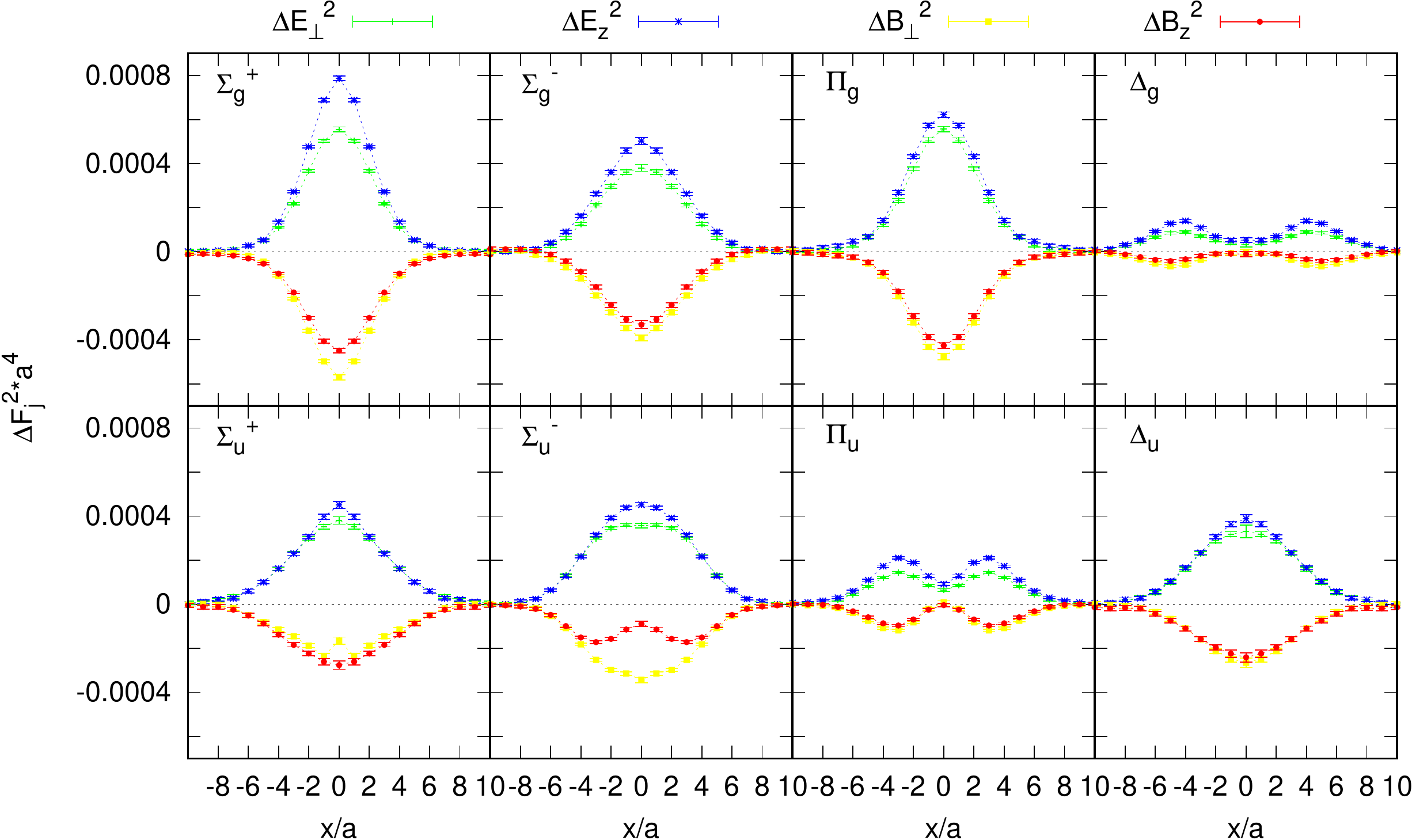}

\vspace{+0.5cm}
\includegraphics[width=15.0cm]{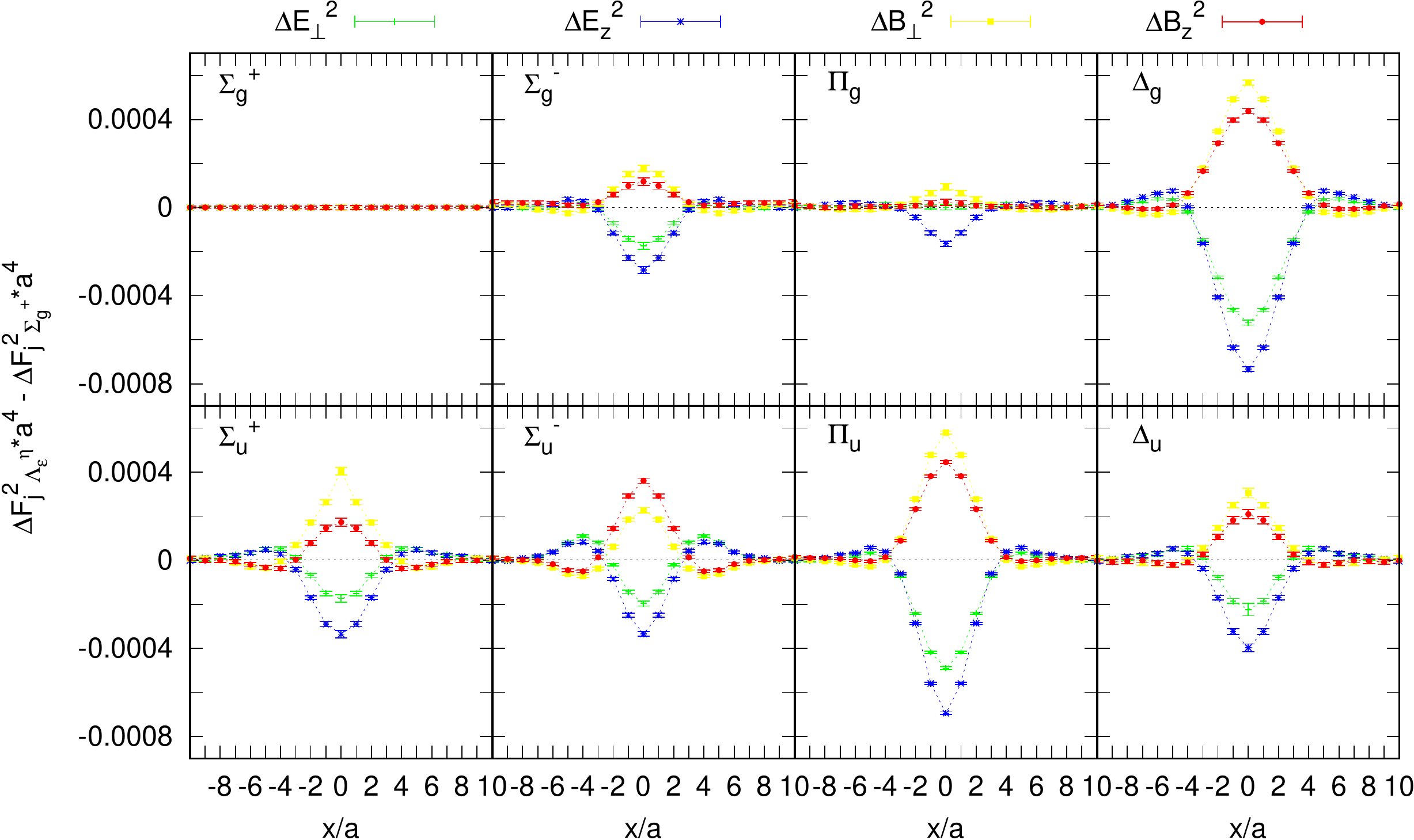}
\end{center}
\caption{\label{FIG_med_10_x_axis_}Flux densities on the $x$ axis for gauge group SU(3), all investigated sectors \\ $\Lambda_\eta^{(\epsilon)} = \Sigma_g^+, \Sigma_u^+, \Sigma_g^-, \Sigma_u^-, \Pi_g, \Pi_u, \Delta_g, \Delta_u$ and $Q \bar{Q}$ separation $r = 10 \, a$. \\ \textbf{(top)} $\Delta F_{j,\Lambda_\eta^{(\epsilon)}}^2(r;\mathbf{x} = (x,0,0))$, $j = \perp,z$. \\ \textbf{(bottom)}~$\Delta F_{j,\Lambda_\eta^{(\epsilon)}}^2(r;\mathbf{x} = (x,0,0)) - \Delta F_{j,\Sigma_g^+}^2(r;\mathbf{x} = (x,0,0))$, $j = \perp,z$.}
\end{figure}

\begin{figure}[p]
\begin{center}
\includegraphics[width=7.0cm]{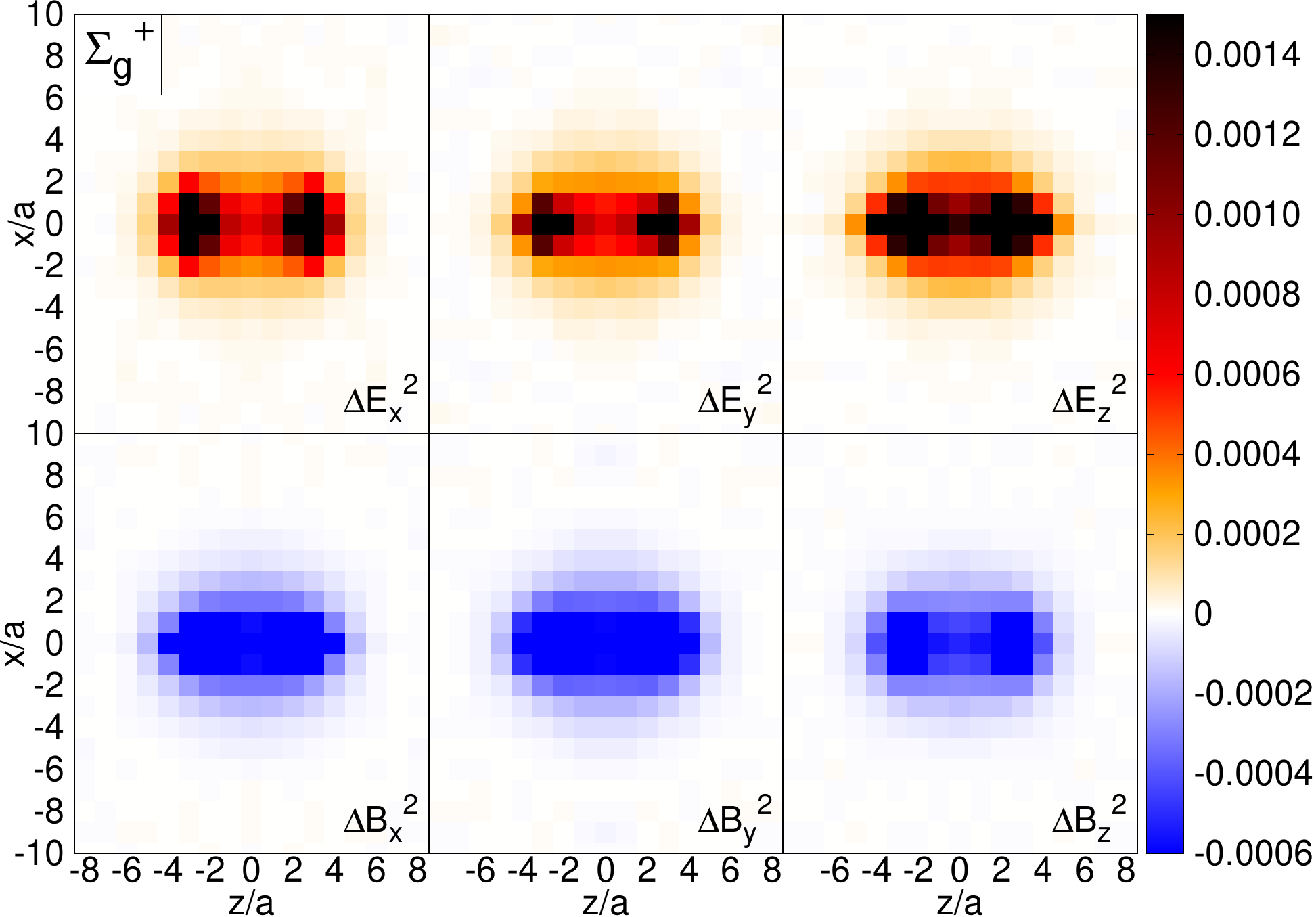}
\includegraphics[width=7.0cm]{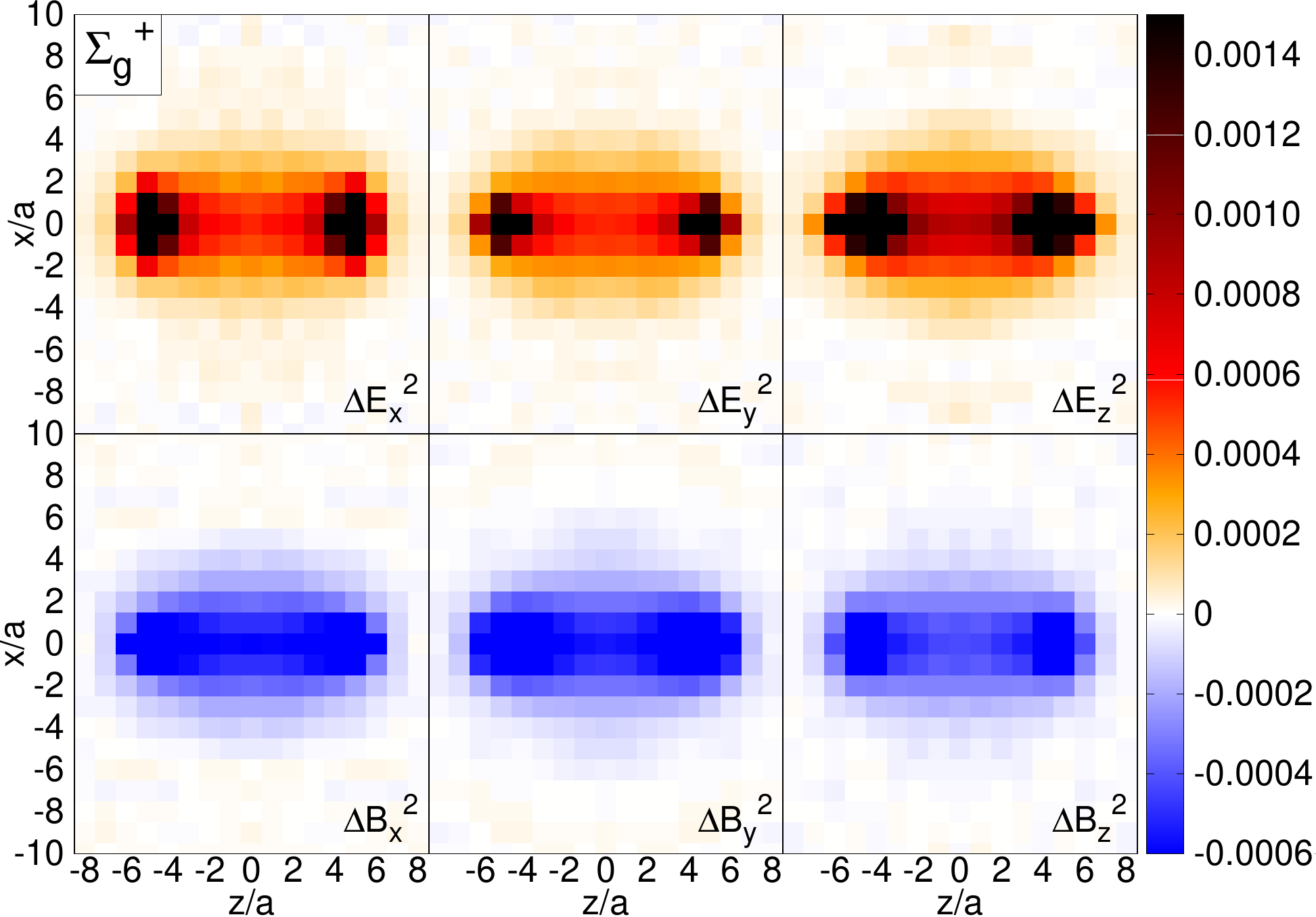}
\includegraphics[width=7.0cm]{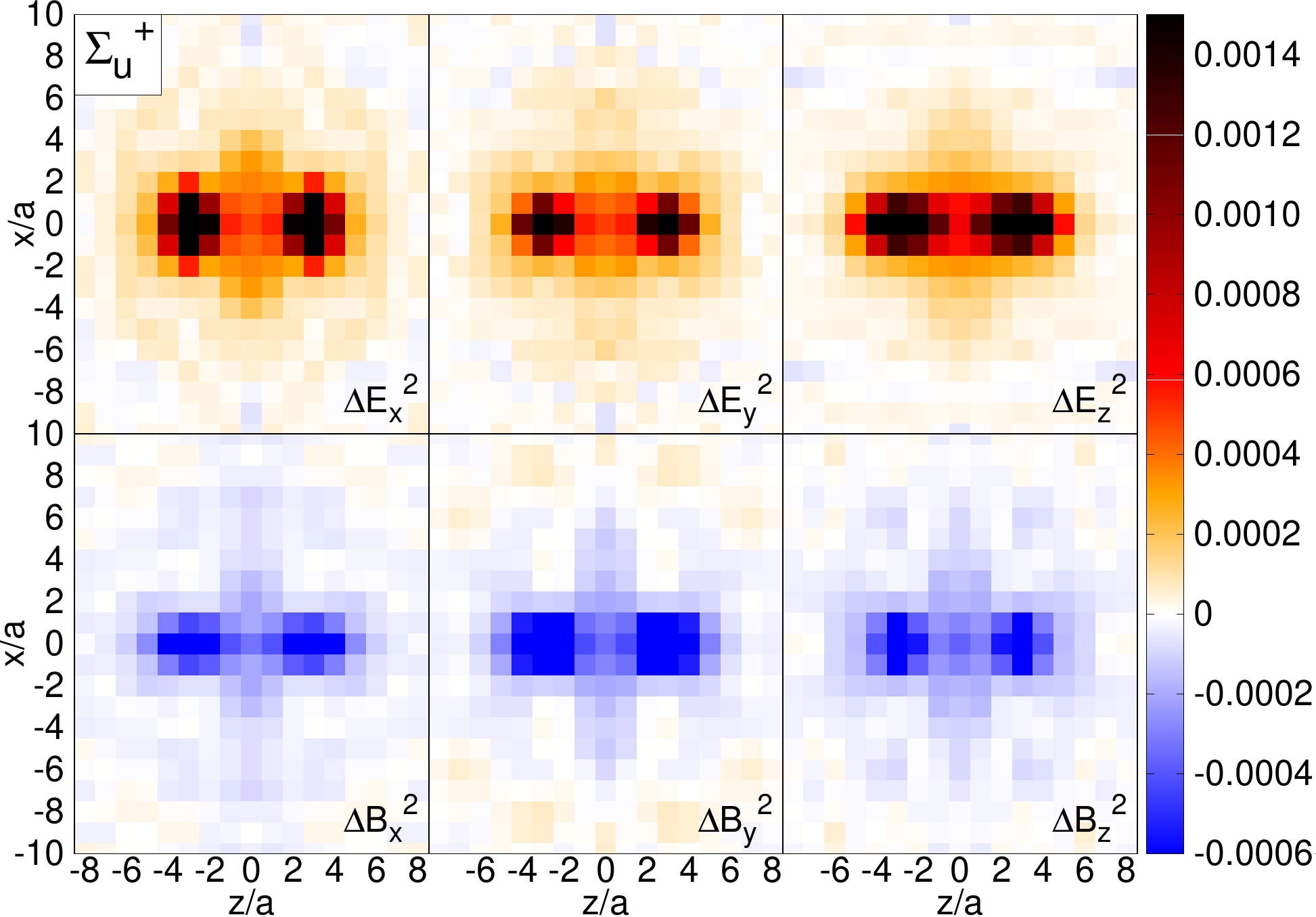}
\includegraphics[width=7.0cm]{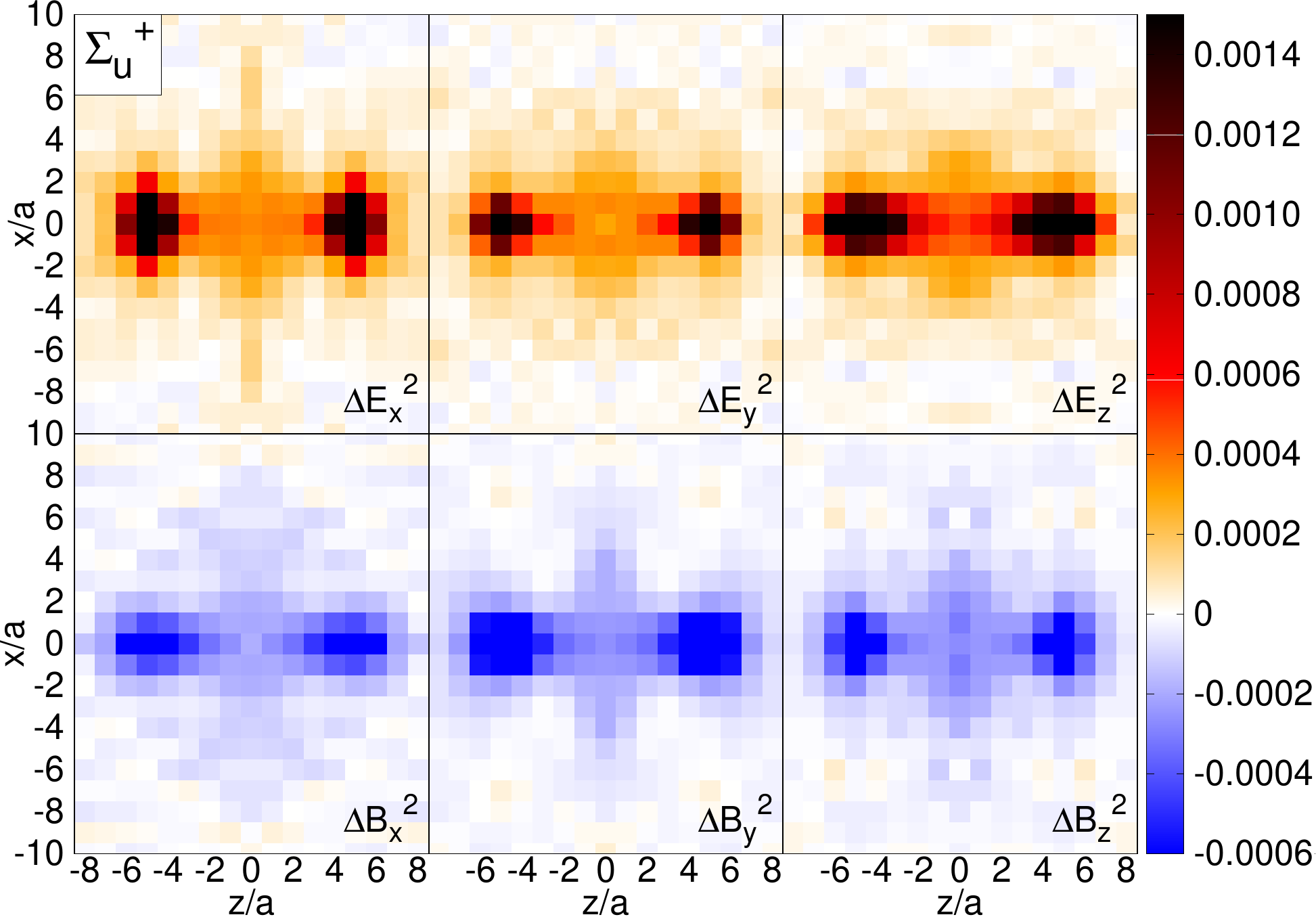}
\includegraphics[width=7.0cm]{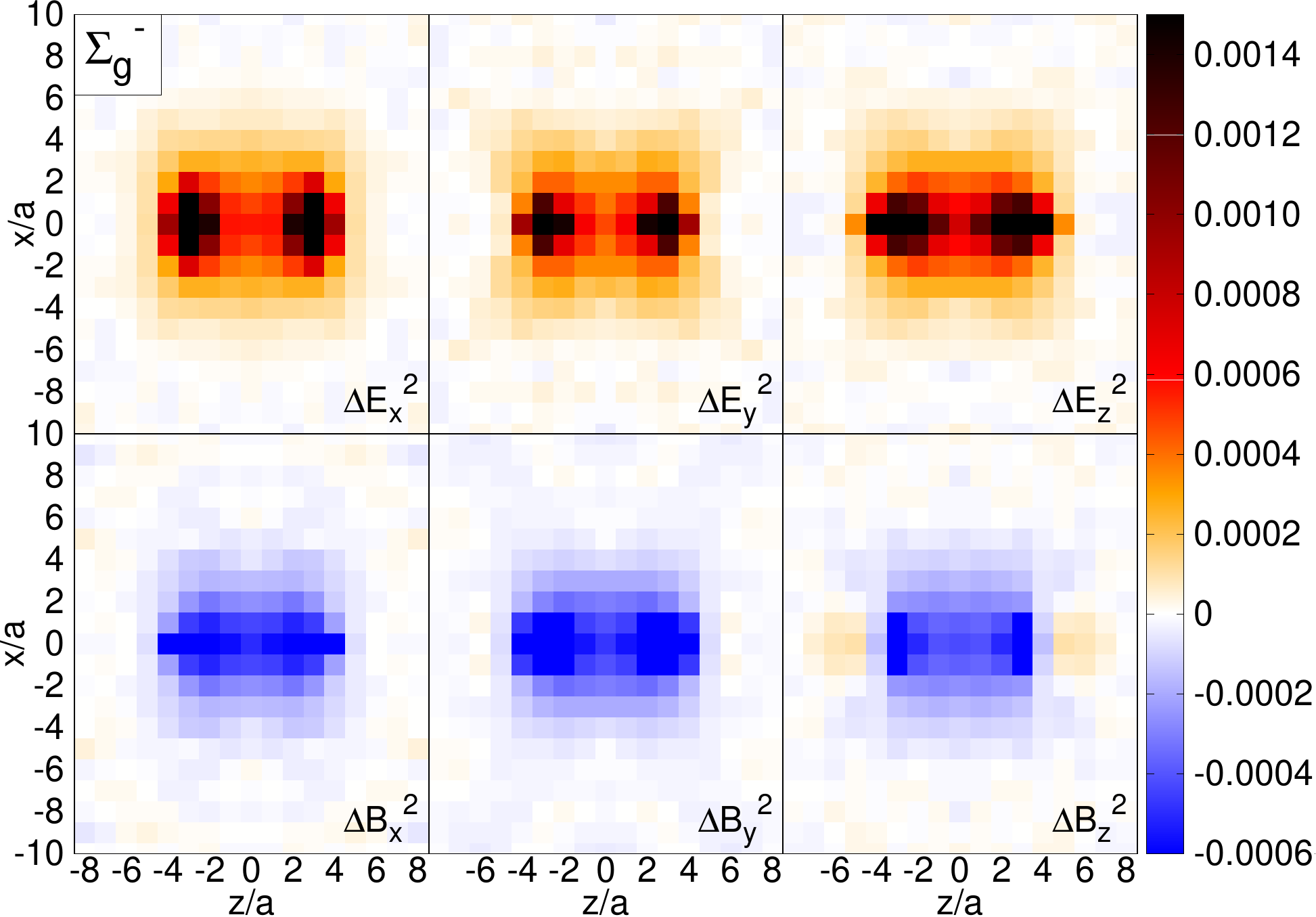}
\includegraphics[width=7.0cm]{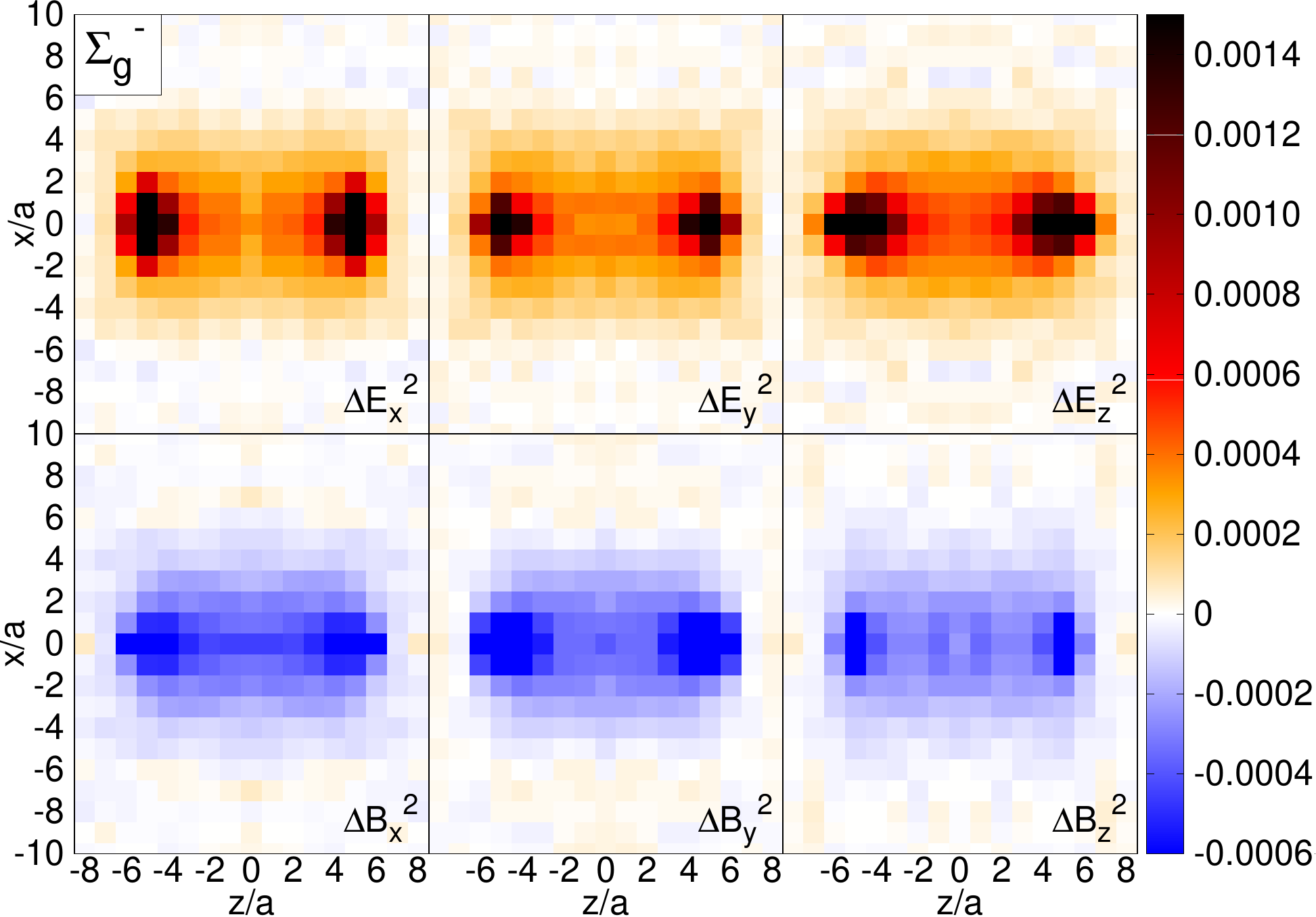}
\includegraphics[width=7.0cm]{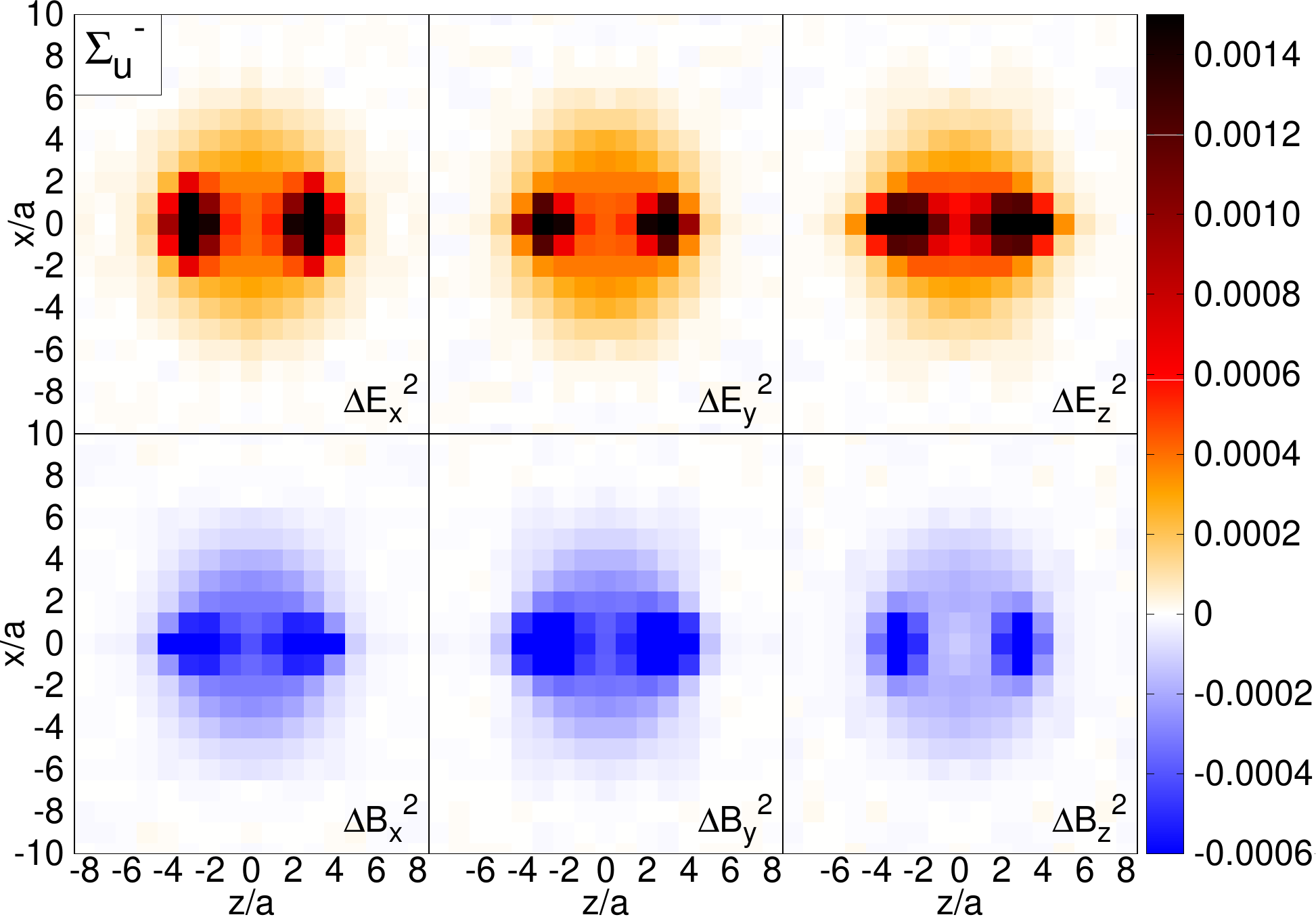}
\includegraphics[width=7.0cm]{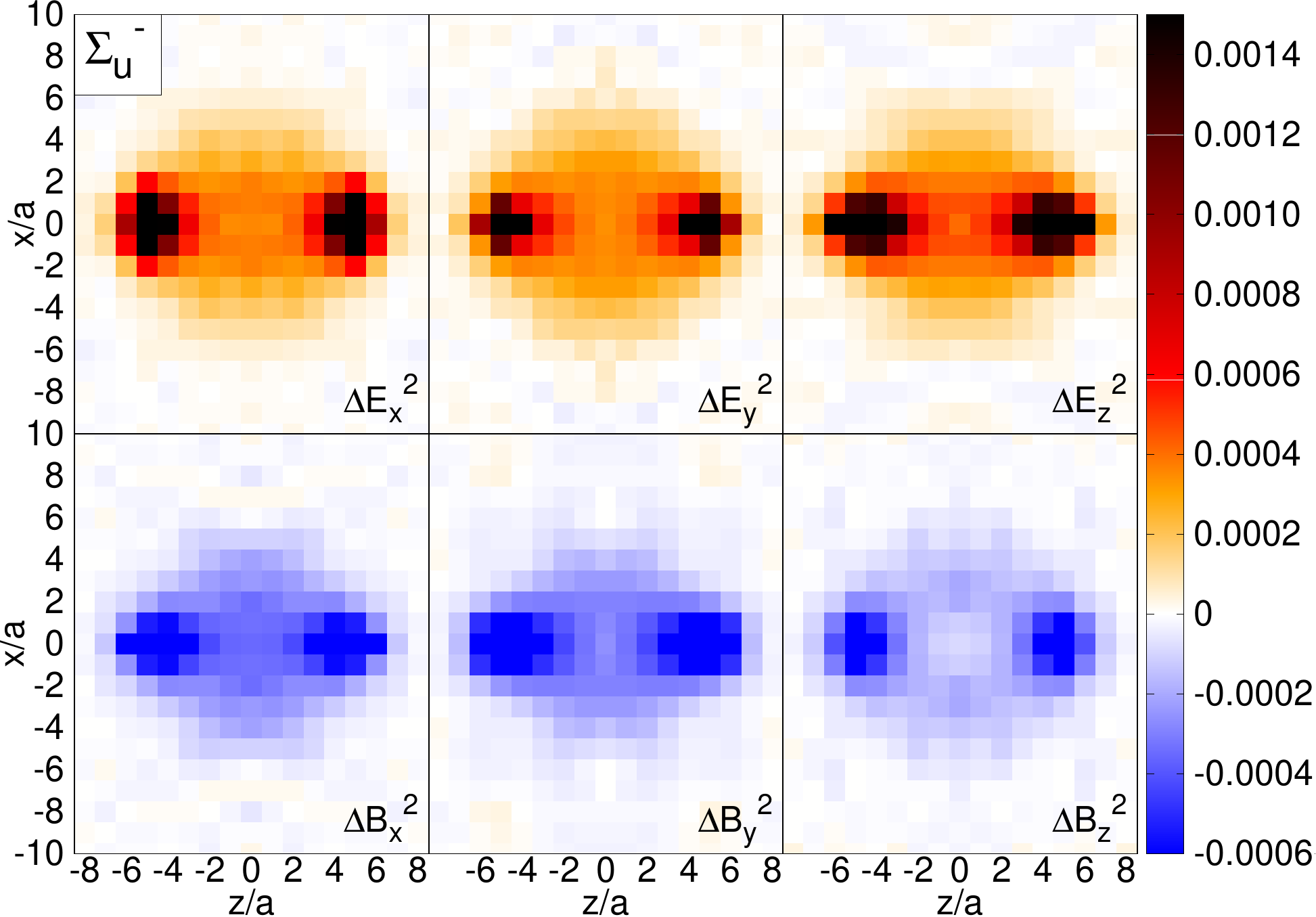}
\end{center}
\caption{\label{FIG_sep_Sigma_}Flux densities $\Delta F_{j,\Lambda_\eta^\epsilon}^2(r;\mathbf{x} = (x,0,z))$, $j = x,y,z$ in the separation plane for gauge group SU(3) and sectors $\Lambda_\eta^\epsilon = \Sigma_g^+, \Sigma_u^+, \Sigma_g^-, \Sigma_u^-$. \textbf{(left)} $Q \bar{Q}$ separation $r = 6 \, a$. \textbf{(right)} $Q \bar{Q}$ separation $r = 10 \, a$.}
\end{figure}

\begin{figure}[p]
\begin{center}
\includegraphics[width=7.0cm]{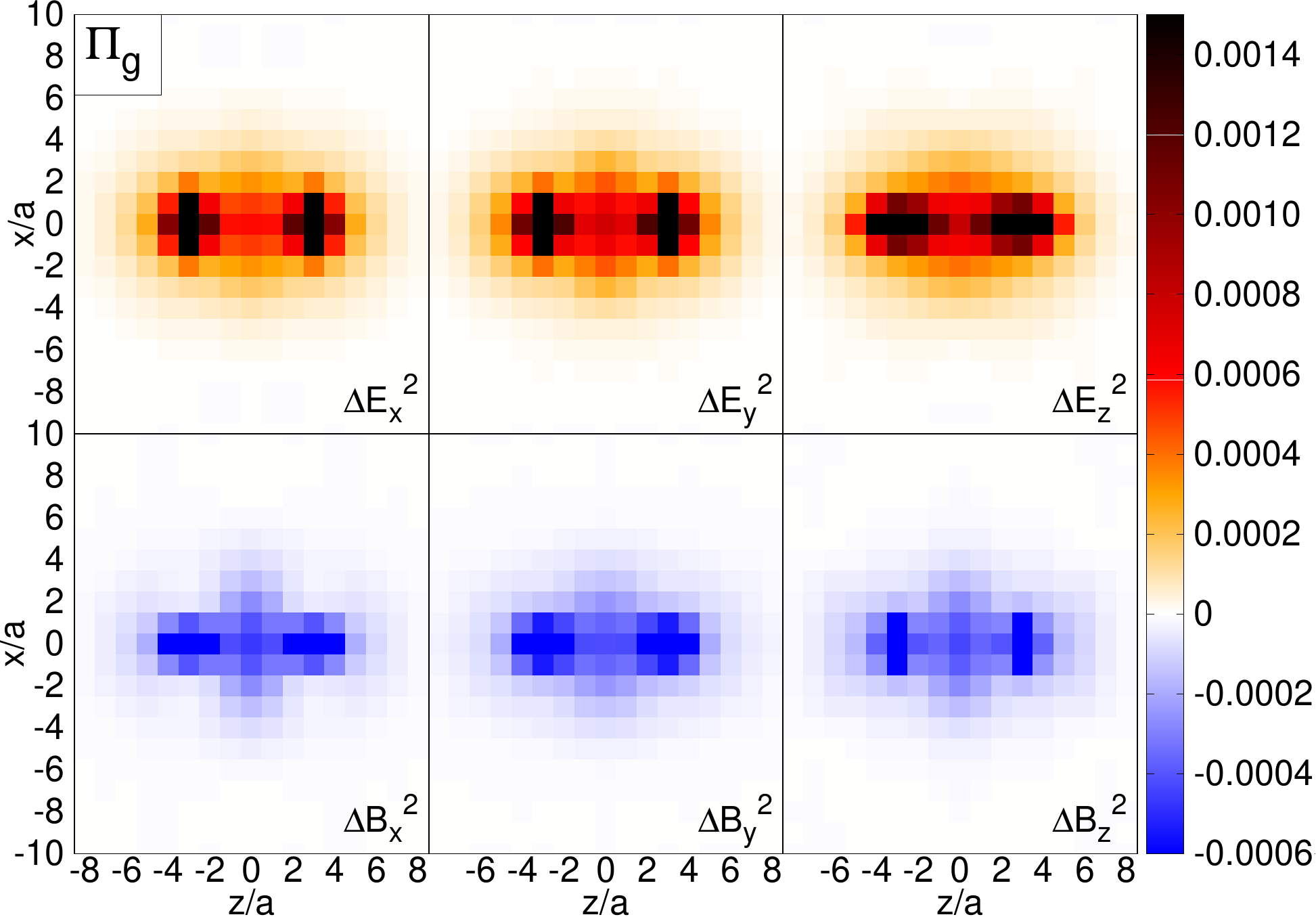}
\includegraphics[width=7.0cm]{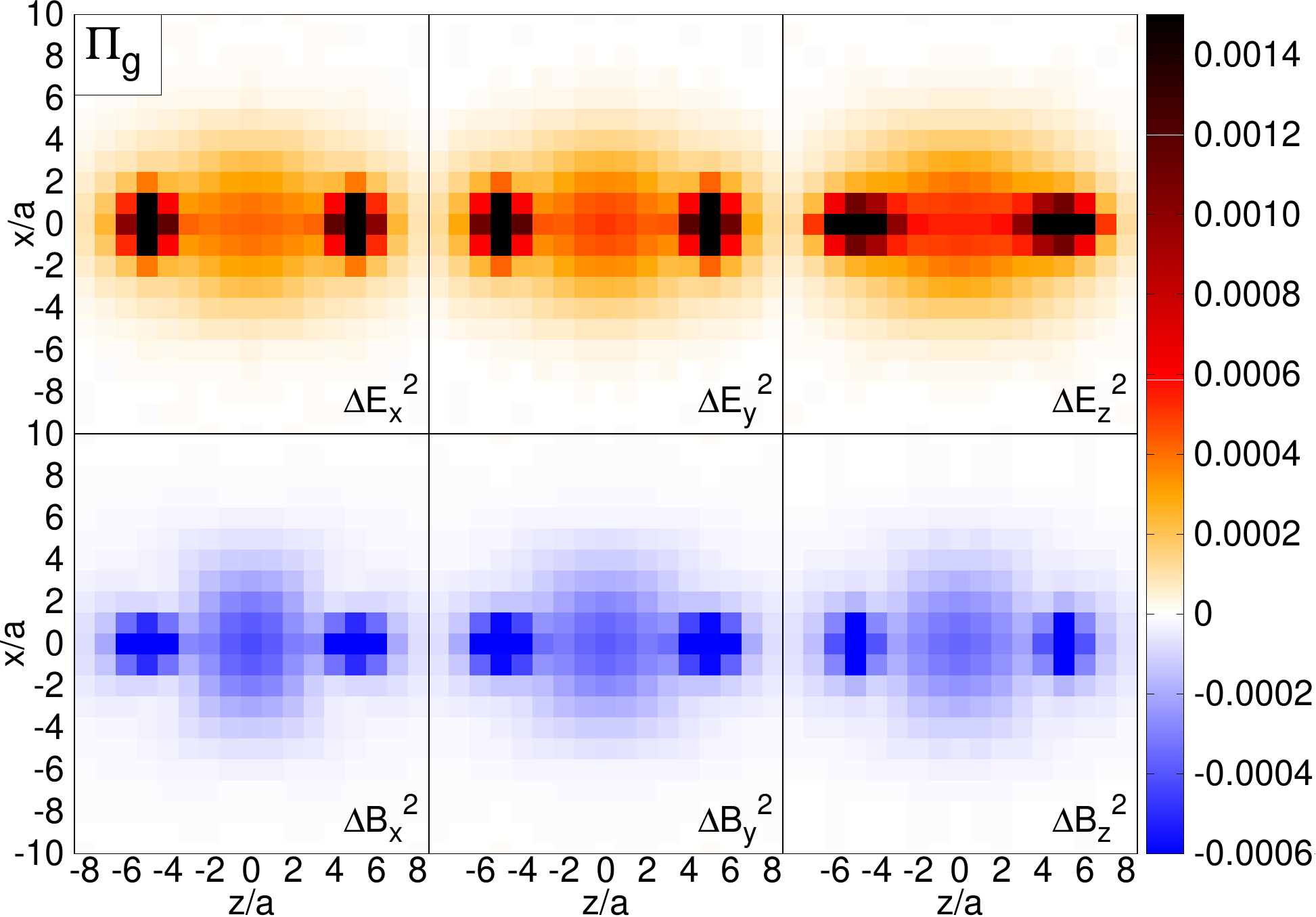}
\includegraphics[width=7.0cm]{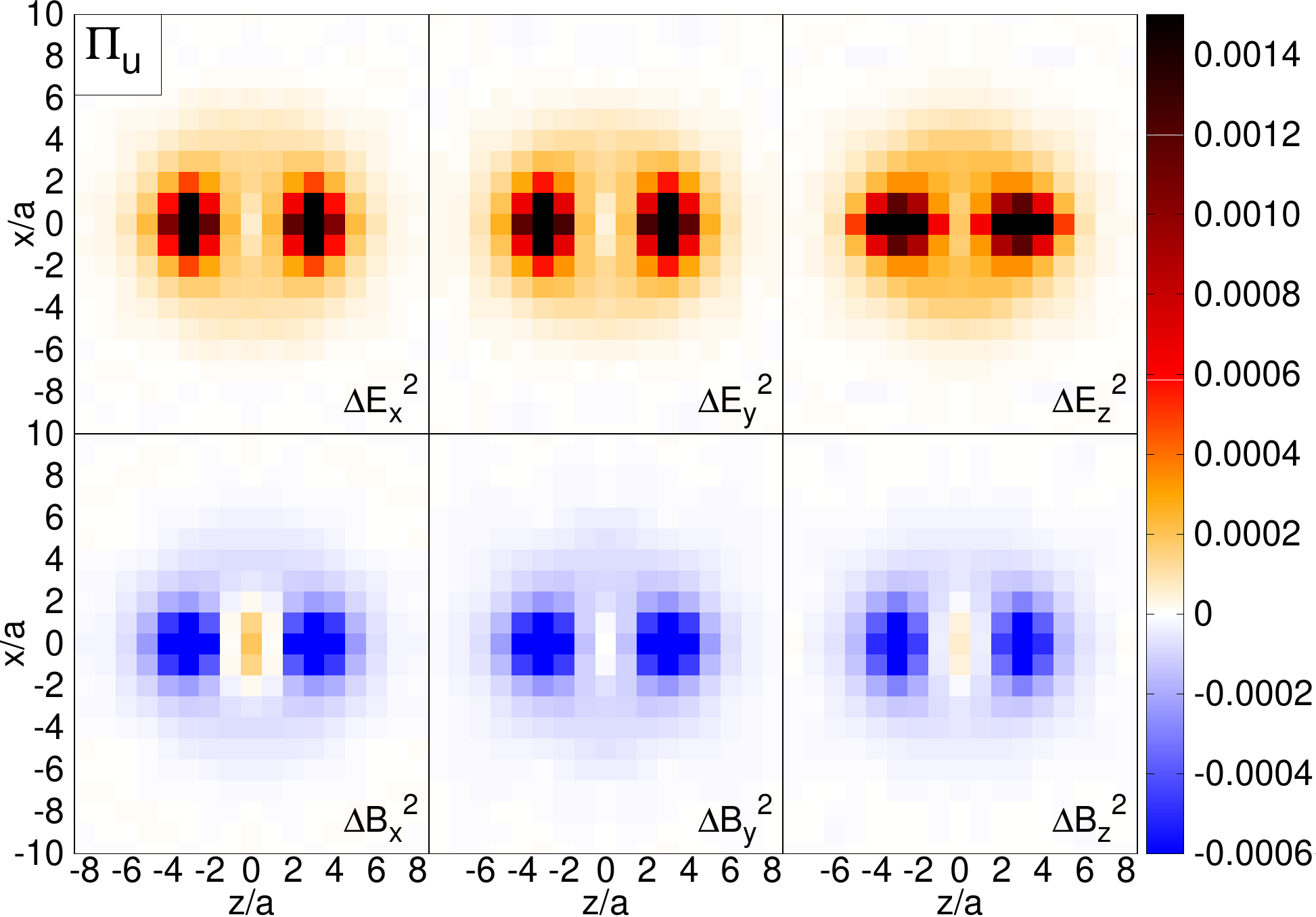}
\includegraphics[width=7.0cm]{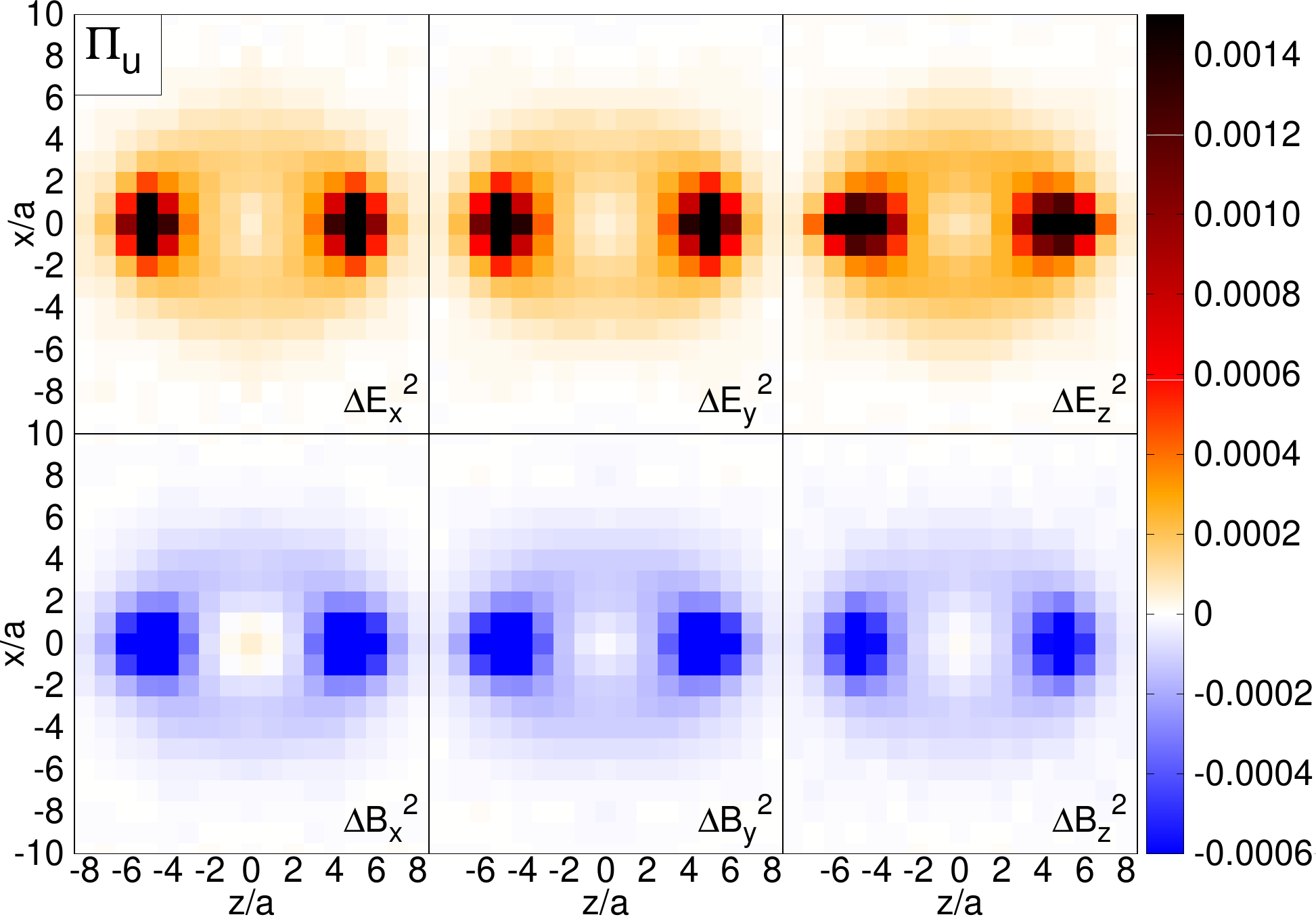}
\includegraphics[width=7.0cm]{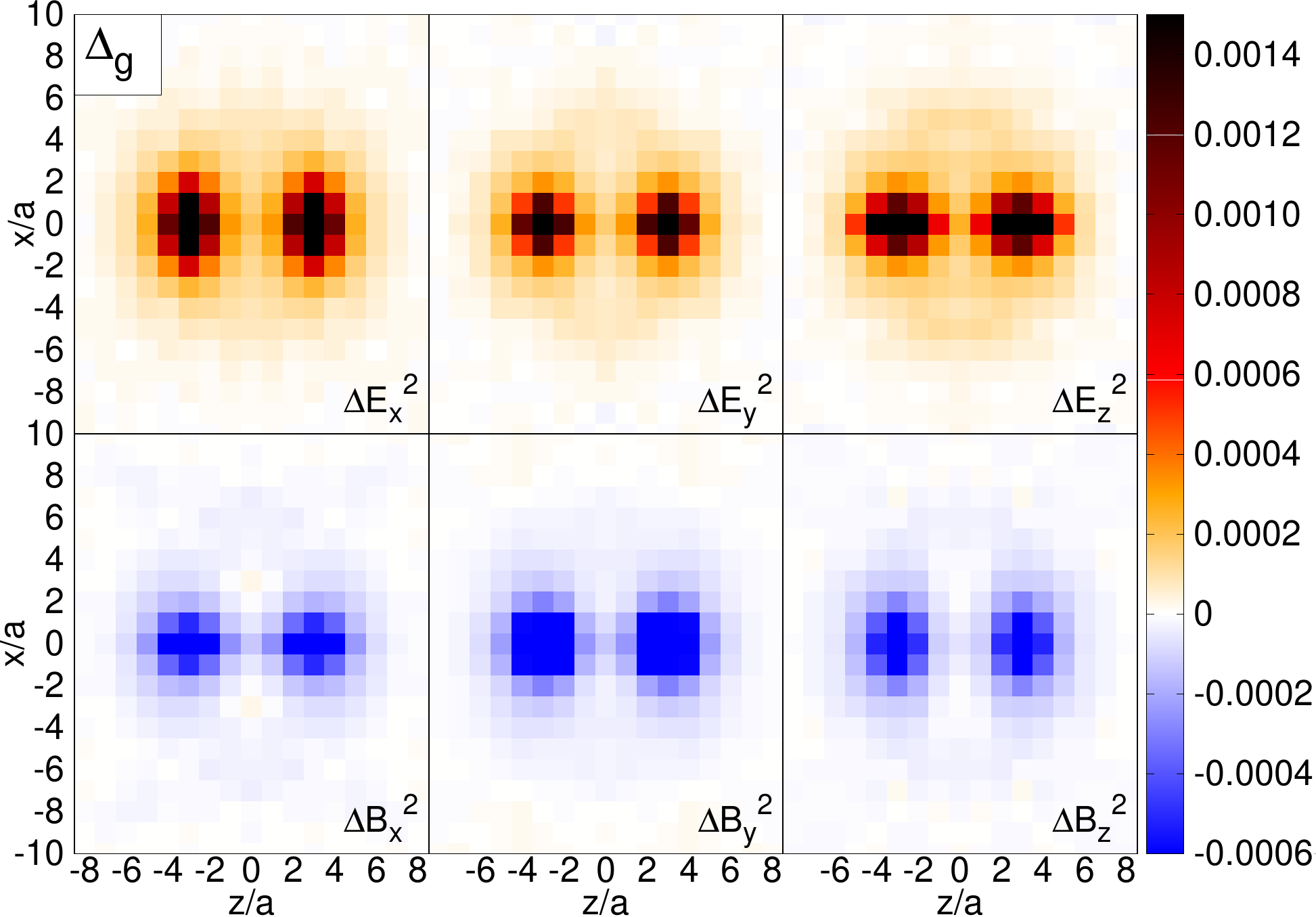}
\includegraphics[width=7.0cm]{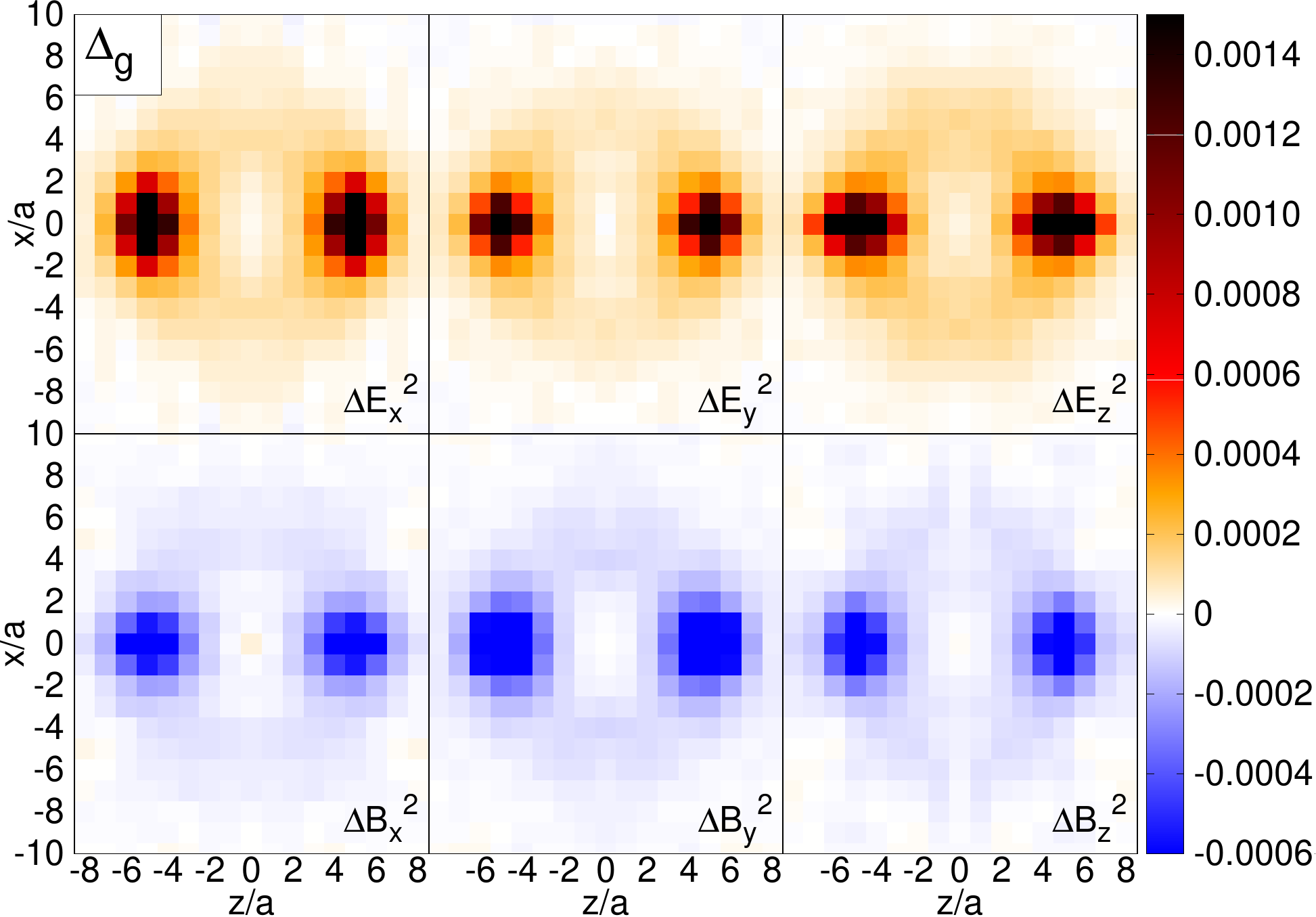}
\includegraphics[width=7.0cm]{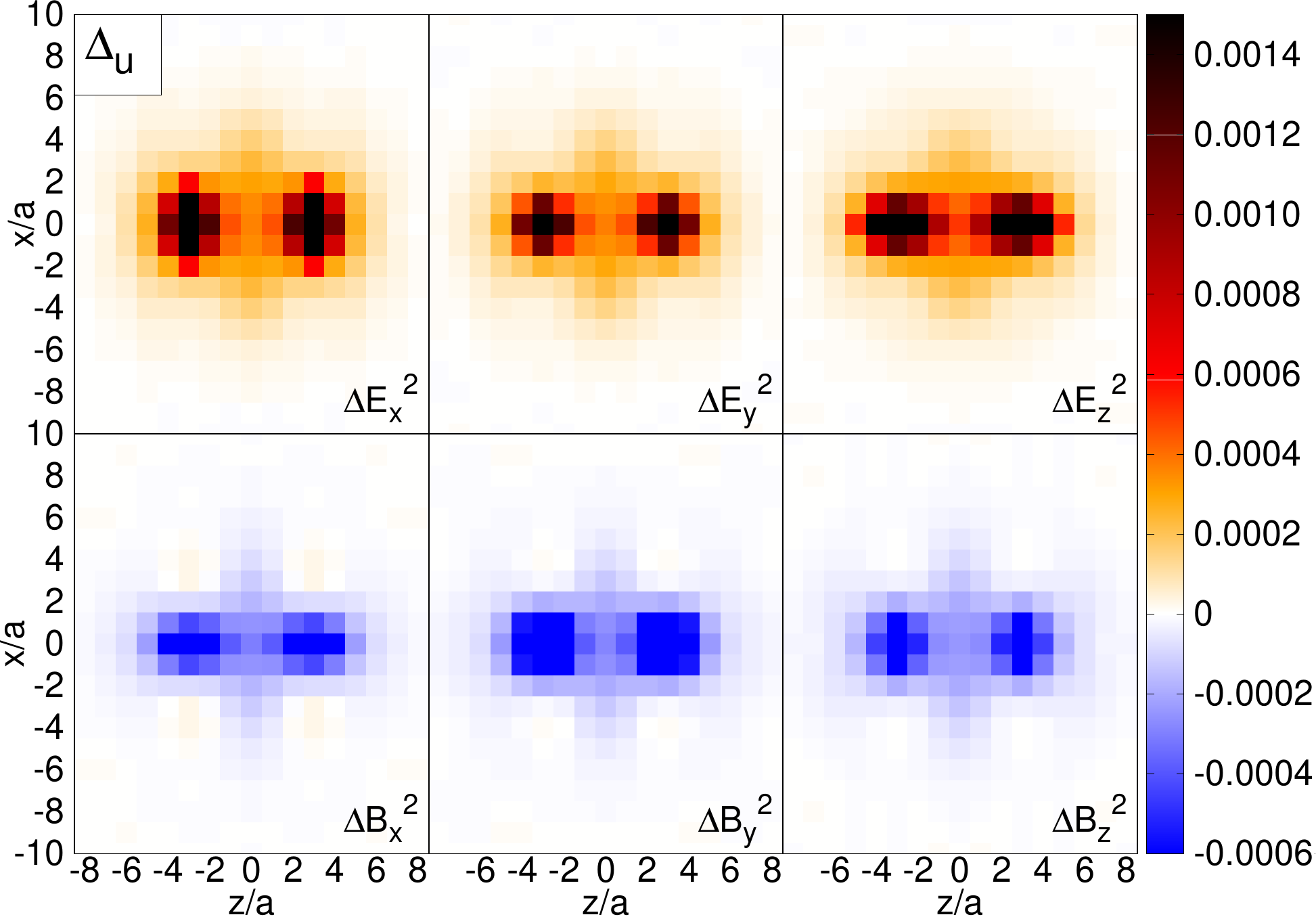}
\includegraphics[width=7.0cm]{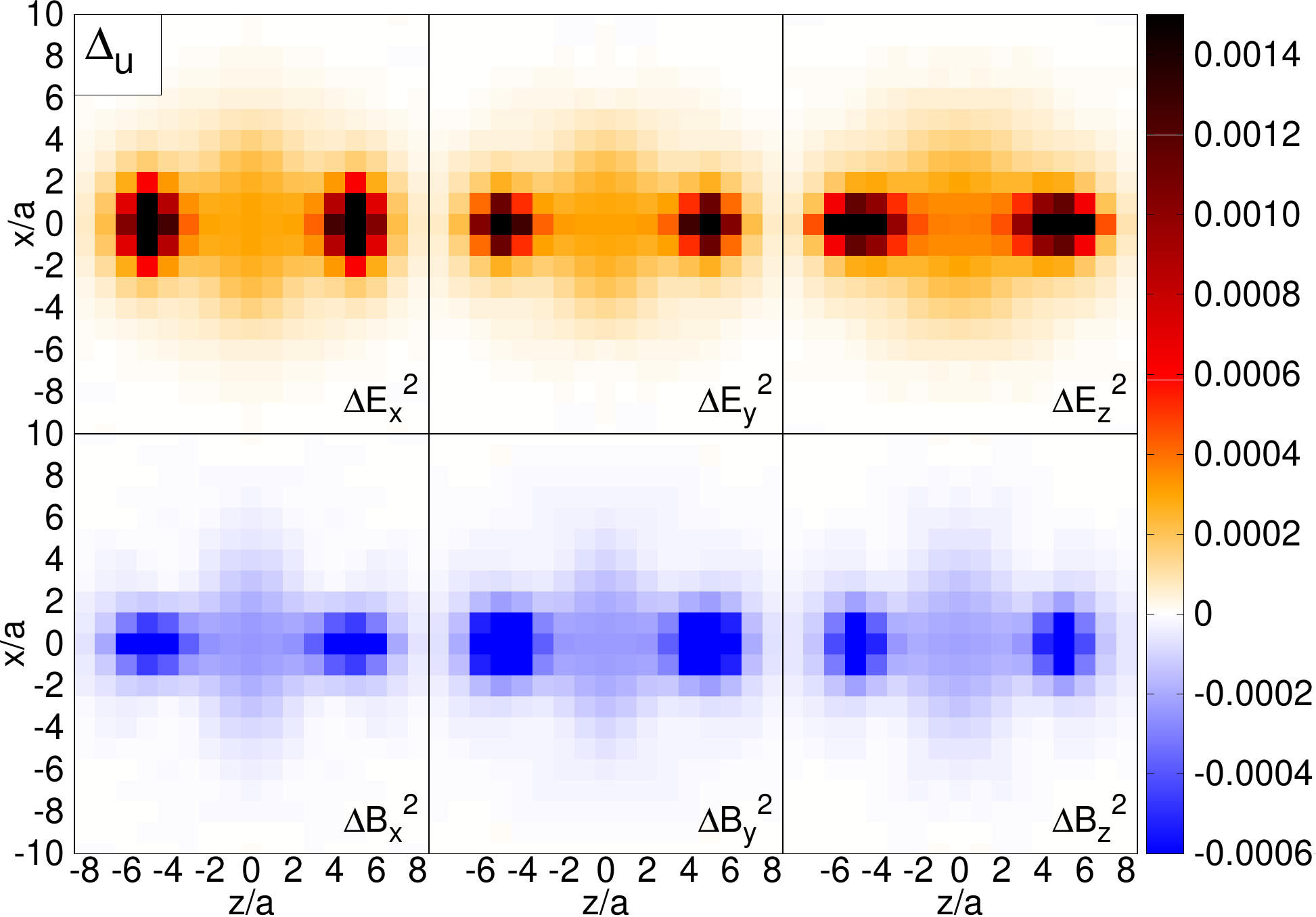}
\end{center}
\caption{\label{FIG_sep_Pi_Delta_}Flux densities $\Delta F_{j,\Lambda_\eta}^2(r;\mathbf{x} = (x,0,z))$, $j = x,y,z$ in the separation plane for gauge group SU(3) and sectors $\Lambda_\eta = \Pi_g, \Pi_u, \Delta_g, \Delta_u$. \textbf{(left)} $Q \bar{Q}$ separation $r = 6 \, a$. \textbf{(right)} $Q \bar{Q}$ separation $r = 10 \, a$.}
\end{figure}

Hybrid static potential flux densities for SU(3) gauge theory are shown in Figure~\ref{FIG_med_10_} to Figure~\ref{FIG_sep_Pi_Delta_}. Qualitatively, these plots are very similar to the corresponding SU(2) plots in Figure~\ref{FIG_med_10} to Figure~\ref{FIG_sep_Pi_Delta}. For a discussion see section~\ref{SEC386}.

\FloatBarrier

% ********************
% ********************
% ********************
% ********************
% ********************

\newpage

\section*{Acknowledgements}

We thank Charlotte Meyerdierks for contributions at an early stage of this work \cite{CM2017}. We acknowledge useful discussions with Pedro Bicudo and Francesca Cuteri.

C.R.\ acknowledges support by a Karin and Carlo Giersch Scholarship of the Giersch foundation. O.P.\ and M.W.\ acknowledge support by the DFG (German Research Foundation), grants PH 158/4-1 and WA 3000/2-1. M.W.\ acknowledges support by the Heisenberg Programme of the DFG (German Research Foundation), grant WA 3000/3-1.

This work was supported in part by the Helmholtz International Center for FAIR within the framework of the LOEWE program launched by the State of Hesse.

Calculations on the Goethe-HLR and on the on the FUCHS-CSC high-performance computer of the Frankfurt University were conducted for this research. We would like to thank HPC-Hessen, funded by the State Ministry of Higher Education, Research and the Arts, for programming advice.

% ********************
% ********************
% ********************
% ********************
% ********************

% ********************

\end{document}